\documentclass[12pt,preprintnumbers,nofootinbib,amsmath,amssymb]{revtex4}
\usepackage{float}
\usepackage{indentfirst}
\usepackage{amsmath,amssymb,epsfig,subfigure}
\usepackage{graphicx}
\usepackage[export]{adjustbox}
\usepackage{color}
\usepackage{multirow}
\usepackage{booktabs}
\usepackage{changes}
\usepackage{footnote}
\usepackage{mathrsfs} 
\usepackage[shortcuts]{extdash}

\usepackage{hyperref}
\hypersetup{colorlinks=true,linkcolor=blue,citecolor=magenta}
\begin{document}
\renewcommand{\baselinestretch}{1.15}
\makeatletter
\renewcommand{\thesubfigure}{\alph{subfigure}}
\renewcommand{\@thesubfigure}{(\thesubfigure)\hskip\subfiglabelskip}
\makeatother

\title{Orbits of photon in Bardeen-boson stars and their frozen states}

\preprint{}

\author{Long-Xing Huang$^{1,2,3}$, Shi-Xian Sun$^{1,2,3}$, Yu-Peng Zhang$^{1,2,3}$, \\Zhen-Hua Zhao$^{4}$, and Yong-Qiang Wang$^{1,2,3,}$\footnote{yqwang@lzu.edu.cn, corresponding author}}

\affiliation{$^{1}$Institute of Theoretical Physics and Research Center of Gravitation, Lanzhou University, Lanzhou 730000, China\\
	$^{2}$Key Laboratory of Quantum Theory and Applications of MoE, Lanzhou University, Lanzhou 730000, China\\
    $^{3}$Lanzhou Center for Theoretical Physics and Key Laboratory of Theoretical Physics of Gansu Province, Lanzhou University, Lanzhou 730000, China\\
     $^{4}$Department of Applied Physics, Shandong University of Science and Technology, Qingdao 266590,  China}


\begin{abstract}
In a recent study~\cite{Wang:2023tdz}, the Bardeen-boson star (BBS) model involving a scalar field minimally coupled to Einstein gravity and a Bardeen's nonlinear electromagnetic field was investigated. It was found that when the magnetic charge $q$ of the electromagnetic field exceeds a certain critical value $q_c$, a frozen Bardeen-boson star (FBBS) can be obtained with the frequency approaching zero. In this paper, we study the null orbits in the background of the general BBS and FBBS. We find that similar to the boson star (BS), all BBSs do not have the event horizon and possess complete null geodesics, allowing photons to move throughout the entire spacetime of BBS. Among these BBSs, the FBBSs whose spacetime is very similar to that of black holes are particularly special. The null orbits around the FBBSs exhibit sharp deflections near the critical horizon while becoming nearly straight inside the critical horizon. Furthermore, the photon in the background of FBBSs moves for a very long time inside the critical horizon from the perspective of an infinity viewer.
\end{abstract}


\maketitle
\newpage

\section{INTRODUCTION}\label{sec: introduction}

    At present, a large number of observations suggest that there are extremely compact and massive objects in our universe~\cite{KAGRA:2021duu}. The compactness of these exotic ultra-compact objects (UCO) exceeds that of the most compact directly observable objects currently known — neutron stars. An example is the bright radio source Sagittarius A* located at the center of other spiral and elliptic galaxies. According to the Keplerian orbits of stars in its vicinity, its mass can be estimated to be about $4.1\times10^6$ $M_\text{sun}$ ($M_\text{sun}$ is the mass of the Sun) with a radius no larger than $6.25$ light-hours~\cite{Ghez:2003qj,Ghez:2008ms} ($6.7$ billion kilometers). However, the mass of neutron stars is below $2.2$ $M_\text{sun}$ and its radius above $9.27\times10^{-9}$ light-hours~\cite{Rhoades:1974fn, Kalogera:1996ci, NANOGrav:2019jur, Kunz:2022wnj} ($10$ - $20$ kilometers). From the well-developed physical models that match the observational data of these ultra-compact objects, the black hole (BH) of general relativity seems to be the most promising theoretical candidate for these compact objects. However, due to the existence of their defining properties — the event horizon and the singularity - BHs face significant challenges in theory and are elusive in observation. This allows for the possibility that these compact objects may, in reality, be some other kind of horizonless exotic ultra-compact objects whose phenomenology is sufficiently similar to that of BHs.

    In this context, the development of astronomical observation technology has provided an important window for us to probe the nature of compact objects, which will help us to clarify what types of compact objects these exotic UCOs are. In this respect, the study of photon orbits around the compact object is crucial. At present, we mainly obtain spacetime geometry close to these exotic compact objects through gravitational wave astronomy~\cite{Hild:2011np,Hobbs:2009yy,eLISA:2013xep,Abbott2016a,LIGOScientific:2018jsj,LIGOScientific:2020iuh,LIGOScientific:2020ufj,LIGOScientific:2020kqk,KAGRA:2021duu} and large baseline interferometry measurements of the center of the galaxy using the Event Horizon Telescope (EHT)~\cite{EventHorizonTelescope:2019dse,EventHorizonTelescope:2022wkp}. Both methods are closely related to photon orbits. In the gravitational wave channel, the unstable circular orbit of the photon determines the early-time ringdown of the perturbed compact objects~\cite{Goebel}, corresponding to the post-merger part of the gravitational wave transients recently detected by LIGO~\cite{LIGOScientific:2016aoc,LIGOScientific:2016sjg,Cardoso:2016rao,Cunha:2017eoe}. The EHT will study the shadows of compact objects~\cite{Falcke:1999pj}: the dark region projected by the compact object for an observer at asymptotic infinity and the radius of the shadow is closely related to the unstable circular photon orbit around the compact object. In recent years, these two experiments have helped us rule out many exotic UCO models that do not match astronomical observations. Yet, despite this, there are still many models (see Ref.~\cite{Cardoso:2019rvt}) of compact objects that conform to astronomical observations.

    Among these compact-object models, because the mechanism of the dynamical generation of boson stars (BS) is known~\cite{Liebling:2012fv,Brito:2015yga,Brito:2015yfh}, the boson star occupies a special place. The concept of BSs can be traced back from J. Wheeler’s geon theory~\cite{Wheeler:1955zz,Power:1957zz}. In the 1960s, the theoretical framework of the boson star was established using the Einstein-scalar model~\cite{Feinblum:1968nwc,Brill:1964zz,Ruffini:1969qy}. As the mimicker of exotic ultra-compact object in our universe~\cite{Guzman:2009zz}, recent studies have shown that the shadow image of the boson star can mimic the shadow image of a BH~\cite{Vincent:2015xta,Olivares:2018abq,Vincent:2020dij,Herdeiro:2021lwl,Pombo:2023ody,Sengo:2024pwk}. In addition, it can serve as a gravitational wave source behind the binary black hole merger events~—~GW190521~\cite{CalderonBustillo:2020fyi,LIGOScientific:2020iuh}. Since the boson star was proposed, it has received a lot of attention, and various BSs have been proposed - see Refs.~\cite{Schunck:2003kk,Liebling:2012fv} for a review.

    In 2023, the Einstein-scalar model coupled with a nonlinear electromagnetic field was studied in Ref.~\cite{Wang:2023tdz}. It includes two special cases: one is the pure boson star solution, and another is the Bardeen spacetime solution~\cite{Bardeen:1968}. The Bardeen spacetime solution was first proposed by Bardeen, and when the magnetic charge $q$ of the nonlinear electromagnetic field (referred to ``Bardeen's nonlinear electromagnetic field") exceeds a certain threshold $q_{bc}$, the solution actually is a regular BH solution~\cite{Ayon-Beato:1998hmi,Ayon-Beato:2000mjt} (see Ref.~\cite{Shirokov1948,Duan:1954bms,Sakharov:1966aja,Gliner1966} for the early studies about the regular BH and see Ref.~\cite{Ansoldi:2008jw,Lan:2023cvz,Torres:2022twv,Carballo-Rubio:2025fnc} for a review). If both the Bardeen electromagnetic field and the scalar field exist simultaneously, this model is essentially a boson star model coupled with a nonlinear electromagnetic field, and the solution is a horizonless solution known as the Bardeen-boson star (BBS). They found that the introduction of a scalar field seems to make the magnetic charge lower than $q_{bc}$. Moreover, in particular, when the magnetic charge exceeds another critical value $q_c$ (where $q_c<q_{bc}$), as the frequency of the scalar field approaches zero, a special solution emerges. This solution has a critical position $r_{cH}$. The scalar field is almost distributed within $r_{cH}$. Moreover, for the metric function, $1/g_{rr}$ is very close to zero at $r_{cH}$, and $-g_{tt}$ approaches zero for $r\leq r_{cH}$. It is worth noting that since at $r_{cH}$, $1/g_{rr}$ is only very close to zero but not exactly zero, this solution does not have an event horizon. Therefore, it is not classified as the BH but rather as the “frozen star”~\cite{Oppenheimer:1939ue,zeldovichbookorpaper,Ruffini:1971bza} (this phenomenon can also be generalized, see Ref.~~\cite{Huang:2023fnt, Yue:2023sep,Ma:2024olw,Huang:2024rbg,Chen:2024bfj, Zhang:2024ljd,Wang:2024ehd,Sun:2024mke,Zhao:2025hdg}). 
    
    Considering the importance of photon orbits for compact stars in astronomical observations, in this paper, we extend the study of the BBS in Ref.~\cite{Wang:2023tdz}, exploring the photon orbits around BBSs by using the null geodesic equation and the radial effective potential. Due to the FBBSs having very similar spacetime properties to BHs, we will mainly focus on the photon orbit of the FBBS.  
    
    The paper is organized as follows. In Sect.~\ref{sec: model}, we provide an introduction to the model of the Einstein-Bardeen theory coupled to a complex scalar field. In Sect.~\ref{sec: geo}, we discuss the null geodesics equation and the radial effective potential. The photon orbits of BBSs are discussed in Sect.~\ref{sec: results}. The conclusion and discussion are given in Sect.~\ref{sec: conclusion}.

\section{The model of Bardeen action with scalar field}\label{sec: model}
\subsection{The action and field equations}
 Firstly, we provide a brief overview of the model of the BBS. It can be described by the following action (we use units with $c = \hbar = 1$)
\begin{equation}
     S=\int d^4 x \sqrt{-g}\left [\frac{R}{16\pi G}+\mathcal{L}^{(1)}+\mathcal{L}^{(2)}\right],
         \label{eq:action}
\end{equation}
where $\mathcal{L}^{(1)}$ and $\mathcal{L}^{(2)}$ are the Lagrangians of the Bardeen's nonlinear electromagnetic field~\cite{Ayon-Beato:2000mjt} and the scalar field, respectively:
\begin{equation}
     \mathcal{L}^{(1)}=-\frac{3}{2s}\left( \frac{\sqrt{q^2F^{\mu\nu}F_{\mu\nu}}}{\sqrt{2}+\sqrt{q^2 F^{\mu\nu}F_{\mu\nu}}} \right)^{\frac{5}{2}},
\end{equation}
\begin{equation}
     \mathcal{L}^{(2)}=-\left(\nabla^{\mu}\psi \right)\left(\nabla_{\mu}\psi \right)-\mu^2\psi\psi^*,
\end{equation}
with $R$ the Ricci curvature, $G$ Newton’s constant. $F^{\mu \nu}$ is the electromagnetic field strength of the nonlinear electromagnetic field $A_{\mu}$, given in terms of the potential 1-form as
\begin{equation}
   F^{\mu \nu}:=\partial_{\mu} A_{\nu}-\partial_{\nu} A_{\mu}.
\end{equation}
The constants $s$, $q$, and $\mu$ are three independent parameters, where $q$ represents the magnetic charge and $\mu$ is the scalar field mass.

Variation of the action (\ref{eq:action}) respect to the metric $g_{\mu \nu}$, nonlinear electromagnetic field $A_{\mu}$, and scalar field $\psi$ results in the following field equations:
\begin{equation}
    R_{\mu \nu} - \frac{1}{2}g_{\mu \nu}R - 8\pi G (T^{(1)}_{\mu \nu}+T^{(2)}_{\mu \nu})=0,
    			\label{eq:einstein}
\end{equation}
\begin{equation}
    \nabla_{\mu}\left(\frac{\partial \mathcal{L}^{(1)}}{\partial \mathcal{F}} F^{\mu \nu}\right)=0,
    			\label{eq:equationBardeen}
\end{equation}
\begin{equation}
    D^{\mu}\psi-\mu^2\psi=0,
    			\label{eq:equationKG}
\end{equation}
where $\mathcal{F}=\frac{1}{4}F^{\mu \nu} F_{\mu \nu}$, the energy–momentum tensor $T_{\mu \nu}=T^{(1)}+T^{(2)}$ is the source of the gravitational field, includes contributions from both the nonlinear electromagnetic fields $T^{(1)}_{\mu \nu}$ and the complex scalar field $T^{(2)}_{\mu \nu}$, in which the nonlinear electromagnetic and scalar contributions are, respectively:
\begin{equation}
    T^{(1)}_{\mu \nu}=-\frac{\partial \mathcal{L}^{(1)}}{\partial \mathcal{F}} F_{\mu \lambda} F_\mu{ }^\lambda+g_{\mu \nu} \mathcal{L}^{(1)},
\end{equation}
\begin{equation}
    T^{(2)}_{\mu \nu}=\frac{1}{2}\left[\left(D_\mu \psi\right)^*\left(D_\nu \psi\right)+\left(D_\nu \psi\right)\left(D_\mu \psi\right)^*-g_{\mu \nu}\left(\left(D_\lambda \psi\right)\left(D^\lambda \psi\right)^*+m^2|\psi|^2\right)\right].
\end{equation}
%

\subsection{The ansatz and the boundary conditions}
In this paper, we shall focus on spherically symmetric configurations, thus we adopt the following spacetime metric ansatz:
\begin{equation}
    d s^2=-n(r) \sigma^2(r) d t^2+\frac{d r^2}{n(r)}+r^2\left(d \theta^2+\sin ^2 \theta d \varphi^2\right),
    \label{eq:metric}
\end{equation}
in which the two functions $n(r)$ and $\sigma(r)$ are radial functions.
Then, for the scalar field and the nonlinear electromagnetic field, we employ the following ansatz:
\begin{equation}
    A_{\mu}dx^{\mu}=p\cos(\theta)d\varphi,\quad \psi=\phi(r) e^{-i\omega t}, 
    \label{eq:ansatz}
\end{equation}
here $\phi(r) $ is a real radial function, and $\omega$ is a real constant corresponding to the frequency of oscillation of the scalar field.

Using the metric (\ref{eq:metric}) and the ansatz (\ref{eq:ansatz}) for the scalar field and the electromagnetic field into the equations (\ref{eq:einstein}~-~\ref{eq:equationKG}), we can get the following equations for $n(r)$, $\sigma(r)$, and $\phi(r)$
\begin{align}
     n^{\prime}+n\left(\frac{1}{r}+8\pi Gr\phi^{\prime 2}\right)+\frac{8\pi Gr \phi^{2}}{n\sigma^2}\omega^2+8\pi Gr\mu^2\phi^{ 2}+\frac{12\pi Gr p^5}{(p^2+r^2)^{5/2}s}-\frac{1}{r}=0,\label{eq:n}
\end{align}
\begin{equation}
    \frac{\sigma^\prime}{\sigma}-8\pi Gr\left(\phi^{\prime 2}-\frac{ \phi^2\omega^2}{n^2\sigma^2} \right)=0,
\end{equation}
 \begin{equation}
\phi^{\prime \prime}+\phi^{\prime}\left( \frac{n^{\prime}}{n}+\frac{\sigma^{\prime}}{\sigma}+\frac{2}{r}\right)+\left(\frac{\omega^2}{n\sigma^2}-\mu^2\right)\frac{\phi}{n}=0.\label{eq:phi}
\end{equation}
The solutions of the four coupled ordinary differential equations (ODEs) can be obtained numerically, after imposing suitable boundary conditions, as we now describe.

The boundary conditions can be obtained by considering the approximate construction of solutions on the boundary of the domain of integration together with the assumptions of regularity and asymptotic flatness. For the metric functions, they are
\begin{equation}
    n(0)=1,\quad \sigma(0)=\sigma_0,\quad n(\infty)=1-\frac{2GM}{r},\quad \sigma(\infty)=1,
\end{equation}
where $M$ is the Arnowitt-Deser-Misner (ADM) mass. The constant $\sigma_0$ and the ADM mass can be fixed by solving the system of ODEs. In addition, we required the following boundary conditions for the complex scalar field:
\begin{equation}
    \left.\frac{d \phi}{d r}\right|_{r=0}=0,\quad \phi(\infty)=0,\quad \left.\frac{dV}{d r}\right|_{r=0}=0,\quad V(\infty)=0.
\end{equation}
\subsection{Solutions}
In order to be able to solve numerically in the full space region, we introduce a new radial coordinate
\begin{equation}
    \bar{r}=\frac{r}{r+1},
\end{equation}
which maps the semi-infinite region $\left[0,\infty\right)$ onto the unit interval $\left[0, 1\right]$. After obtaining the numerical solution, we can use the inverse transformation $r=\frac{\bar{r}}{1-\bar{r}}$ to replace the $\bar{r}$ coordinates into the $r$ coordinates. 
In this work, we use the finite element method to solve nonlinear ordinary differential equations, and the number of grid points in the integration region $[0, 1]$ is $1000$. The iterative method we use is the Newton-Raphson method, and to ensure the accuracy of the computed results, we require the relative error to be below $10^{-5}$.

In addition, to facilitate numerical computations, we take the following dimensionless variables:
\begin{equation}
    r\rightarrow \frac{r}{\mu},\quad \omega\rightarrow\omega\mu,\quad q\rightarrow\frac{q}{\mu},\quad s\rightarrow\frac{4\pi  s}{\mu^2M_{Pl}^2},  \quad \phi\rightarrow\frac{M_{Pl}}{\sqrt{4\pi }}\phi,                     
    \end{equation}
where $M_{Pl}=1/\sqrt{G}$ is the Planck mass. During the solution process, this scaling is equivalent to selecting $\mu=1$ and $G=1/(4\pi)$. Furthermore, without loss of generality, we set $s=0.2$ (except where specifically indicated). Consequently, the solution of the Baredeen boson star is controlled only by the parameter $\omega$.

Before numerically solving these equations, it is worth noting that there are two special solutions for these equations, which can be obtained from simple analysis. Firstly, when $q=0$, the action (\ref{eq:action}) reduces to Einstein-scalar theory, and the corresponding soliton solution of these equations is called the BS. Secondly, in the case of a vanishing complex scalar field with $q\neq0$, these equations provide the solutions of the Bardeen spacetime, in which the BH solutions are also known as the Bardeen BH~\cite{Bardeen:1968}. The metric of the Bardeen spacetime is
\begin{equation}
d s^2=-f(r) d t^2+f(r)^{-1} d r^2+r^2\left(d \theta^2+\sin ^2(\theta) d \varphi^2\right)
\end{equation}
here, 
\begin{equation}
    f(r)=1-\frac{q^3r^2}{s(r^2+q^2)^{3/2}}.
\end{equation}
The asymptotic behavior of the metric function $f(r)$ at infinity is given by
\begin{equation}
    f(r)=1-2\frac{q^3}{2s}/r+O(1/r^3).
\end{equation}
From the second term, one can deduce the ADM mass $M$ of this configuration as $M = q^3/2s$ from the term $1/r$. In addition, the function $f(r)$ exhibits a local minimum value at $ r=\sqrt{2}q$. Therefore, in case where $q<3^{3/4}\sqrt{\frac{s}{2}}$, no horizons are present. When $q=3^{3/4}\sqrt{\frac{s}{2}}$, degenerate horizons are observed, and for $q>3^{3/4}\sqrt{\frac{s}{2}}$, two distinct horizons exist. 

\begin{figure}[!htbp]
		\begin{center}
		\subfigure{ 
			\includegraphics[height=.28\textheight,width=.30\textheight, angle =0]{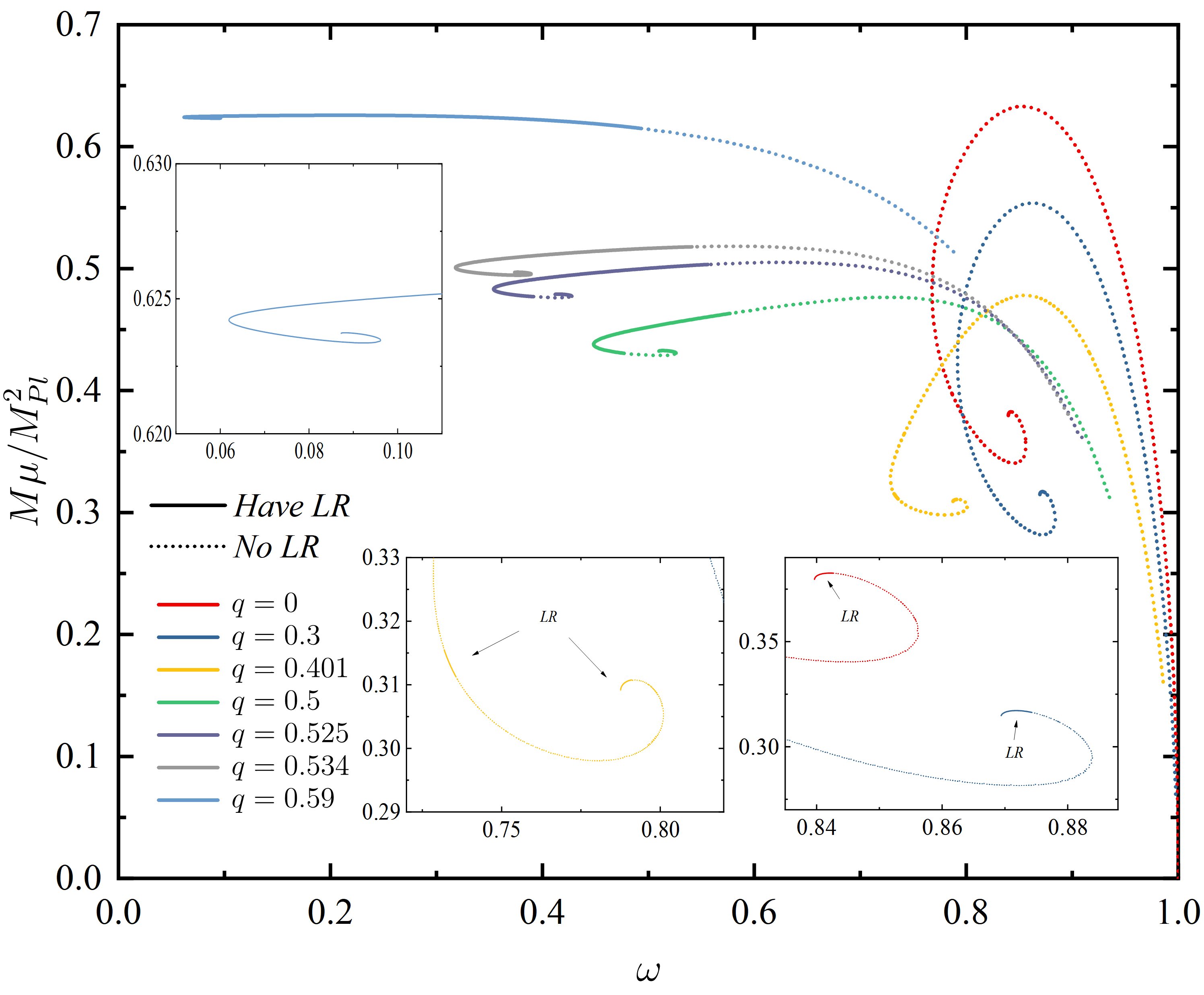}
			\label{fig:smallmatter}
		}	 
  		\subfigure{  
			\includegraphics[height=.28\textheight,width=.30\textheight, angle =0]{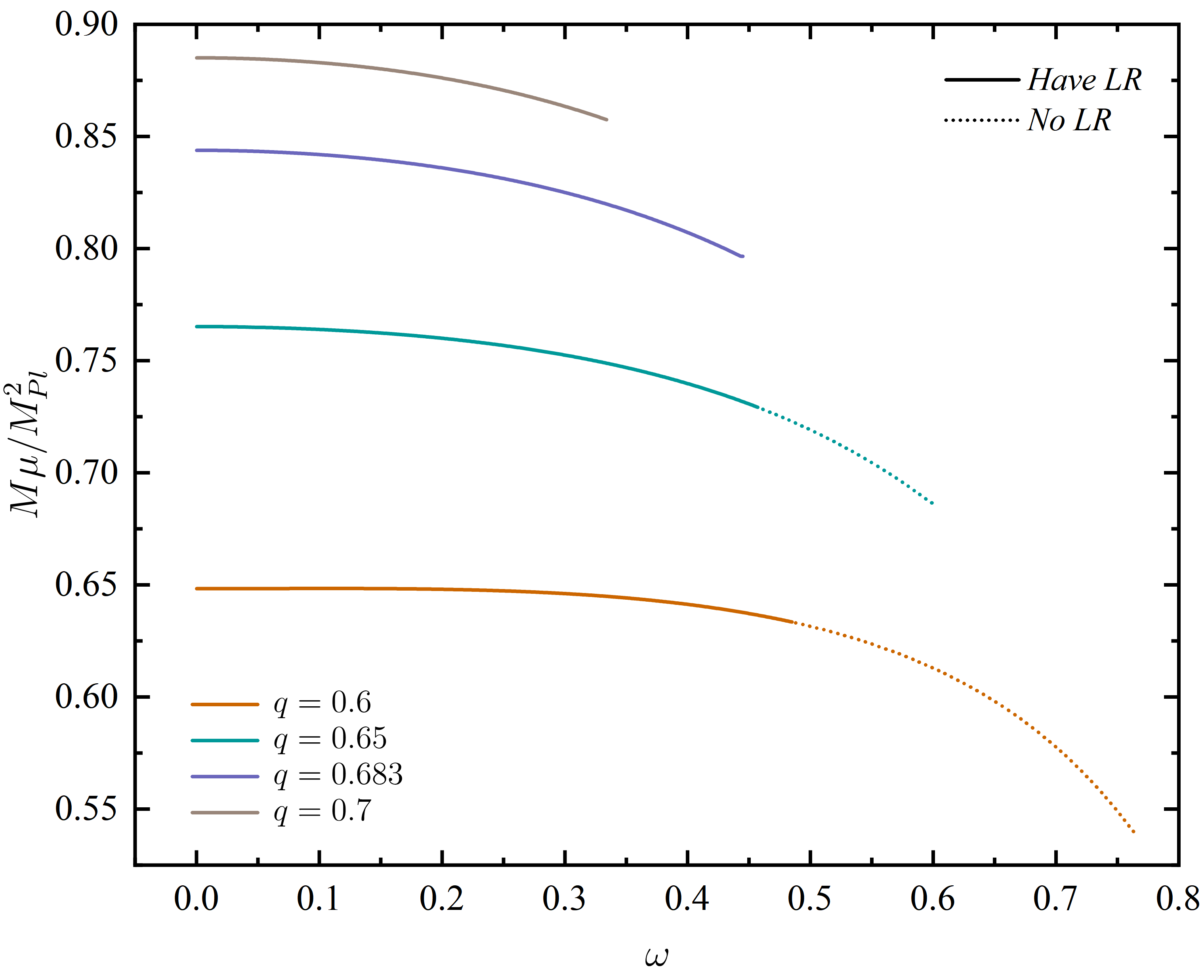}
			\label{fig:largematter}
		}	 
  		\end{center}	
		\caption{ The mass $M$ versus the frequency $\omega$ with the different values of magnetic charges $q$. The solid line indicates the solution exists the LR.}
	\label{fig:matter}	
		\end{figure}

We now show the numerical solution of the BBS model. Figure~\ref{fig:matter} shows the ADM mass $M$ of BBSs vs. frequency, $\omega$, diagram. For convenience, we call the curve in this diagram ``the $M(\omega)$ curve". The domain of existence of the BBS solution can correspond to these smooth $M(\omega)$ curves in this diagram. These curves start from the position of maximum frequency $\omega_{max}$ (corresponding to the minimal mass), in which limit the scalar fields become very diluted and the solution is almost pure Bardeen solution. From Fig.~\ref{fig:matter}, we can first observe that as the magnetic charge increases, $\omega_{max}$ continuously decreases. Secondly, contrasting between the left panel and right panel of Fig.~\ref{fig:matter} shows that the $M(\omega)$ curve of the BBSs with magnetic charges less than $0.6$ are different from those with magnetic charges greater than $0.6$. In the case of $q<0.6$, the BBS is the same as BS, the relationship between the ADM mass and the frequency is not monotonic, and the solutions form a characteristic spiraling curve. Following the spiral, one finds several backbendings, each defining the end of a branch or the beginning of another branch. As seen in the left panel of Fig.~\ref{fig:matter}, we find three branches in this case. We conjecture that, as the computational accuracy increases, these spirals may obtain further backbends and eventually, at their center, describe approach a critical singular solution. For the case $q\geq0.6$, from the right panel of Fig.~\ref{fig:matter}, it can be seen that the $M(\omega)$ curve has no spiral structure and only the first branch exists. In addition, unlike the case of $q<0.6$, the minimum value of the frequency $\omega_{min}$ of these $M(\omega)$ curves can be very close to zero. These minimum frequencies corresponding to the left endpoints of the $M(\omega)$ curves with different $q$ in the right panel of Fig.~\ref{fig:matter} are all $0.0001$. Likely, the frequency can be smaller. However, due to the limitation of the calculation accuracy, it will be very difficult to obtain solutions with a smaller frequency, and we have not been able to follow the $M(\omega)$ curve that far.

\begin{figure}[!htbp]
		\begin{center}
		\subfigure{ 
			\includegraphics[height=.28\textheight,width=.30\textheight, angle =0]{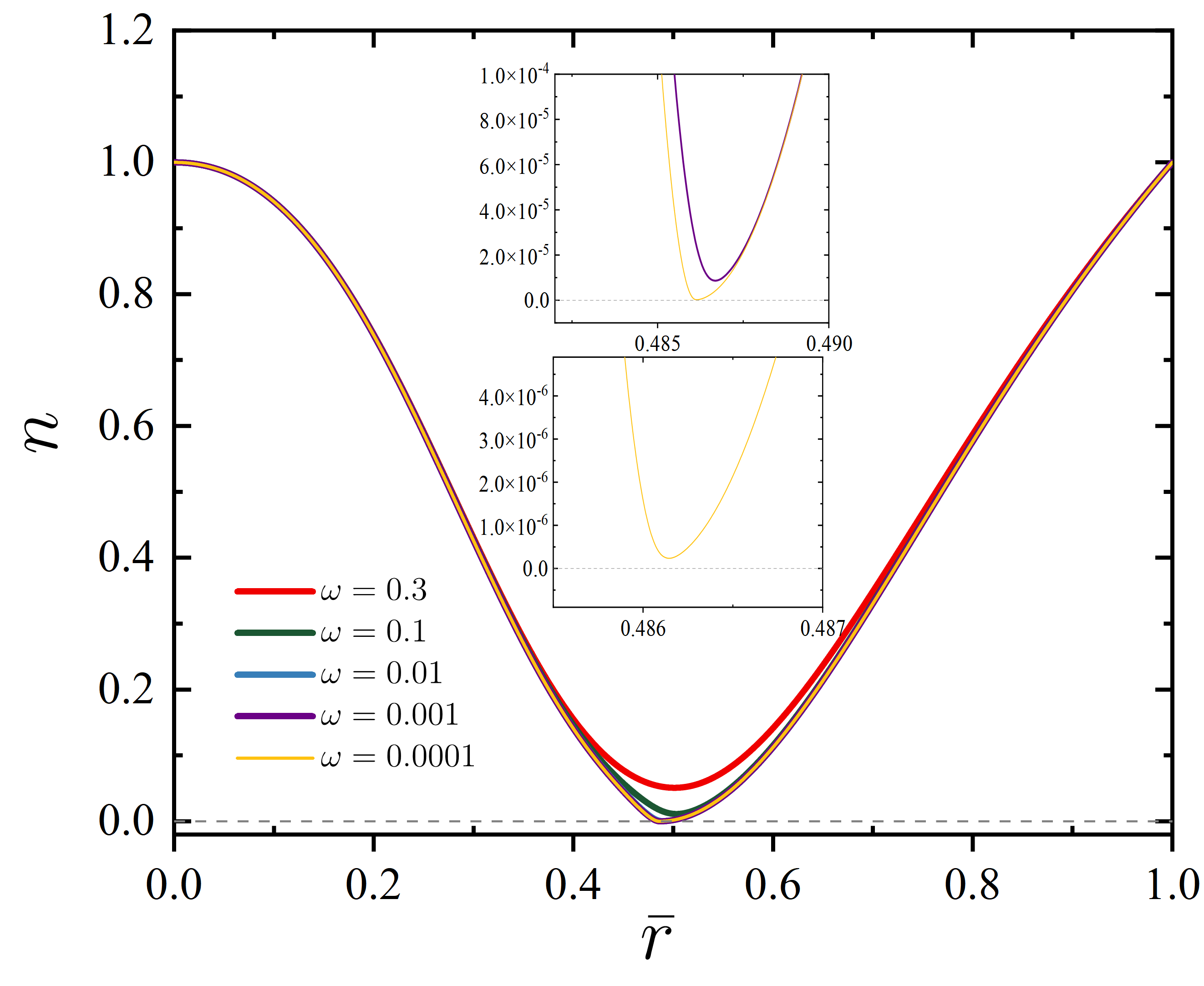}
			\label{fig:n}
		}	 
  		\subfigure{  
			\includegraphics[height=.28\textheight,width=.30\textheight, angle =0]{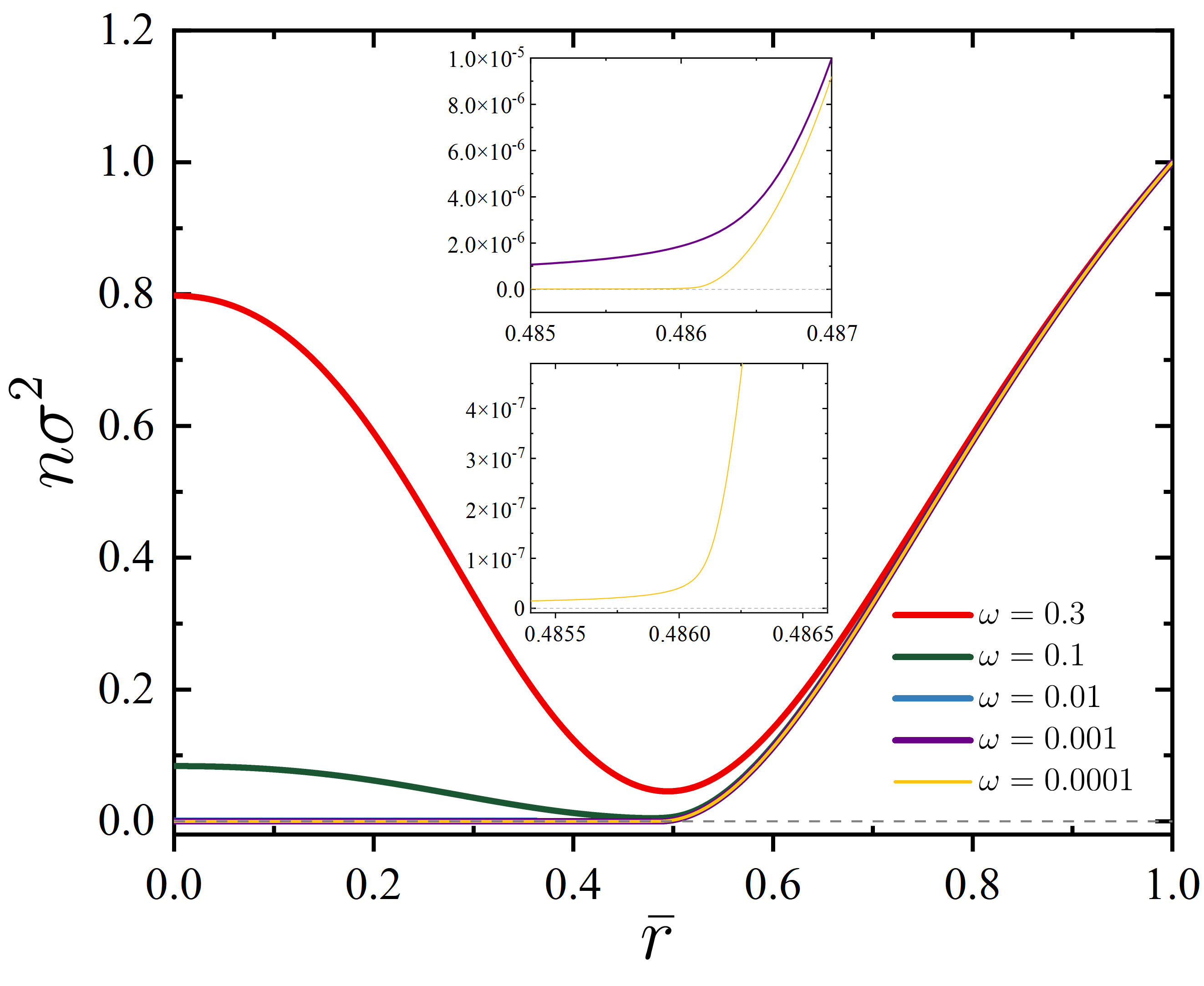}
			\label{fig:no2}
		}	
    \quad
        \subfigure{ 
			\includegraphics[height=.28\textheight,width=.30\textheight, angle =0]{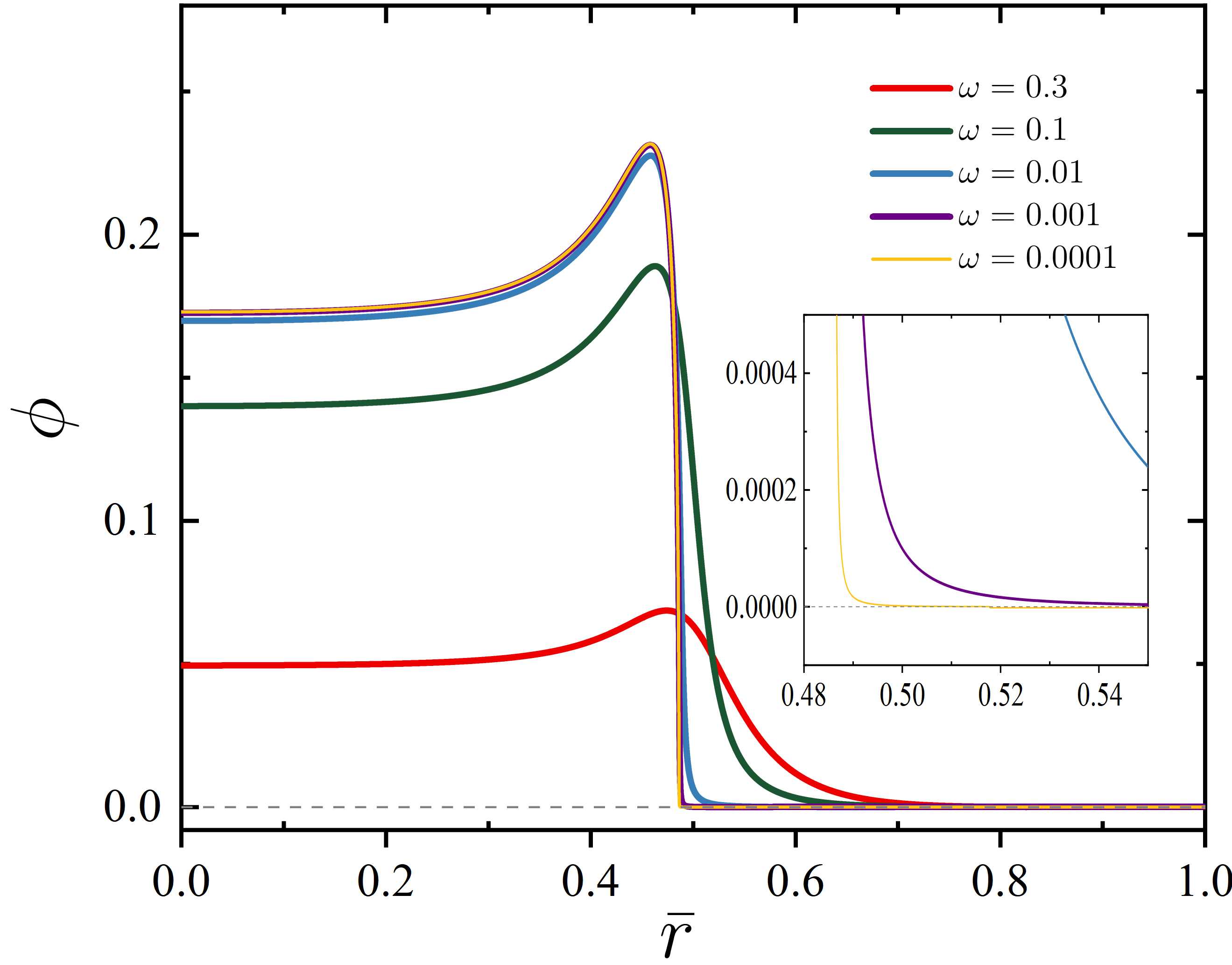}
			\label{fig:phi}
		}	 
  		\subfigure{  
			\includegraphics[height=.28\textheight,width=.30\textheight, angle =0]{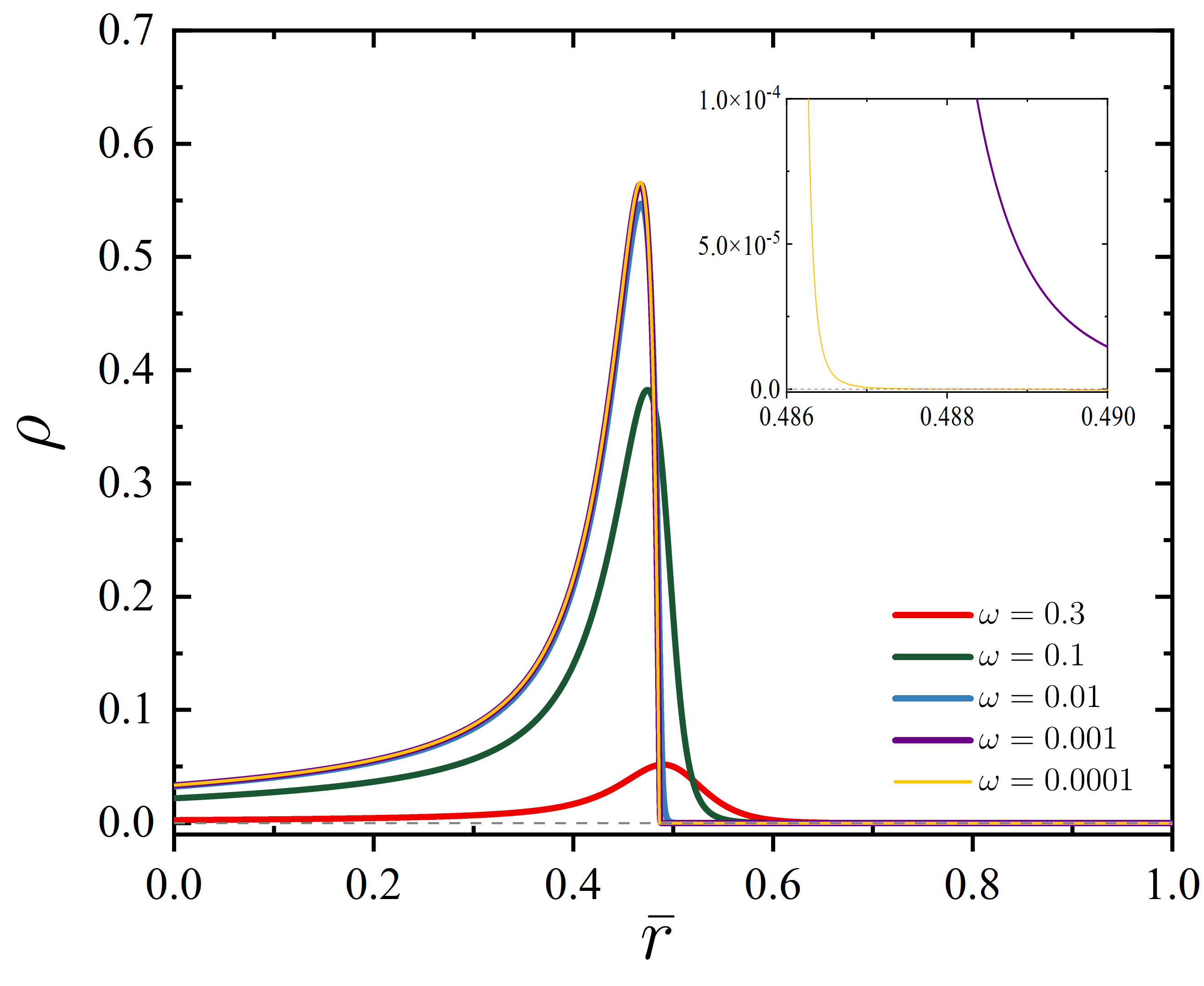}
			\label{fig:rhowith07}
		}	
  		\end{center}	
		\caption{Top panel: The metric field function $g^{rr}=n$ (left) and $-g_{tt}=n\sigma^2$ (right) as a function of $\bar{r}$. Bottom panel: The radial distribution of the field functions and the energy density. All solutions have $q=0.7$.}
	\label{fig:function}	
		\end{figure}

Figure~\ref{fig:function} shows the distribution of the metric field function $g^{rr}=1/g_{rr}=n$, $-g_{tt}=n\sigma^2$, the scalar field function $\phi$, and energy density $\rho=(T^{(2)})^0_0$ of the scalar field. As can be seen from this figure, as the frequency decreases, the minimum value of the metric function $g^{rr}$ and $-g_{tt}$ decreases (denoted as $g^{rr}_{min}$ and $-(g_{tt})_{min}$), while the curve of the function of the scalar field the energy density become steeper and steeper, and when the frequency decreases to $0.0001$, a particular solution appears\footnote{In fact, this is a continuous process, and as can be seen from the Fig.~\ref{fig:function} the solution with $\omega=0.001$ also has some of the properties of this type of solution. However, for convenience in the discussion, we take $\omega\leq0.0001$ as the criterion for distinguishing this special class of solutions.}. On the one hand, for the metric function (from the top panel), $g^{rr}$ of this solution is very close to zero (below $10^{-5}$) near a certain critical position $\bar{r}_{cH}$ (at $\bar{r}_{cH}$, $g^{rr}$ reaches its minimum value), while another component of the metric fired $-g_{tt}$ approaches zero within $\bar{r}_{cH}$ (below $10^{-7}$). On the other hand, for the matter field, it can be seen from the bottom panel that the scalar field function of this solution is also almost distributed in the range of $\bar{r}<\bar{r}_{cH}$, and the energy density has a very steep wall near the $\bar{r}_{cH}$ region.\footnote{It should be noted that these functions at $\bar{r}_{cH}$ are still smooth (see inset of Fig.~\ref{fig:function}).} Since the properties of the metric function, the object inside $\bar{r}_{cH}$ will take an extremely long time to move a very short distance as if they are frozen leading from the view of an observer at infinity and the solution with $\omega=0.0001$ is called the frozen Bardeen-boson star (FBBS). In addition, considering that the FBBS is similar to the BH and $\bar{r}_{cH}$ is very close to forming the event horizon, we refer to this position $\bar{r}_{cH}$ as a critical location and can be termed as a ``critical horizon".

\section{NULL GEODESICS AROUND THE BBS} \label{sec: geo}
Studying the null geodesics in a given spacetime geometry is very important to understand the properties of the spacetime, and we can obtain the astrophysical observable effects from the information of the null geodesics. In this section, we will investigate the motion of a test photon around BBSs by using the null geodesics.

The test photon motion in the geometry (\ref{eq:metric}) is ruled by the following geodesics equation 
\begin{equation}
    \mathcal{L}_{photon}=0
\end{equation}
where $\mathcal{L}_{photon}=\frac{1}{2}g_{\mu \nu}\dot{x}_\mu\dot{x}_\nu$ is the effective Lagrangian of the photon~\cite{Delgado:2021jxd} (dots denote derivatives with respect to the affine parameter $\lambda$). Due to spherical symmetry, for simplicity, we only consider the photon of BBSs lying on the equatorial plane (i.e., $\theta=\pi/2$). Thus, the effective Lagrangian can be simplified as
\begin{equation}
    \mathcal{L}_{photon}=\frac{1}{2}\left[g_{t t}(r,\theta=\pi/2)\dot{t}^2+g_{r r}(r,\theta=\pi/2)\dot{r}^2+g_{\varphi \varphi}(r,\theta=\pi/2)\dot{\varphi}^2\right].\label{eq:effectivelagrangian}
\end{equation}
The effective Lagrangian can define the four-momentum 
\begin{equation}
    p^{\mu}:=\frac{\partial\mathcal{L}_{photon}}{\partial x_{\mu}}.
\end{equation}
Since the static spherical symmetry means that the spacetime can possess a timelike killing vector $\xi^{\mu}=(\partial_t)^{\mu}$ and a spacelike Killing vector $\eta^\mu = (\partial_\varphi)^\mu$, we can identify two motion constants associated with the Killing vectors
\begin{equation}
    E\equiv -p_{t}=-g_{t\mu}(r,\theta=\pi/2)\dot{x}^{\mu},
\end{equation}
\begin{equation}
    L\equiv p_{\varphi}=g_{\varphi\mu}(r,\theta=\pi/2)\dot{x}^{\mu}=g_{\varphi\varphi}(r,\theta=\pi/2)\dot{\varphi}
\end{equation}
in which $E$ and $L$ are the energy per unit mass and angular momentum, respectively.
Considering these two constants and the effective Lagrangian (\ref{eq:effectivelagrangian}), the orbit of a photon around the BBSs is governed by the following set of geodesic equations
\begin{equation}
    \dot{t}=-\frac{E}{g_{tt}(r,\theta=\pi/2)}=\frac{E}{n(r)\sigma^2(r)},\label{eq:dott}
\end{equation}
\begin{align}
    \dot{r}&=\pm\sqrt{-\frac{1}{g_{rr}(r,\theta=\pi/2)}\left[\frac{E^2}{g_{tt}(r,\theta=\pi/2)}+\frac{L^2}{g_{\varphi \varphi}(r,\theta=\pi/2)}\right]}\notag\\&=\pm\sqrt{n(r)\left[\frac{E^2}{n(r)\sigma^2(r)}-\frac{L^2}{r^2}\right]},\label{eq:dotr}
\end{align}
\begin{equation}
    \dot{\varphi}=\frac{L}{g_{\varphi\varphi} (r,\theta=\pi/2)}=\frac{L}{r^2}.\label{eq:dotphi}
\end{equation}
Substituting the equations (\ref{eq:dott}) into (\ref{eq:dotr}) and (\ref{eq:dotphi}), eliminating the affine parameter $\lambda$, we can obtain the following two equations
\begin{align}
    \frac{dr}{dt}&=\frac{dr}{d\lambda}\frac{d\lambda}{dt}=\pm\sqrt{n(r)\left[\frac{E^2}{n(r)\sigma^2(r)}-\frac{L^2}{r^2}\right]}\frac{n(r)\sigma^2(r)}{E}\notag\\
    &=\pm n(r)\sigma^2(r)\sqrt{n(r)\left[\frac{1}{n(r)\sigma^2(r)}-\frac{b^2}{r^2}\right]},\label{eq:dtr}
\end{align}
\begin{equation}
    \frac{d\varphi}{dt}=\frac{d\varphi}{d\lambda}\frac{d\lambda}{dt}=\frac{L}{r^2}\frac{n(r)\sigma^2(r)}{E}=b\frac{n(r)\sigma^2(r)}{r^2}.\label{eq:dtphi}
\end{equation}
with $b\equiv L/E$ the impact parameter of the photon. Thus, when $L\neq0$ (or equivalent, $b\neq0$), the orbit equation of the photon can be expressed as:
\begin{equation}
    \frac{dr}{d\varphi}=\frac{dr}{dt}\frac{dt}{d\varphi}=\pm\frac{r^2}{b}\sqrt{n(r)\left[\frac{1}{n(r)\sigma^2(r)}-\frac{b^2}{r^2}\right]}.\label{eq:orbit}
\end{equation}
When $L=0$, considering that equation (\ref{eq:dtphi}) gives $\phi = constant$, the orbit of a photon can only be a straight line passing through the origin of the coordinates.

It is worth noting that one can rewrite (\ref{eq:dotr}) in the form as follows
\begin{equation}
    \sigma^2(r)\dot{r}^2+L^2\frac{n(r)\sigma^2(r)}{r^2}=E^2.\label{eq:energy}
\end{equation}
Therefore, it is convenient to introduce the radial effective potential~\cite{Alcubierre:2021psa}
\begin{equation}
    V_{eff}(r)\equiv\frac{n(r)\sigma^2(r)}{r^2},
\end{equation}
which is an easy way to determine the radial motion of a photon. Using the radial effective potential and the impact parameter, equation (\ref{eq:energy}) can be rewritten as
\begin{equation}
    \frac{\sigma^2(r)\dot{r}^2}{L^2}=\frac{1}{b^2}-V_{eff}.\label{eq:latterenergy}
\end{equation}
Since $\sigma^2(r)\dot{r}^2/L^2\geq0$, the photon is restricted to the regions where $V_{eff}\leq E^2/L^2=1/b^2$. When there exists a certain value $r_c$ such that $V_{eff}(r_{c})=n(r_{c})\sigma^2(r_{c})/r_c^2=1/b_{c}^2$, then according to the equation (\ref{eq:dotr}), the radial velocity of the photon is zero. Meanwhile, in particular, if the radial effective potential has an extremum at $r_c$ (i.e., $dV(r_c)/dr=0$), the radial acceleration of the photon will be equal to zero, and the photon will remain in $r=r_{LR}$. The corresponding circular orbits (i.e., $r=r_{LR}$) of the photon are referred to as the light ring (LR). Further, if the second derivative of the radial effective potential $d^2V(r_c)/dr^2<0$, the $r_c$ is the maximum point of the radial effective potential, which indicates that the LR is unstable. Conversely, if $d^2V(r_c)/dr^2>0$,  $r_c$ is a minimum point, then $r=r_c$ corresponds to a stable LR~\cite{Cunha:2017qtt}.

\section{ORBITS OF THE PHTON}\label{sec: results}
Now, we have all the information to obtain the orbits of the photon (null orbit) in the BBSs. Next, we will give the orbits around the photon particle in the following part. Before this, it is worth mentioning that circular orbits (i.e., LRs), as mentioned earlier, can be easily obtained through an analysis of the radial effective potential. Similar to boson stars~\cite{Cunha:2015yba}, solutions with the LR exist in parts of their domain of existence of the BBSs solution. Since the LR can only appear for very compact objects generally~\cite{Grandclement:2016eng}, these BBSs with the LR can be referred to as ``ultracompact objects" (UCO) in this respect~\cite{1985CQGra...2..219I,Cardoso:2014sna}.\footnote{The study and discussion about the compactness of BBSs can be seen in Ref.~\cite{Zhao:2025hdg}.}

In Fig.~\ref{fig:matter}, we use solid and dashed lines to mark the solutions with the LR and those without the LR, respectively. It can be seen that for BBSs with a smaller magnetic charge (left panel), similar to the boson star (red line), the first solution with the LR appears in the third branch. However, as the magnetic charge increases, the first solution with the LR can appear in the second branch or the first branch. When the magnetic charge is large enough, the entire $M(\omega)$ curve represents solutions of BBSs with the LR.

	\begin{figure}[!htbp]
		\begin{center}
		\subfigure[]{ 
			\includegraphics[height=.28\textheight,width=.30\textheight, angle =0]{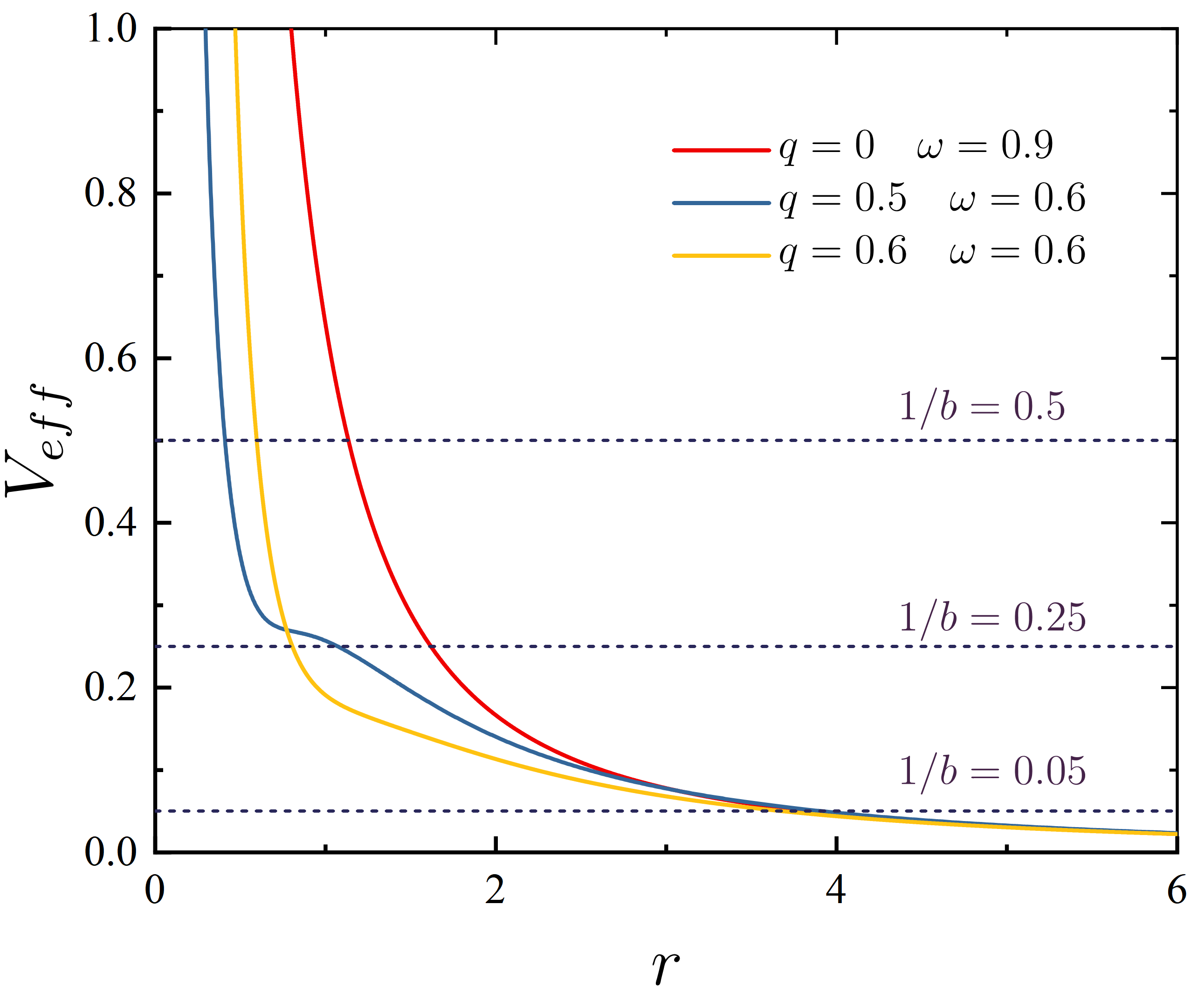}
			\label{fig:EFFUNBdNLR}
		}	 
  		\subfigure[]{  
			\includegraphics[height=.28\textheight,width=.30\textheight, angle =0]{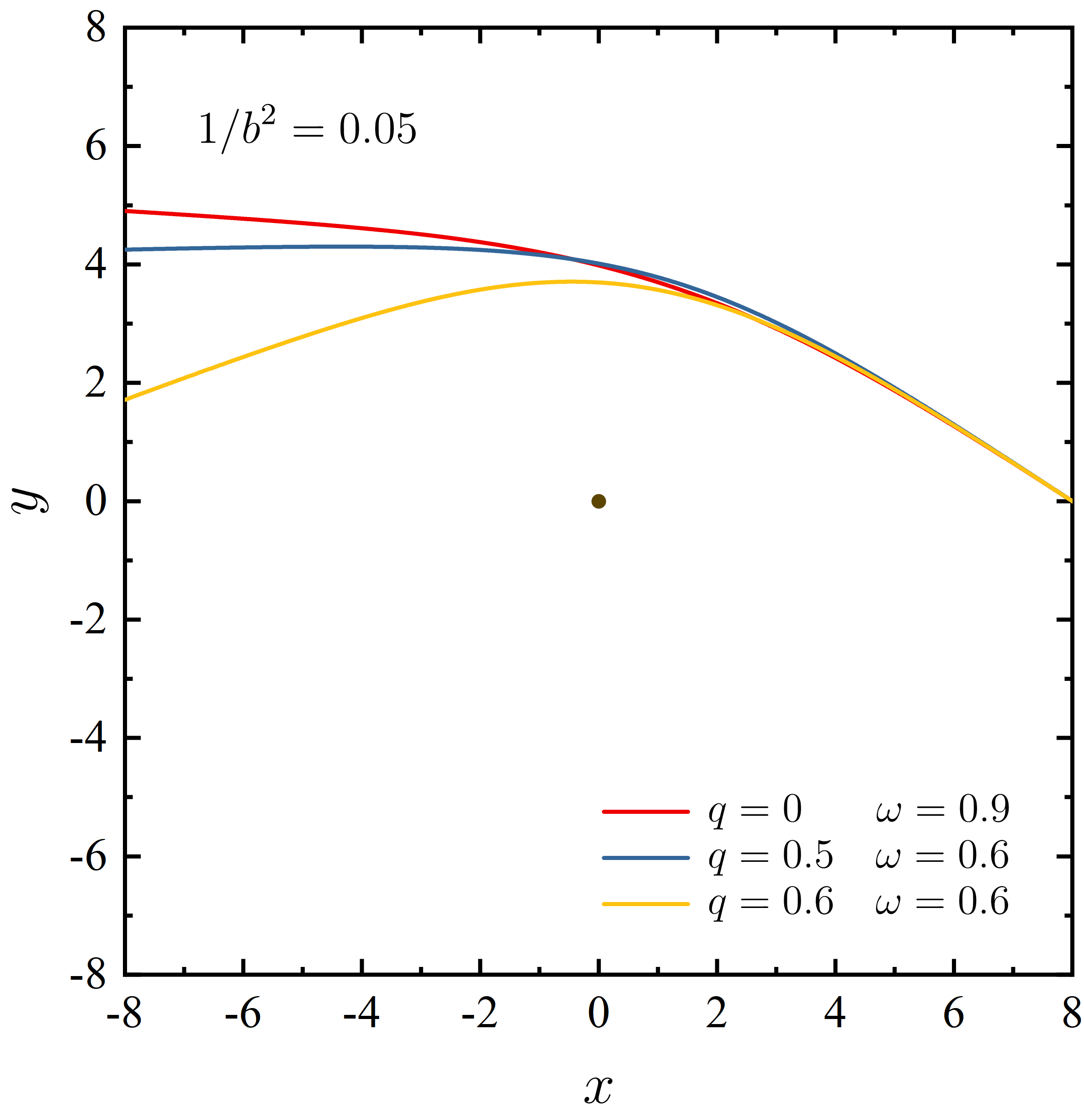}
			\label{fig:NLR005}
		}	
    \quad
        \subfigure[]{ 
			\includegraphics[height=.28\textheight,width=.30\textheight, angle =0]{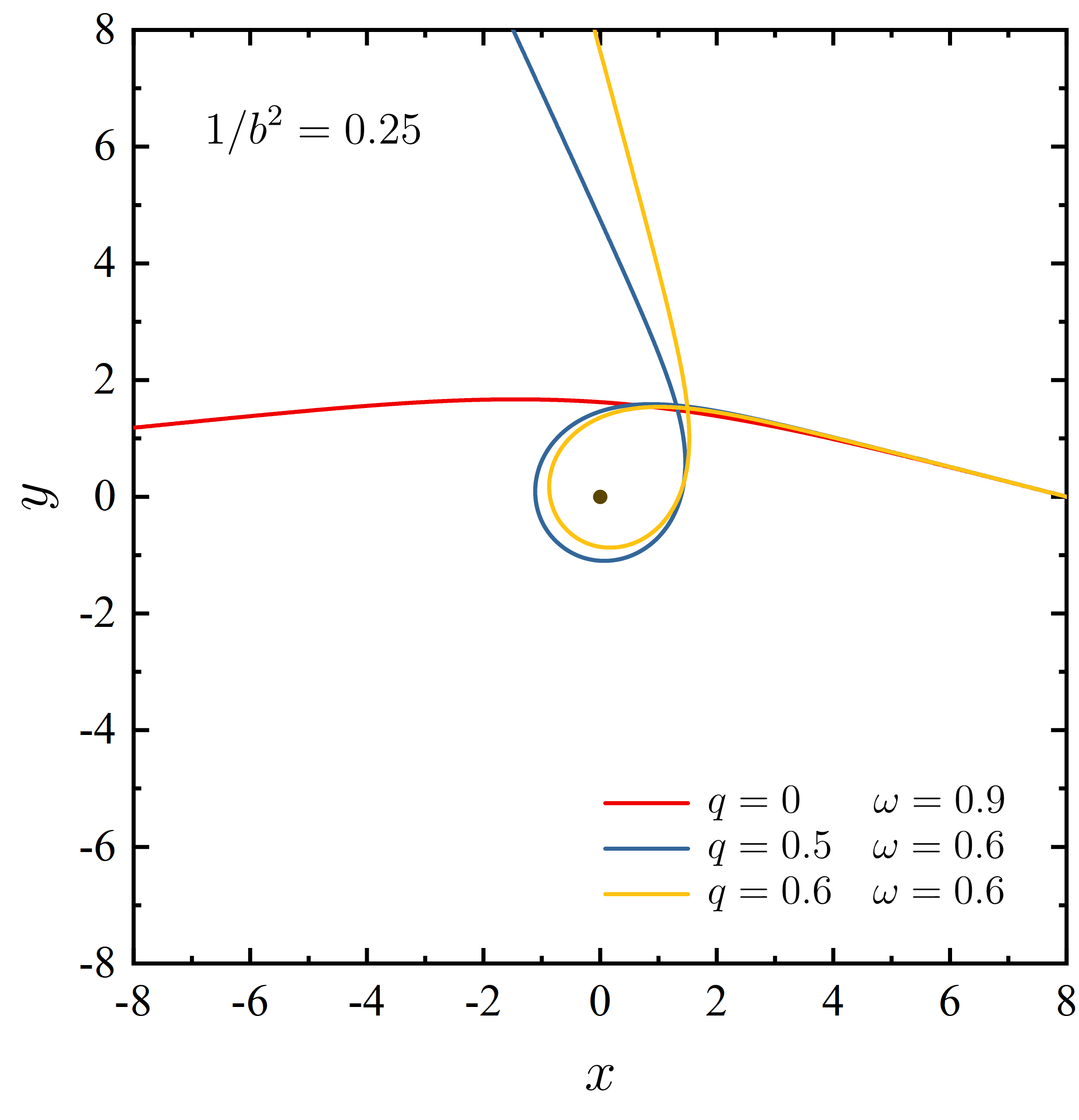}
			\label{fig:NLR025}
		}	 
  		\subfigure[]{  
			\includegraphics[height=.28\textheight,width=.30\textheight, angle =0]{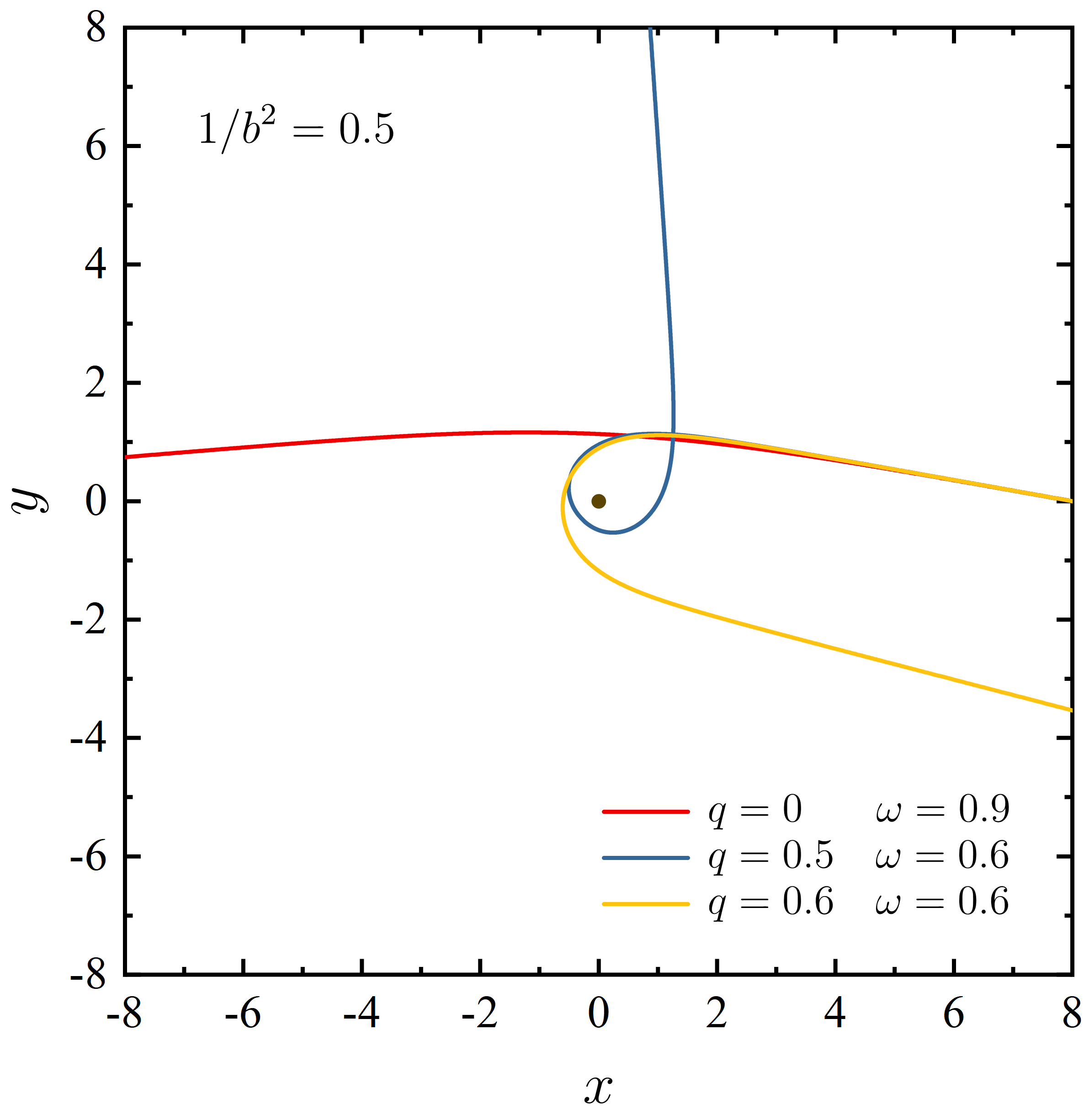}
			\label{fig:NLR05}
		}	
  		\end{center}	
		\caption{The radial effective potential and orbit of the photon for BBSs without the LR.}
	\label{fig:NBNLR}	
		\end{figure}

For solutions without the LR, since equation (\ref{eq:dotr}) (or equation (\ref{eq:dtr})) has at most one root that makes the radial velocity of the particle zero, only unbound orbits exist. However, for solutions with the LR, due to the radial effective potential exist the extremum, when the impact parameter is appropriate, equation (\ref{eq:dotr}) has two different roots, $r_1$, $r_2$ ($r_1$ and $r_2$ are also referred to as the periapsis and apoapsis, respectively), making the radial velocity of the particle zero. Therefore, the photon will move between $r_1$ and $r_2$, meaning that bound orbits can also exist for solutions with the LR. Considering the differences between these two types of orbits, we will analyze each type separately. Furthermore, since photons are less affected when farther away from the BBSs and using the $\bar{r}$ coordinate will deform the shape of the orbit of the photon, we adopt the coordinate $r$ (dimensionless) before the conformal transformation or the standard Cartesian coordinates ordinary $x$, $y$, where $r=\bar{r}/(1-\bar{r})$, $x=r\cos\theta$ and $y=r\sin\theta$.\footnote{Noting that $r$ is dimensionless, $x$ and $y$ are therefore also dimensionless.}

\subsection{Unbound orbit}

We first show the unbound orbital. As mentioned before, they can exist in both BBSs with the LR and those without the LR.
\subsubsection{Solutions without light ring}\label{sec: nolr}
Figure~\ref{fig:EFFUNBdNLR} shows the radial effective potential as a function of $r$ for BBSs without the LR (blue and yellow lines) and the BS (red line). In Fig.~\ref{fig:NLR005}, Fig.~\ref{fig:NLR025} and Fig.~\ref{fig:NLR05}, we show the trajectories of photons passing through the point ($x=8$, $y=0$) around these BBSs (The values of the photon impact parameter $b$ correspond to the dashed lines in Fig.~\ref{fig:EFFUNBdNLR}). From this figure, we can first see that for the same impact parameter, the deflection angle of the photons in the BBS solution in both cases is larger than that of the BSs, which means that the BBSs are more compact than the BSs. Second, for a given solution, as $1/b$ increases, the photon can reach further away from the origin ($0$,$0$). This corresponds to the root of $V_{eff}=1/b$ (that is, the position where the radial velocity is zero) increases as $1/b$ increases.

\subsubsection{Solutions with light ring}\label{sec: havelr}
For BBSs with the LR, the properties of the unbound orbits formed by photons near them are mostly similar to those of BBSs without the light ring. However, as we now show, the nature of the orbits becomes very special in the following two cases: (1) when the inverse of the impact parameter $1/b$ is very close to the extremum of the radial effective potential, and (2) when the frequency is very close to zero (i.e., the frozen star solution).
%
	\begin{figure}[!htbp]
		\centering	
        \subfigure{  
			\includegraphics[height=.20\textheight,width=.20\textheight, angle =0]{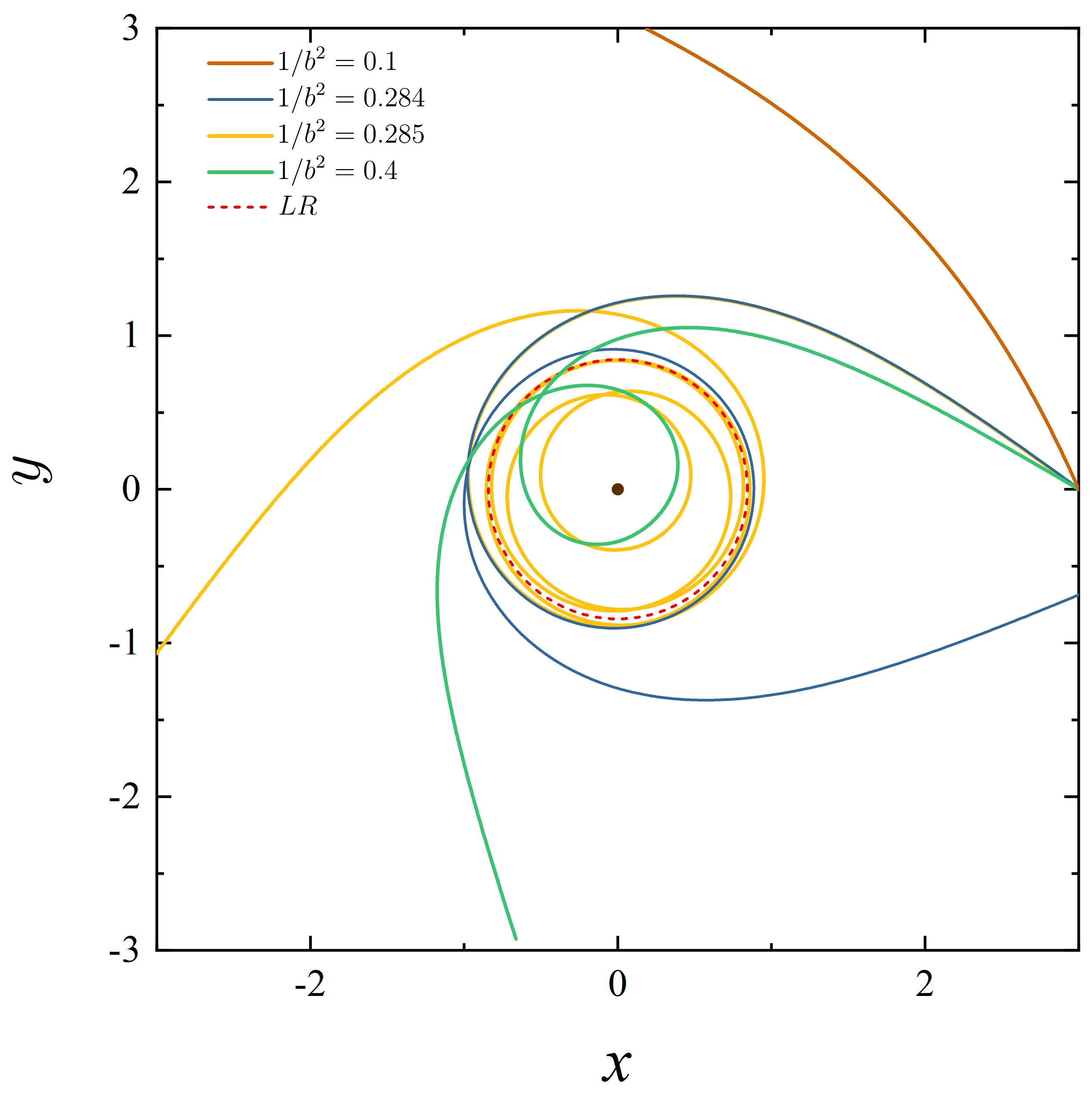}
		}
		\subfigure{
			\includegraphics[height=.20\textheight,width=.20\textheight, angle =0]{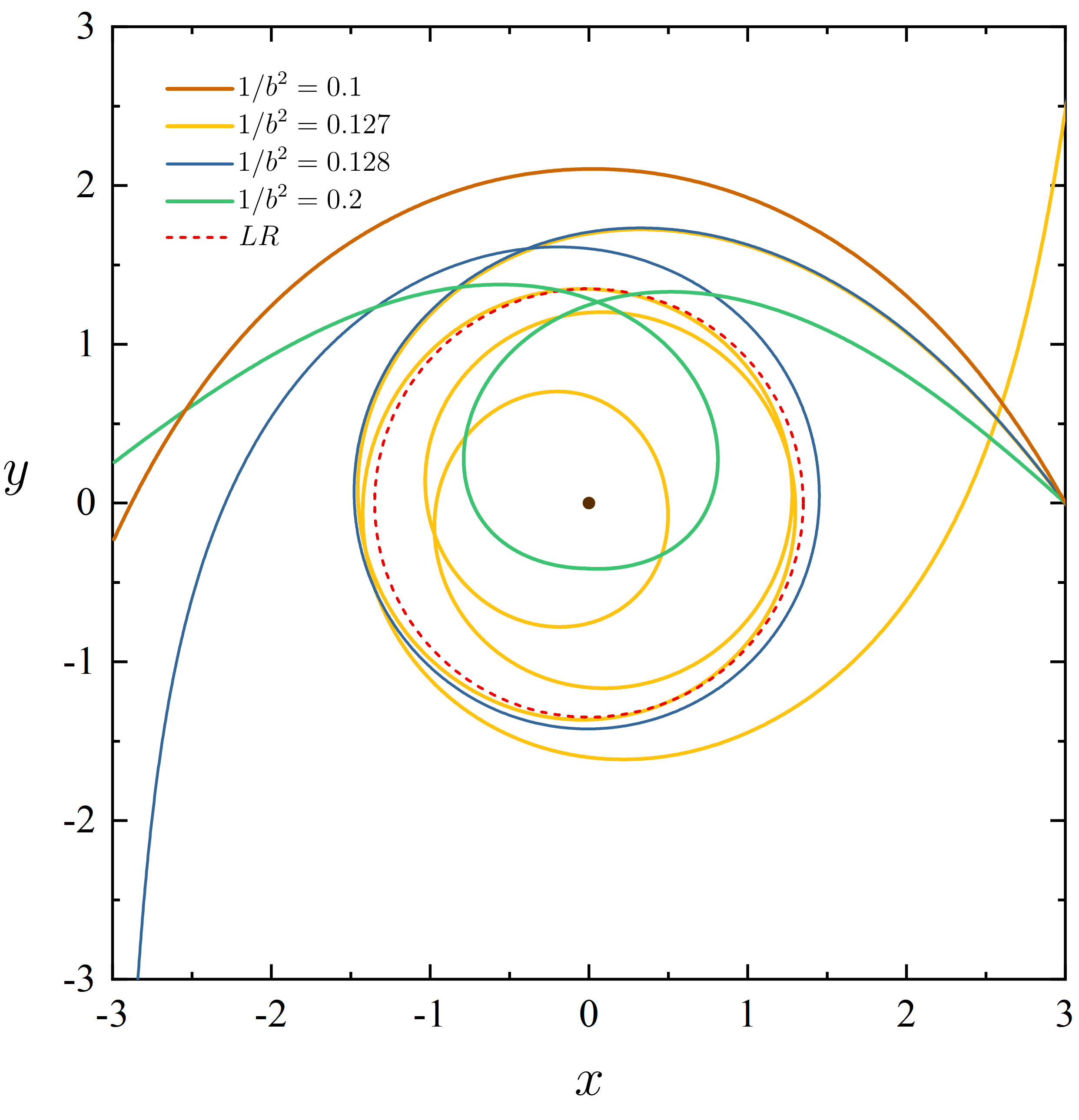}
   		} 	
            \subfigure{
			\includegraphics[height=.20\textheight,width=.20\textheight, angle =0]{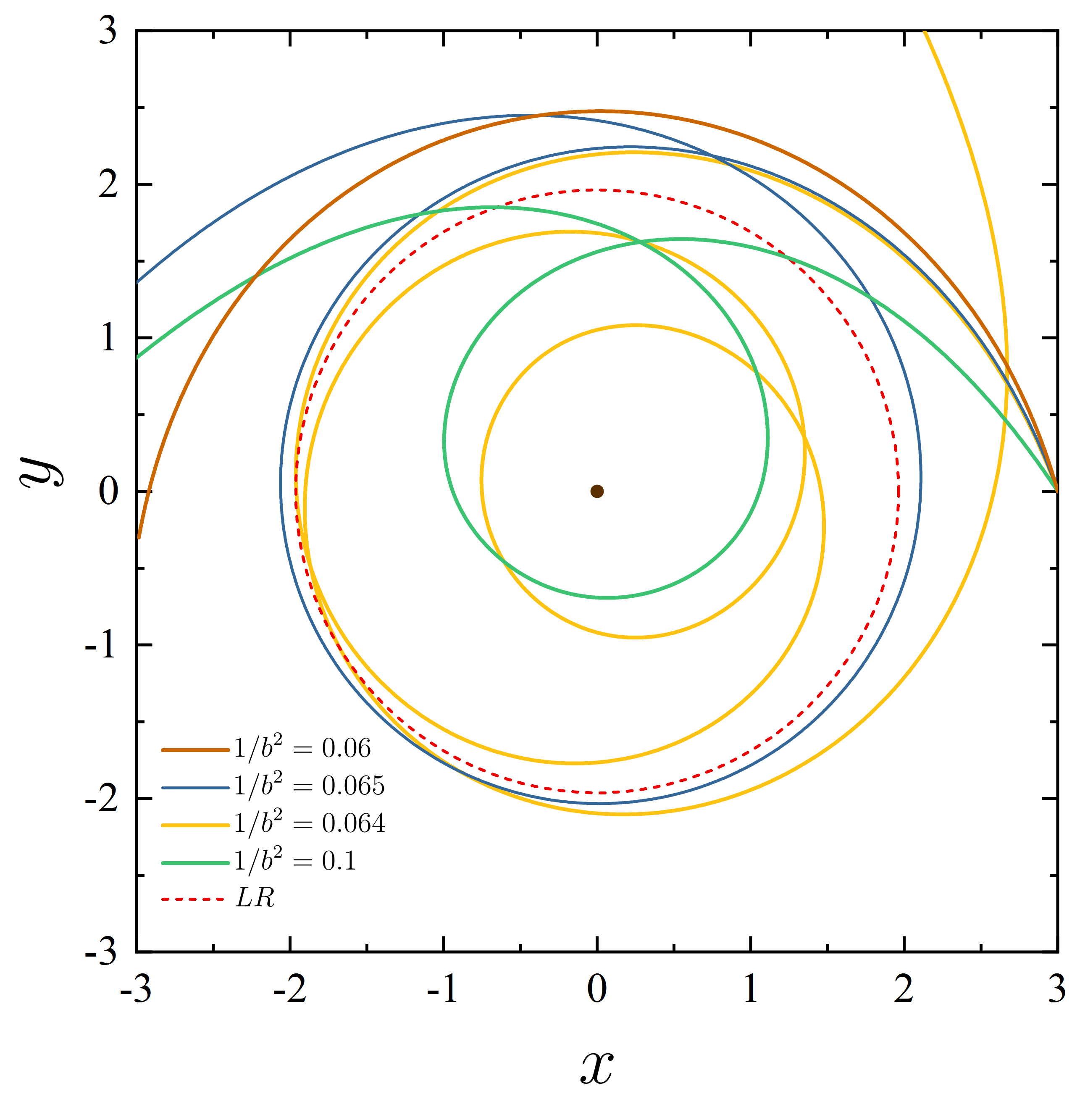}
   		} 
    \quad
        \subfigure{  
			\includegraphics[height=.20\textheight,width=.20\textheight, angle =0]{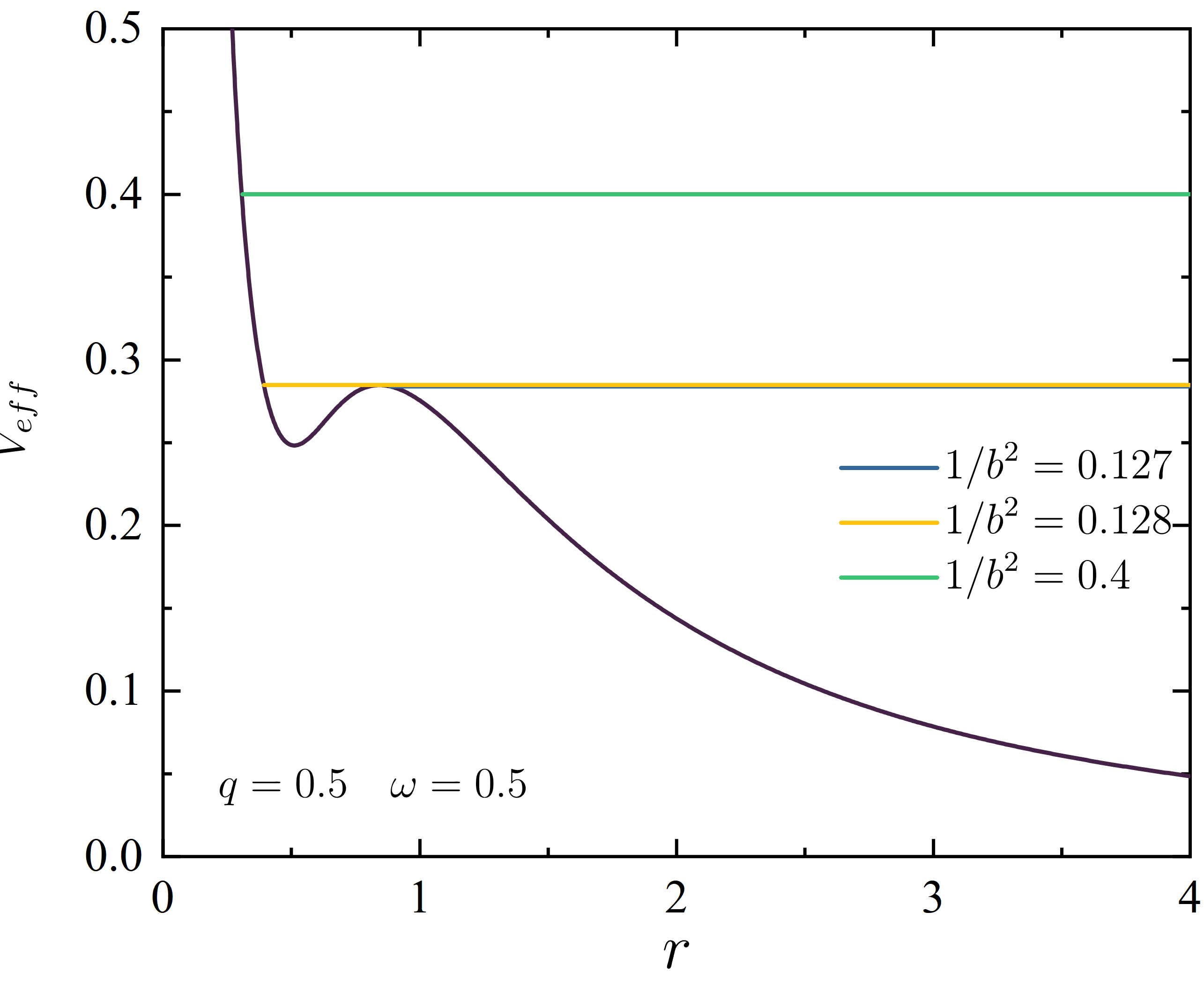}
		}
	\subfigure{
			\includegraphics[height=.20\textheight,width=.20\textheight, angle =0]{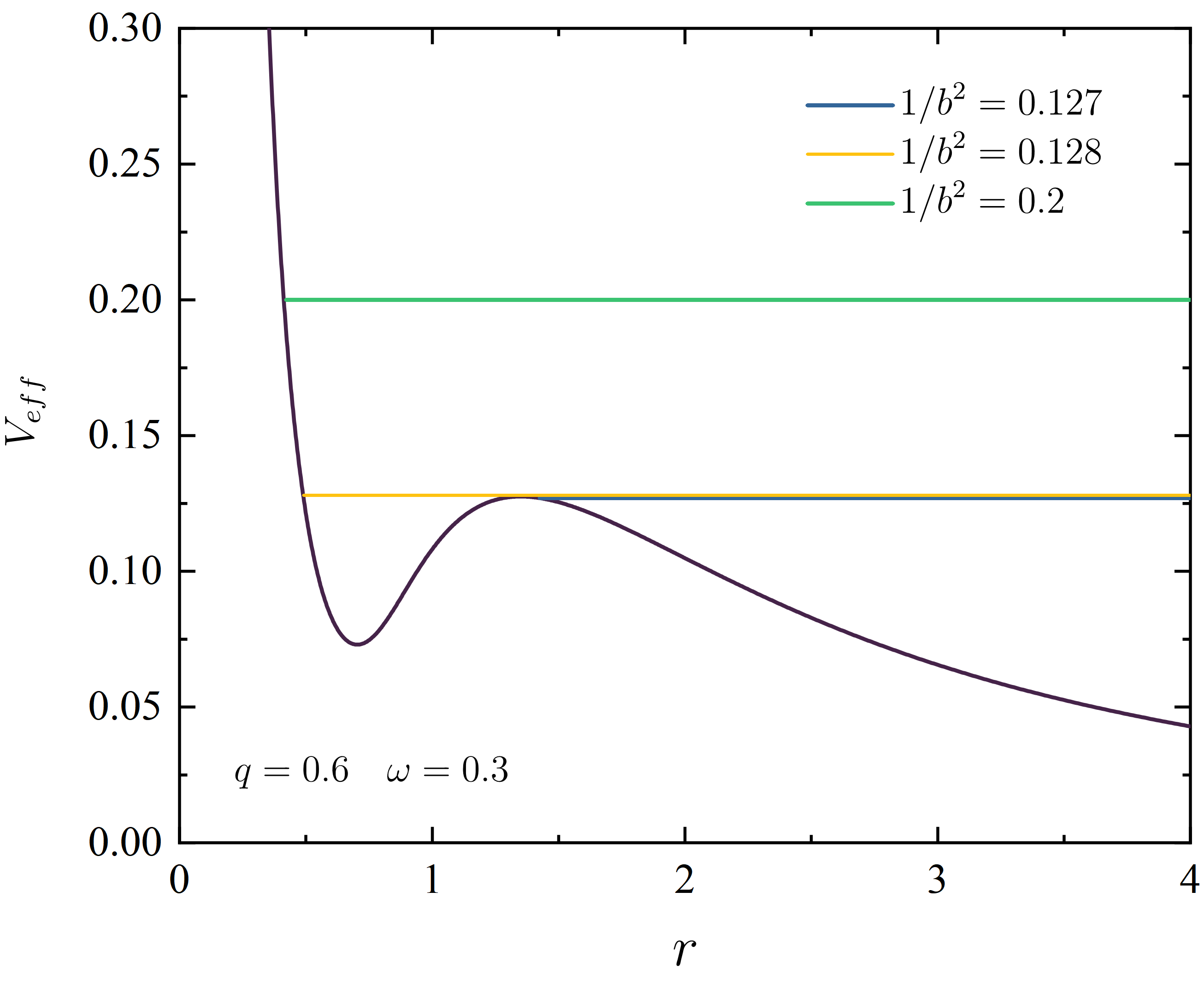}
   		} 	
        \subfigure{
			\includegraphics[height=.20\textheight,width=.20\textheight, angle =0]{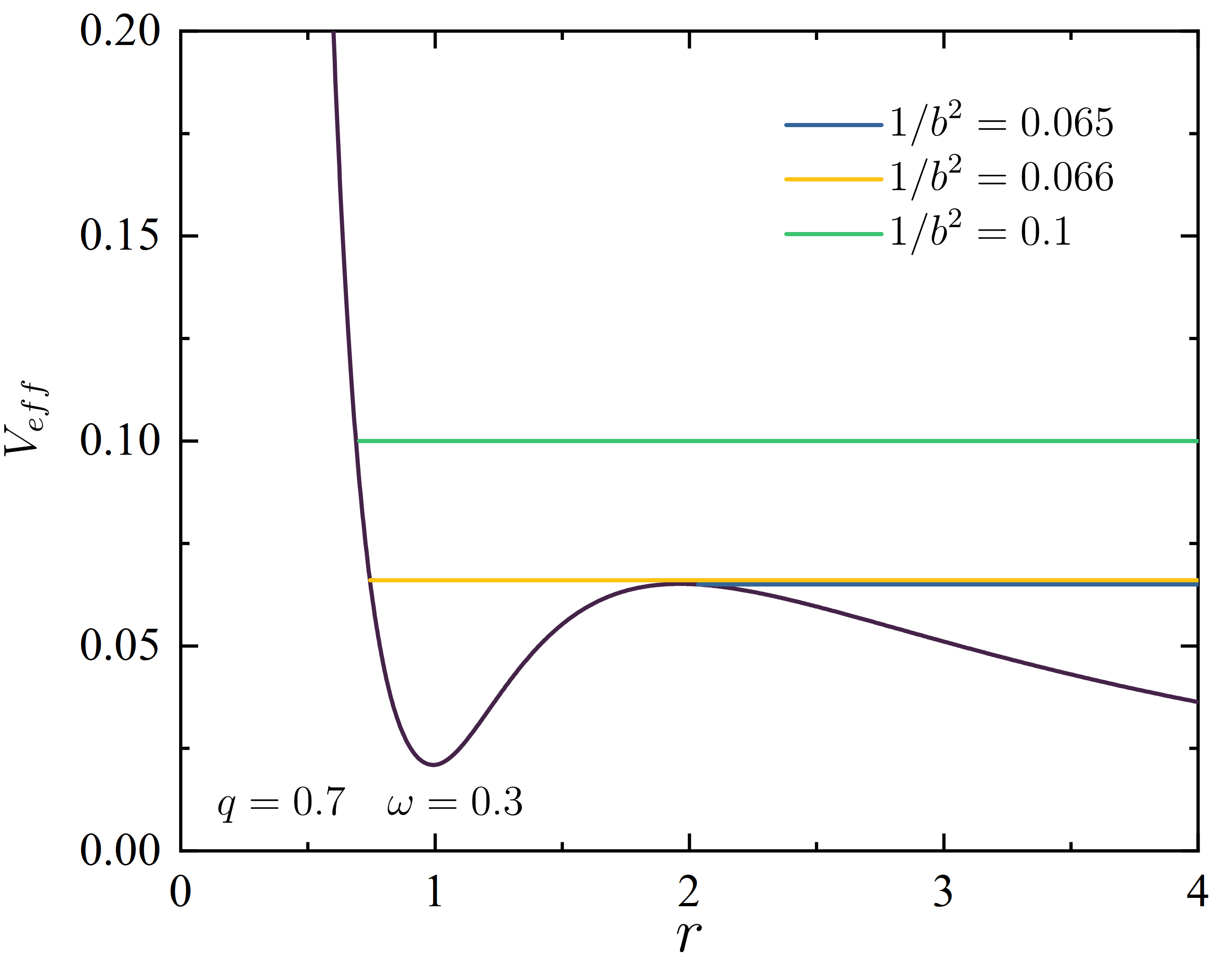}
   		}	
		\caption{The orbit of the photon and the radial effective potential for BBSs with the LR. The curves of different colors in the top panel correspond to $1/b$ represented by the same colored lines in the bottom panel.}
		\label{fig:NBHLR}		
		\end{figure}
%

%
	\begin{figure}[!htbp]
		\centering	
        \subfigure{  
			\includegraphics[height=.20\textheight,width=.20\textheight, angle =0]{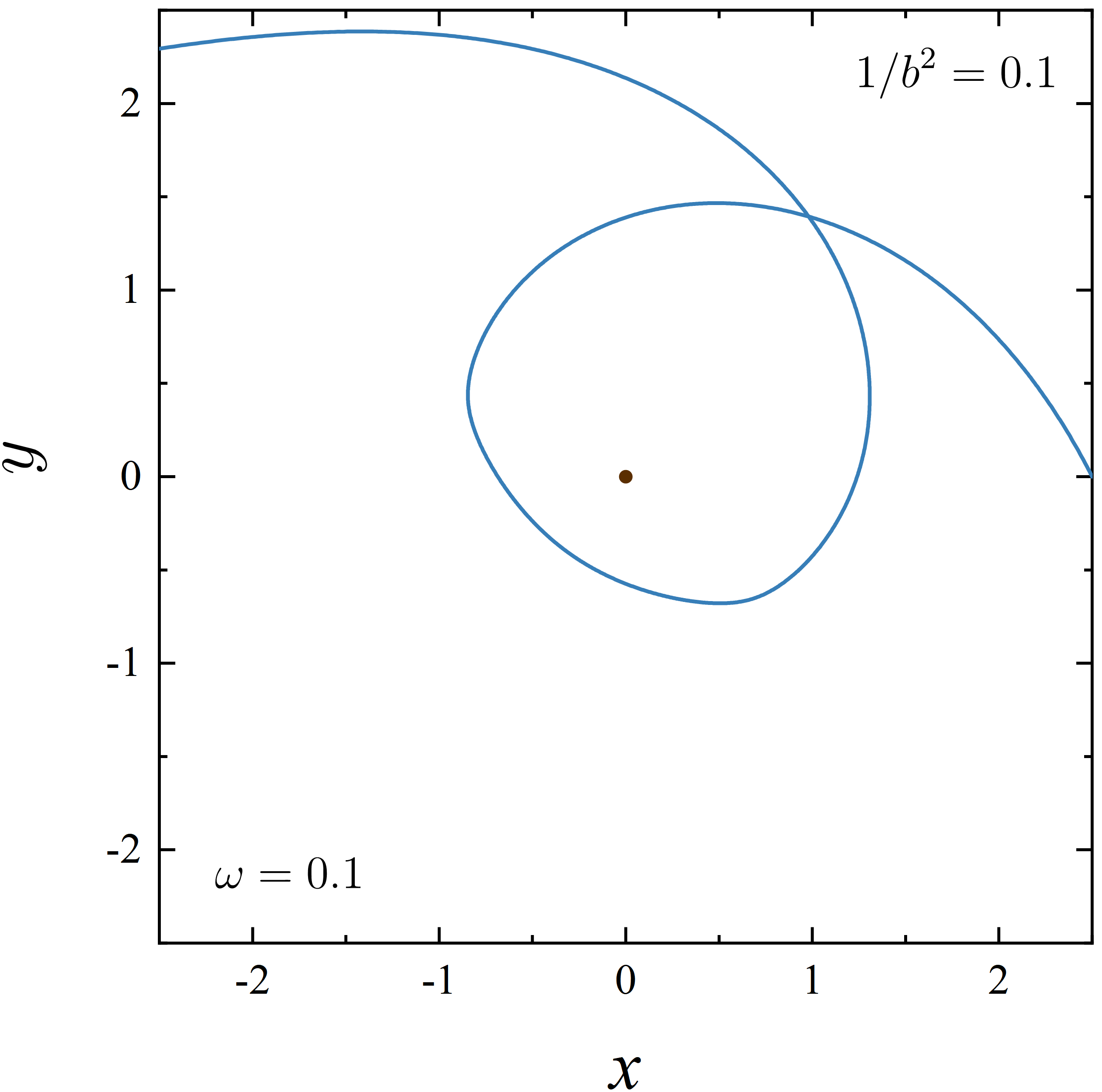}
		}
		\subfigure{
			\includegraphics[height=.20\textheight,width=.20\textheight, angle =0]{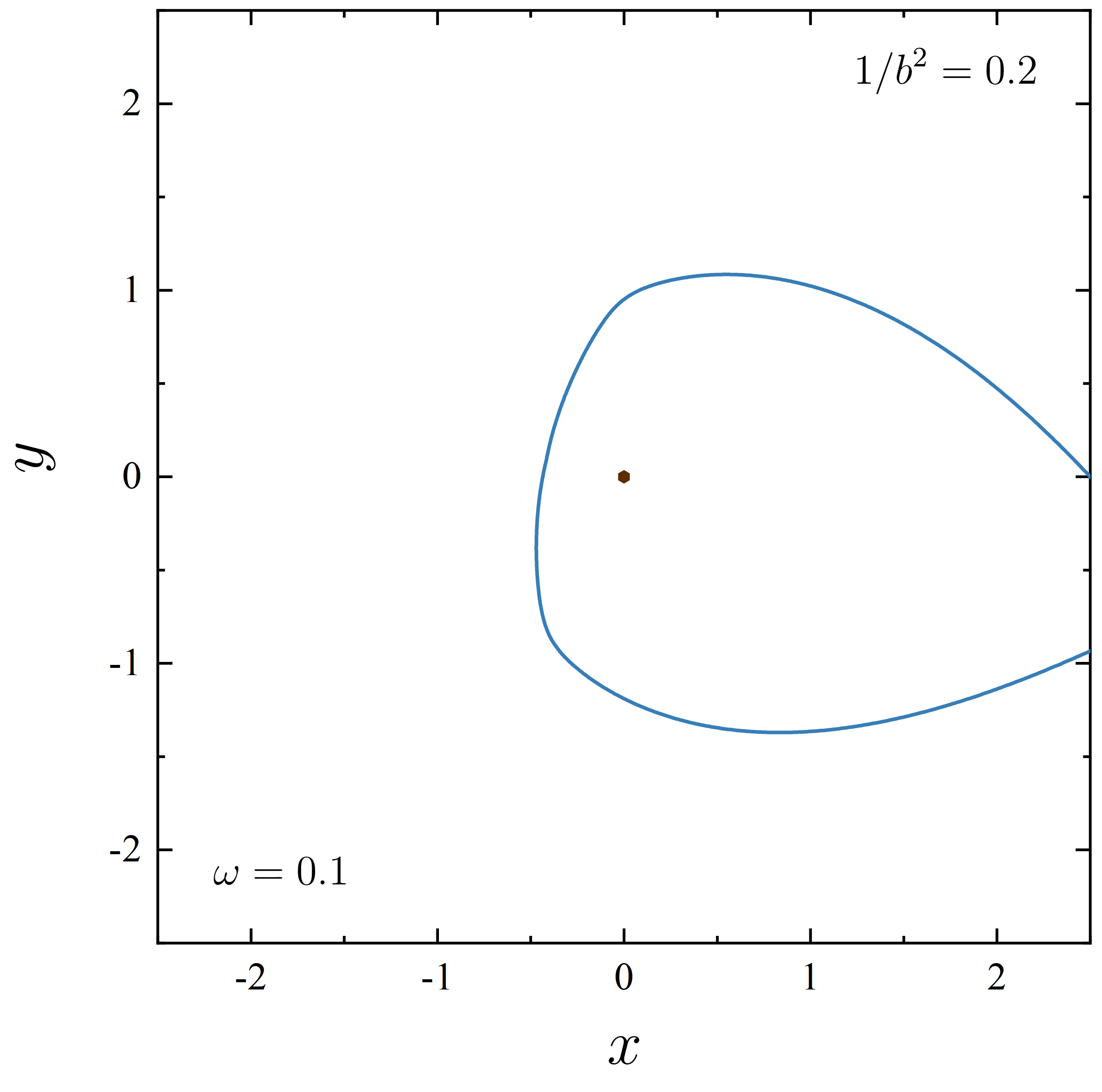}
   		} 	
            \subfigure{
			\includegraphics[height=.20\textheight,width=.20\textheight, angle =0]{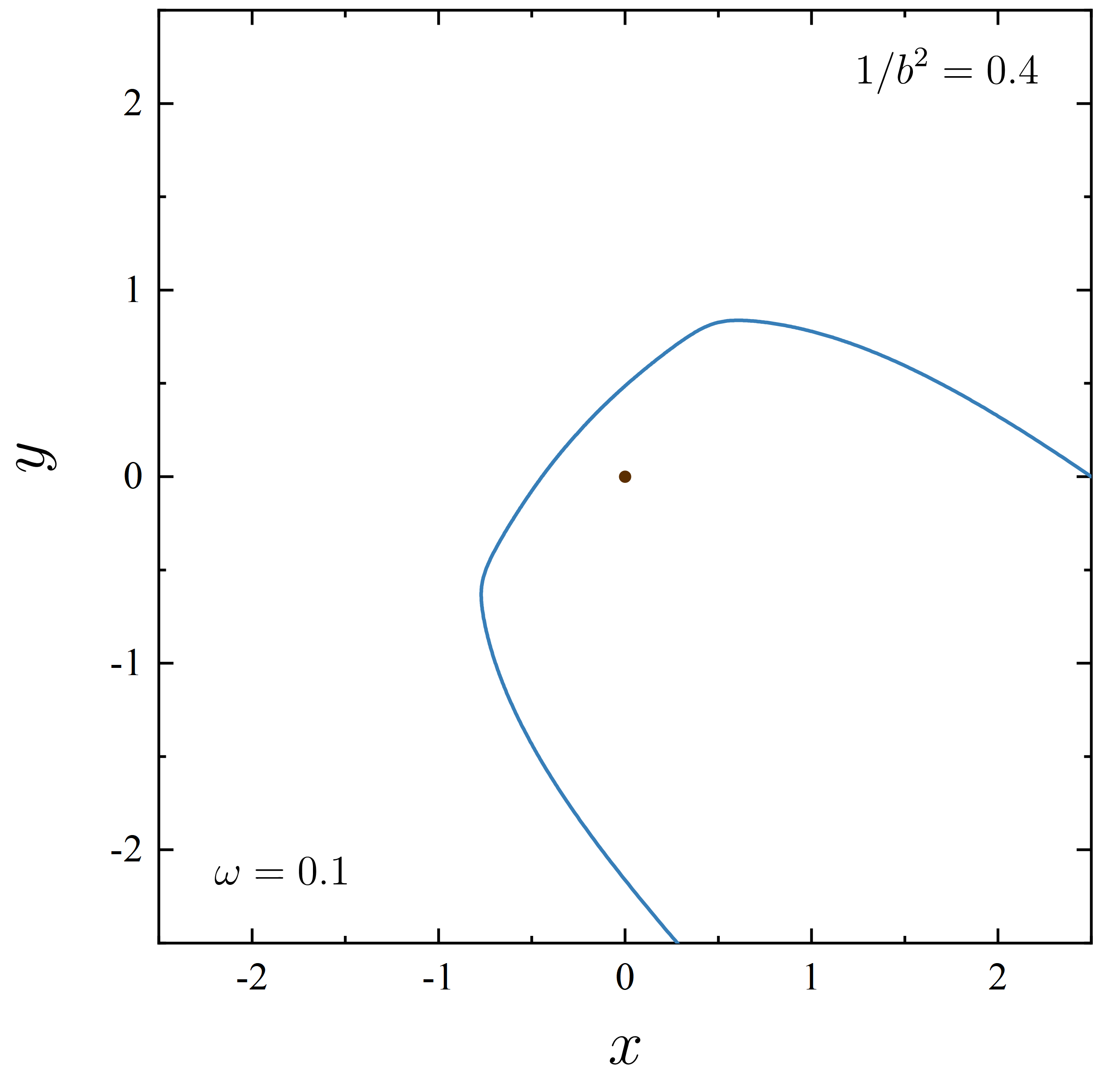}
   		} 
    \quad
        \subfigure{  
			\includegraphics[height=.20\textheight,width=.20\textheight, angle =0]{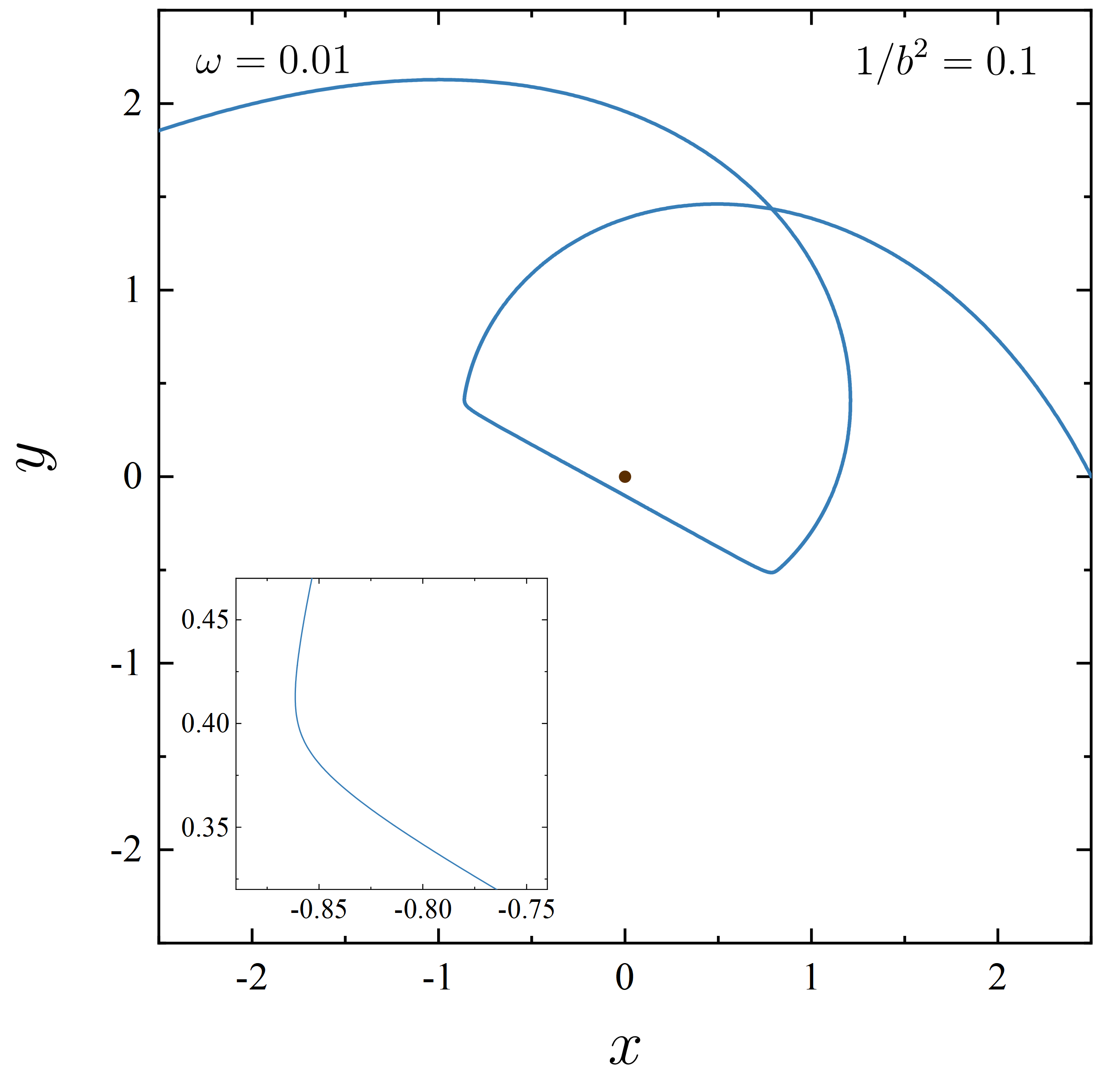}
		}
	\subfigure{
			\includegraphics[height=.20\textheight,width=.20\textheight, angle =0]{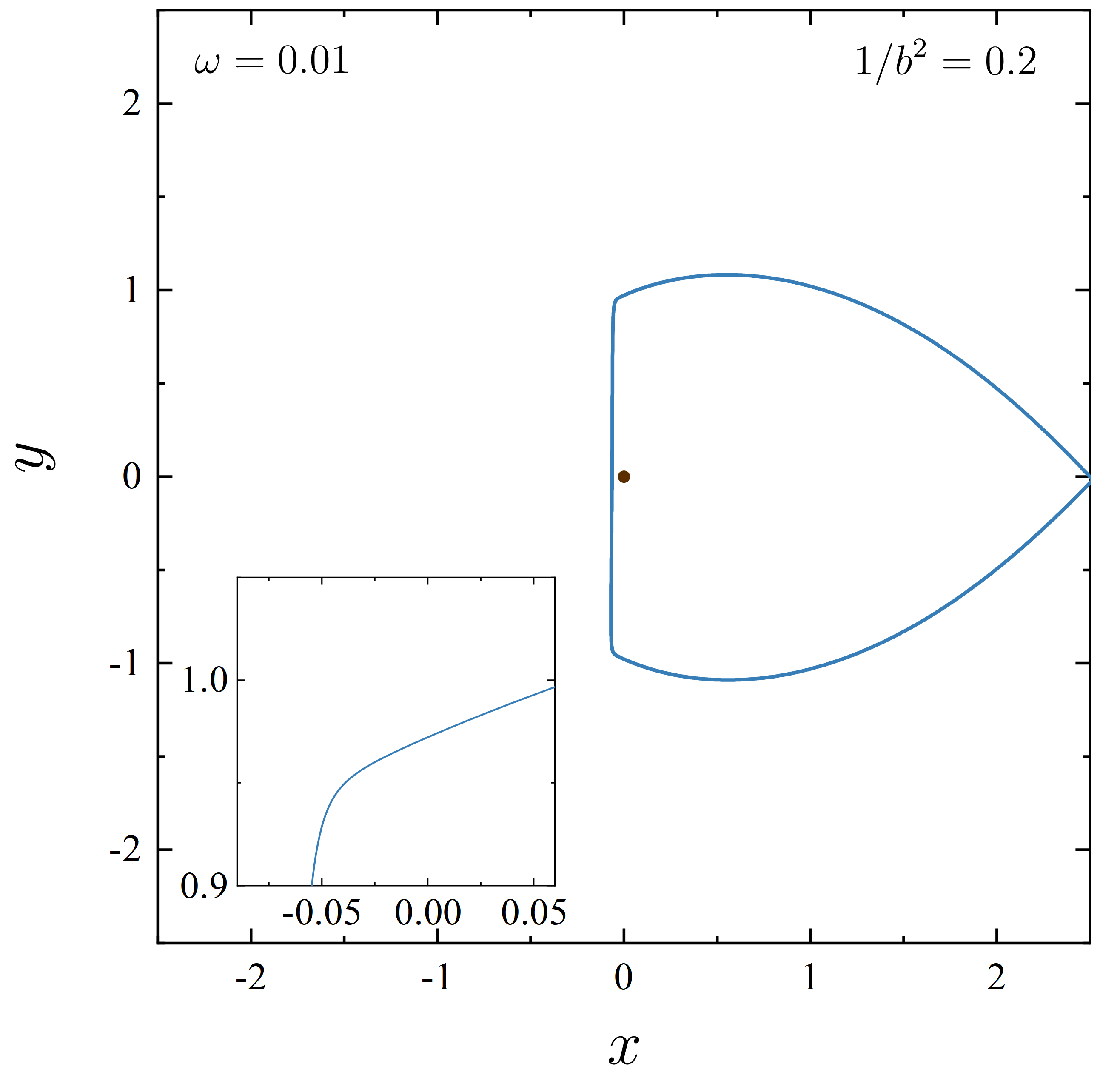}
   		} 	
        \subfigure{
			\includegraphics[height=.20\textheight,width=.20\textheight, angle =0]{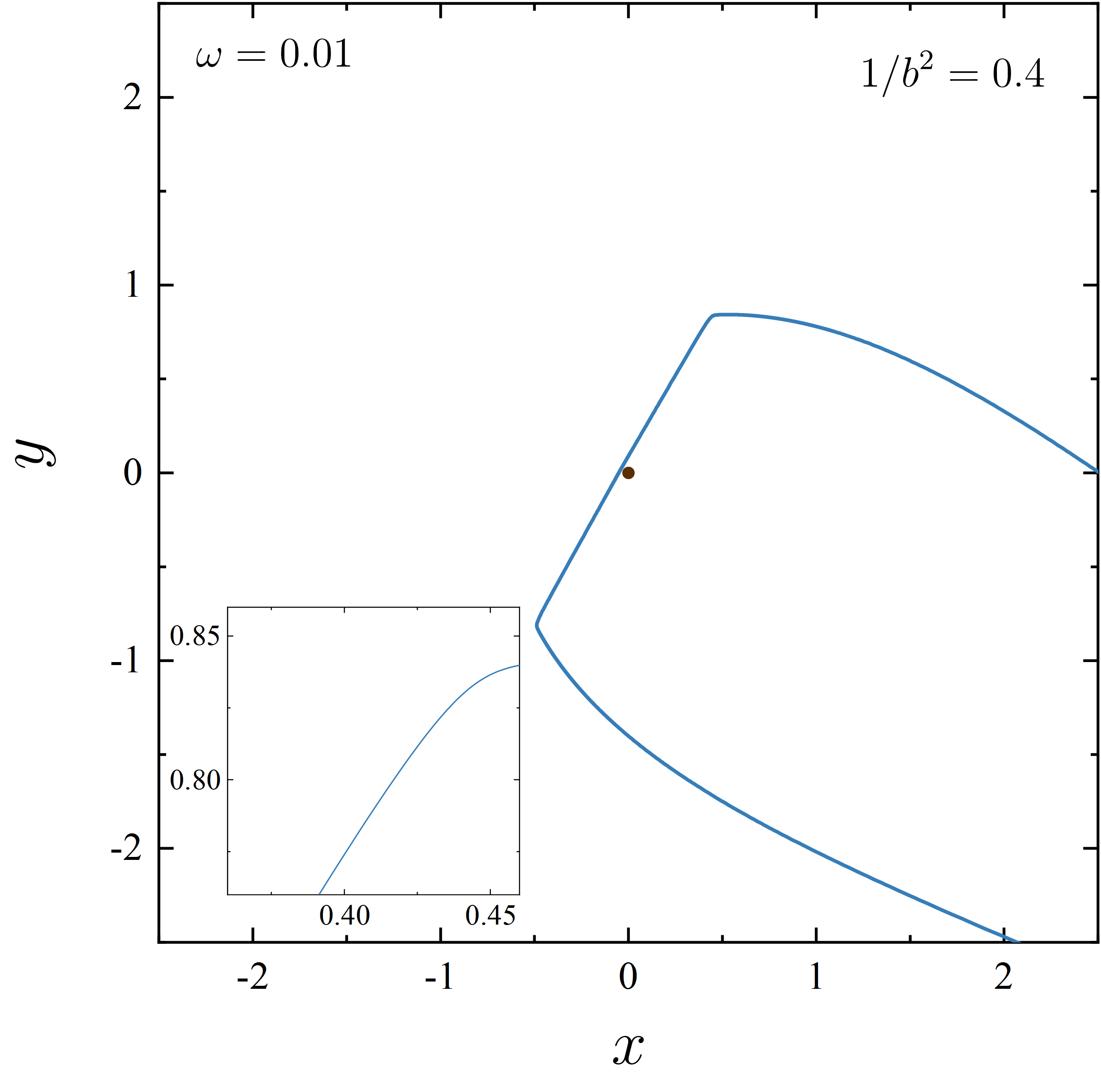}
   		}
            \quad
        \subfigure{  
			\includegraphics[height=.20\textheight,width=.20\textheight, angle =0]{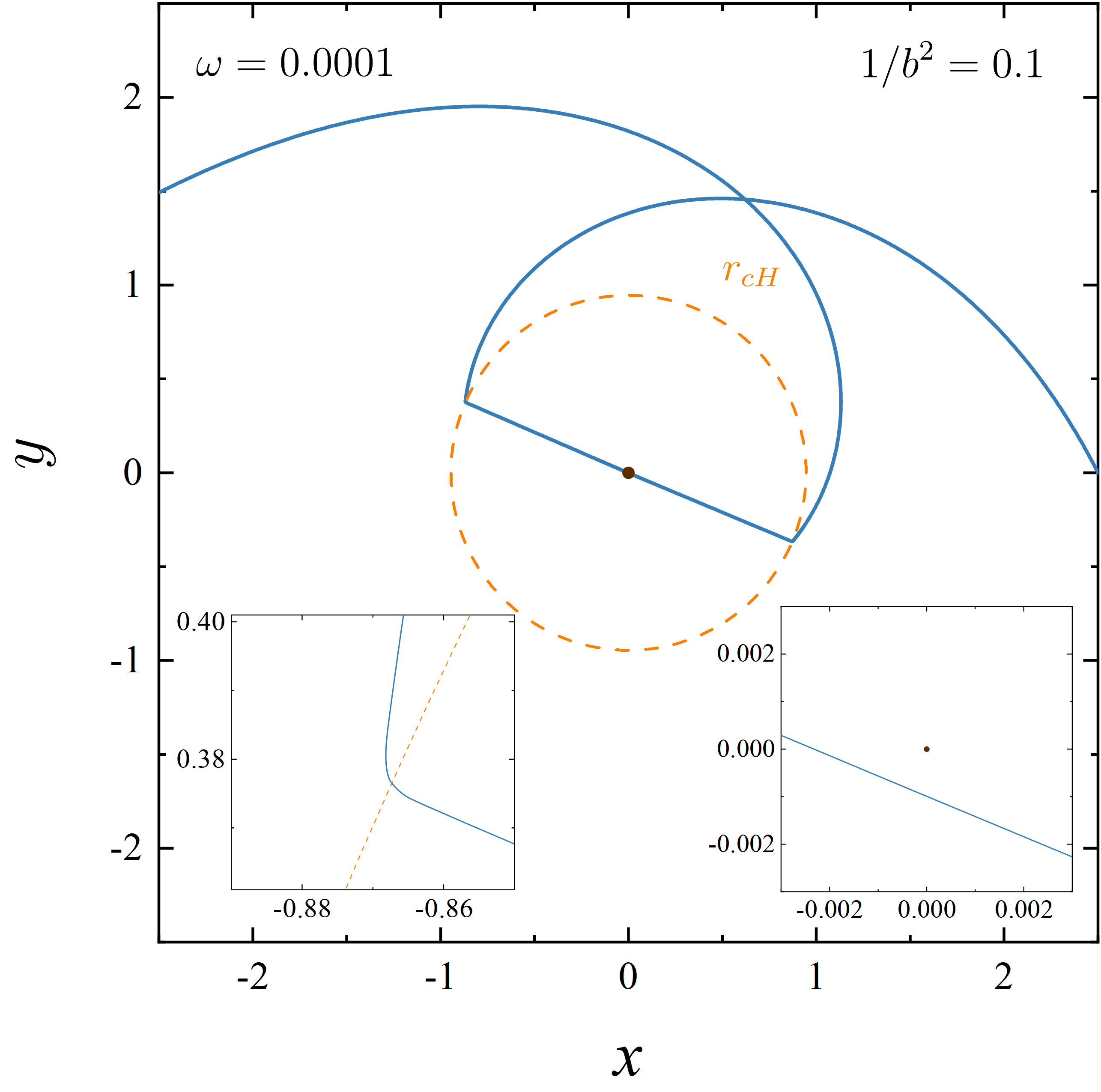}
		}
	\subfigure{
			\includegraphics[height=.20\textheight,width=.20\textheight, angle =0]{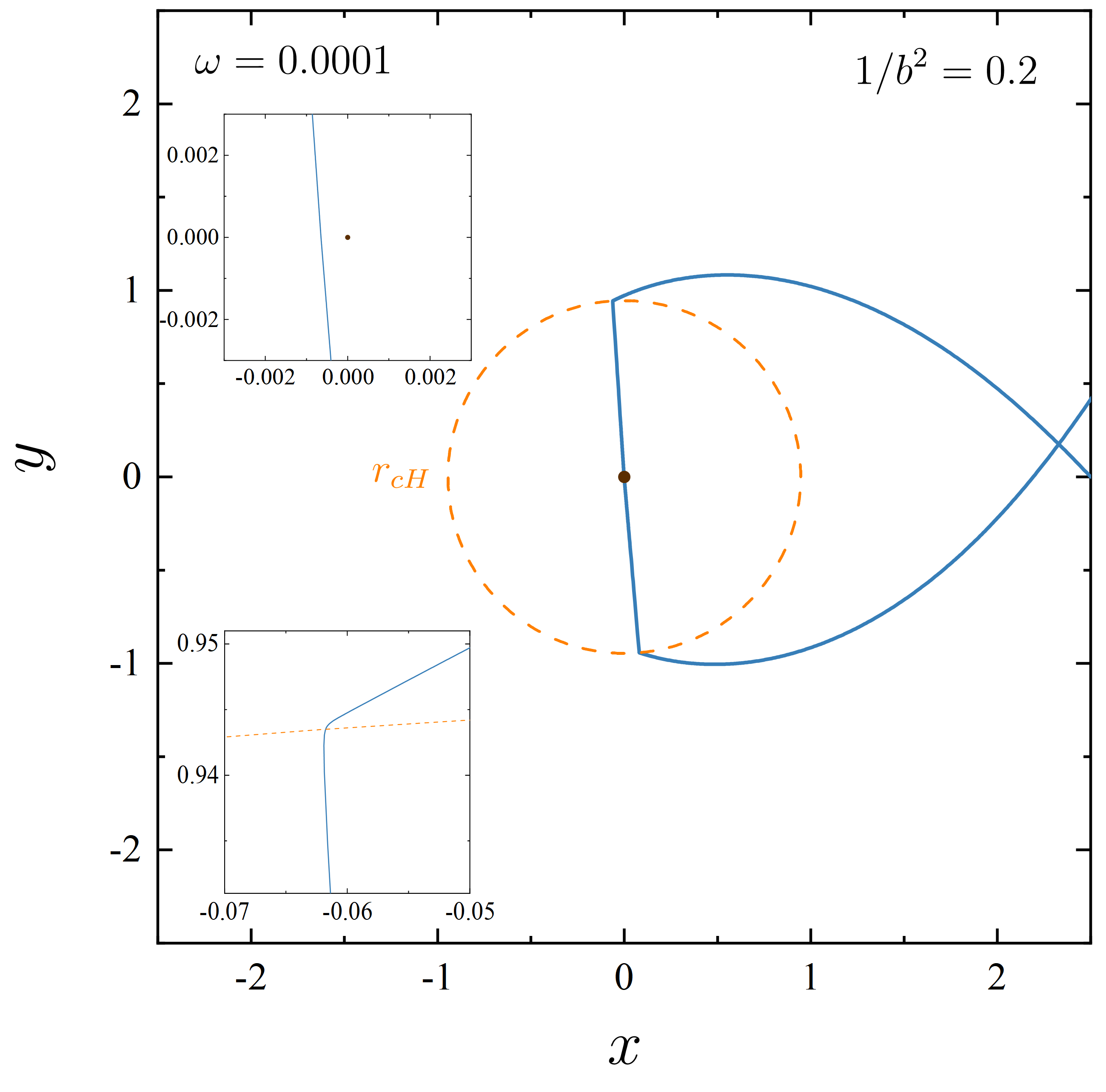}
   		} 	
        \subfigure{
			\includegraphics[height=.20\textheight,width=.20\textheight, angle =0]{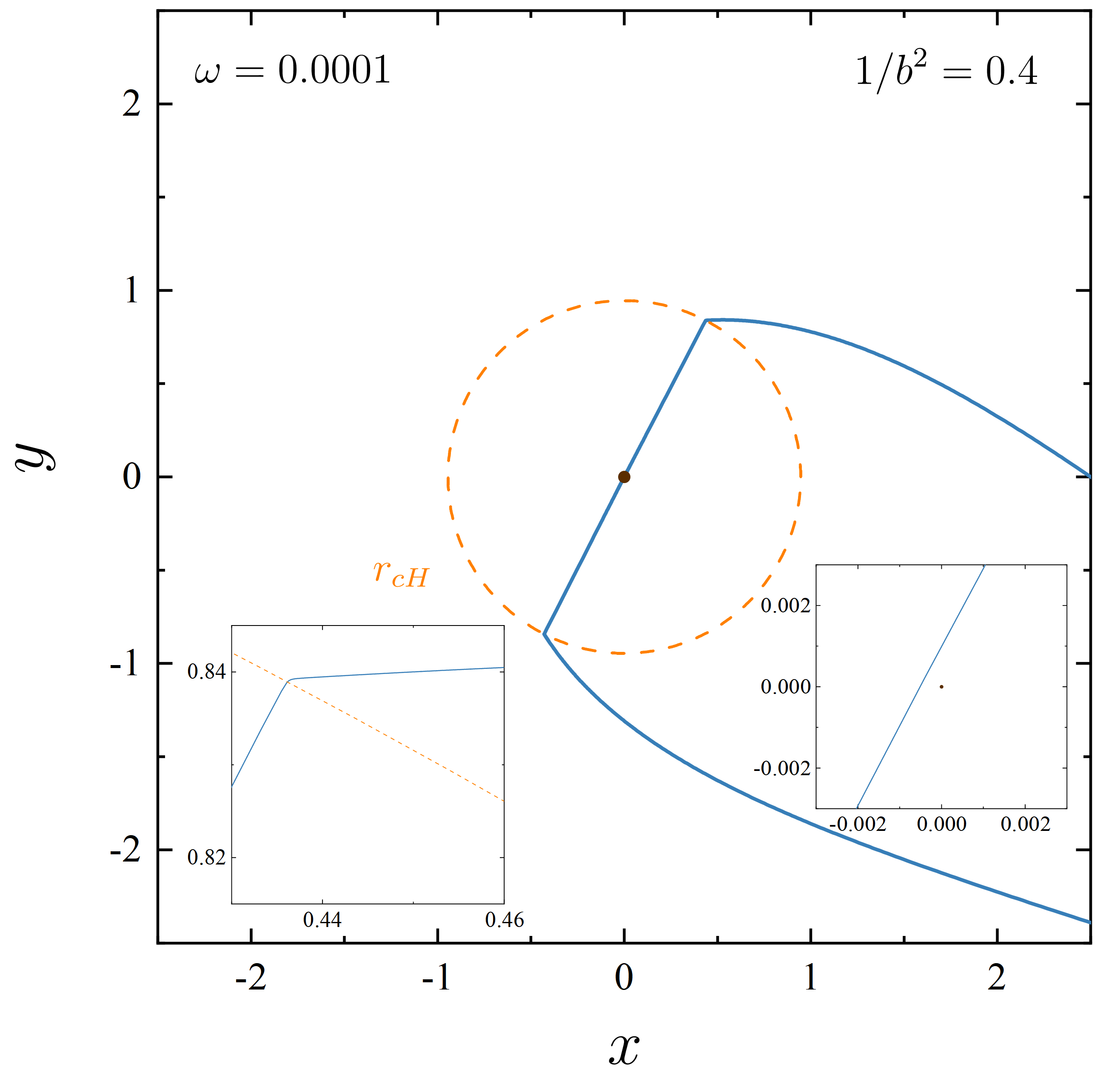}
   		}
		\caption{The orbit of the photon for BBSs with $q=0.7$ and $\omega=0.1$ (top panel), $\omega=0.1$ (middle panel) and $\omega=0.0001$ (bottom panel). The red dot is the initial position.}
		\label{fig:FNBHLR}		
		\end{figure}
%

	\begin{figure}[!htbp]
		\centering	
        \subfigure{  
			\includegraphics[height=.28\textheight,width=.30\textheight, angle =0]{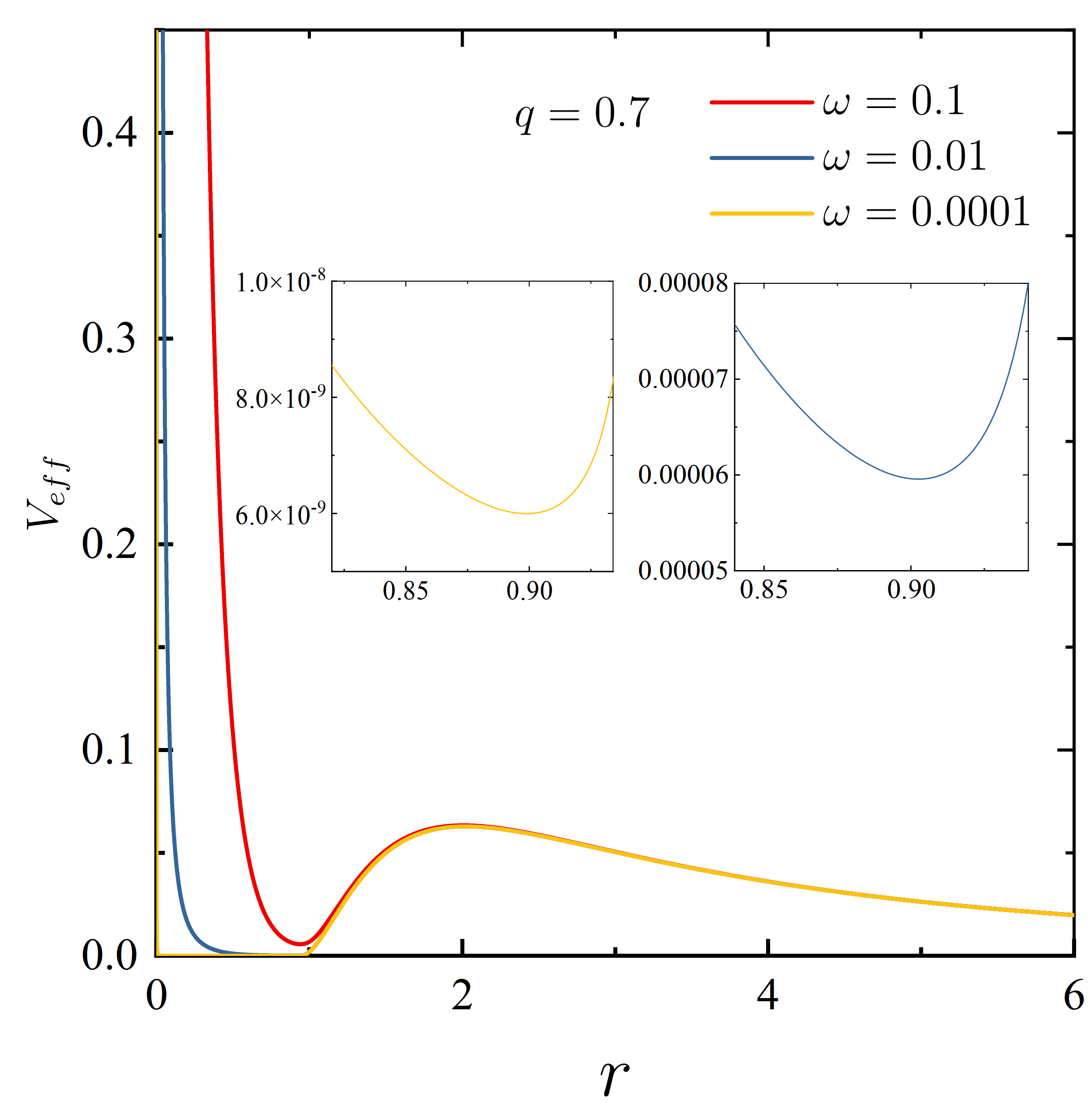}
		}

		\caption{The radial effective potential for BBSs with $q=0.7$.}
		\label{fig:EFFFNBHLR}		
		\end{figure}
%

We use $(q=0.5$, $\omega=0.5)$, $(q=0.6$, $\omega=0.3)$, and $(q=0.7$, $\omega=0.3)$ as examples to illustrate the first case. Figure~\ref{fig:NBHLR} shows the images of the photon orbits for these solutions (top panel), along with the corresponding radial effective potentials (bottom panel). From this figure, it can be seen that the effective potential has two extremums, which means BBSs have two LRs: the inner LR is stable, and the outer one is unstable. When the inverse of the impact parameter $1/b$ is far from the extremum value of $V_{eff}$, the orbit around BBSs with the LR is similar to that around BBSs without the LR. The photons all move in the direction away from the BBSs after some deflection near the BBSs. However, when $1/b$ is very close to the extremum value of the effective potential, although, as in other cases (e.g., Fig~\ref{fig:NBNLR}), the photons eventually move away from the LR and head toward infinity, the photon rotates around the LR by a very large angle close to the LR. In the case where $1/b$ is slightly larger (blue) than the extremum, the photon always moves outside the LR and forms an orbit nearly to a circle near the LR. In the case where $1/b$ is slightly smaller (yellow) than the extremum value of $V_{eff}$, the photon can form many nearly circular orbits near the LR and it can enter the interior of the LR. Inside the LR, the photon rotates through a considerable angle around the origin (that is, $r=0$ or $(x=0$, $y=0)$). Theoretically, the closer $1/b$ is to the extremum, the greater the angle at which the photon can rotate near the LR, and when the $1/b$ is strictly equal to the extremum $V(r_e)$, the photon will rotate an infinite angles at $r=r_e$ after reaching $r_e$, which corresponds to the circular orbit or, in other words, the LR. However, this cannot be done due to numerical calculations.

	\begin{figure}[!htbp]
		\centering	
        \subfigure{  
			\includegraphics[height=.20\textheight,width=.20\textheight, angle =0]{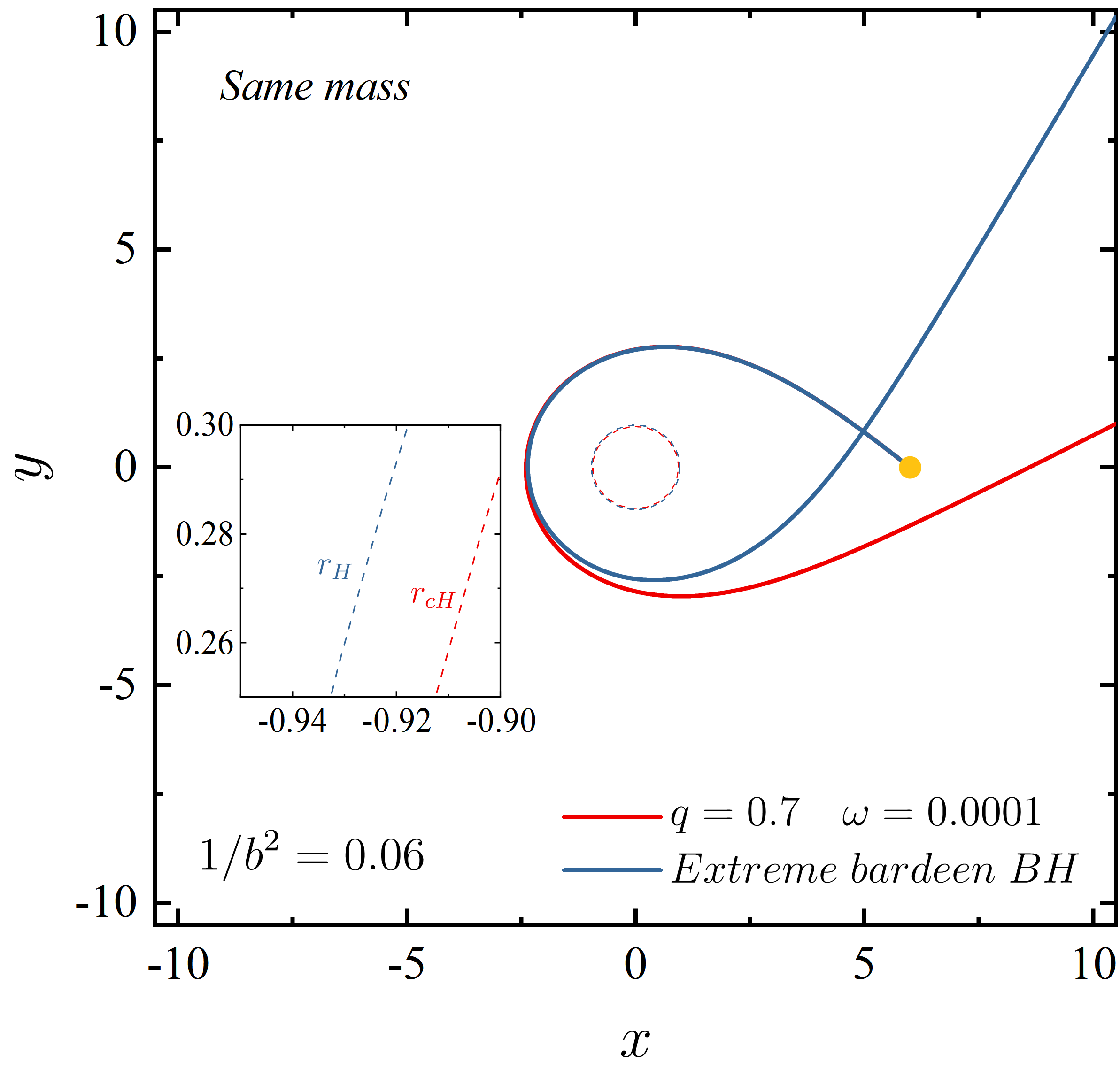}
		}
		\subfigure{
			\includegraphics[height=.20\textheight,width=.20\textheight, angle =0]{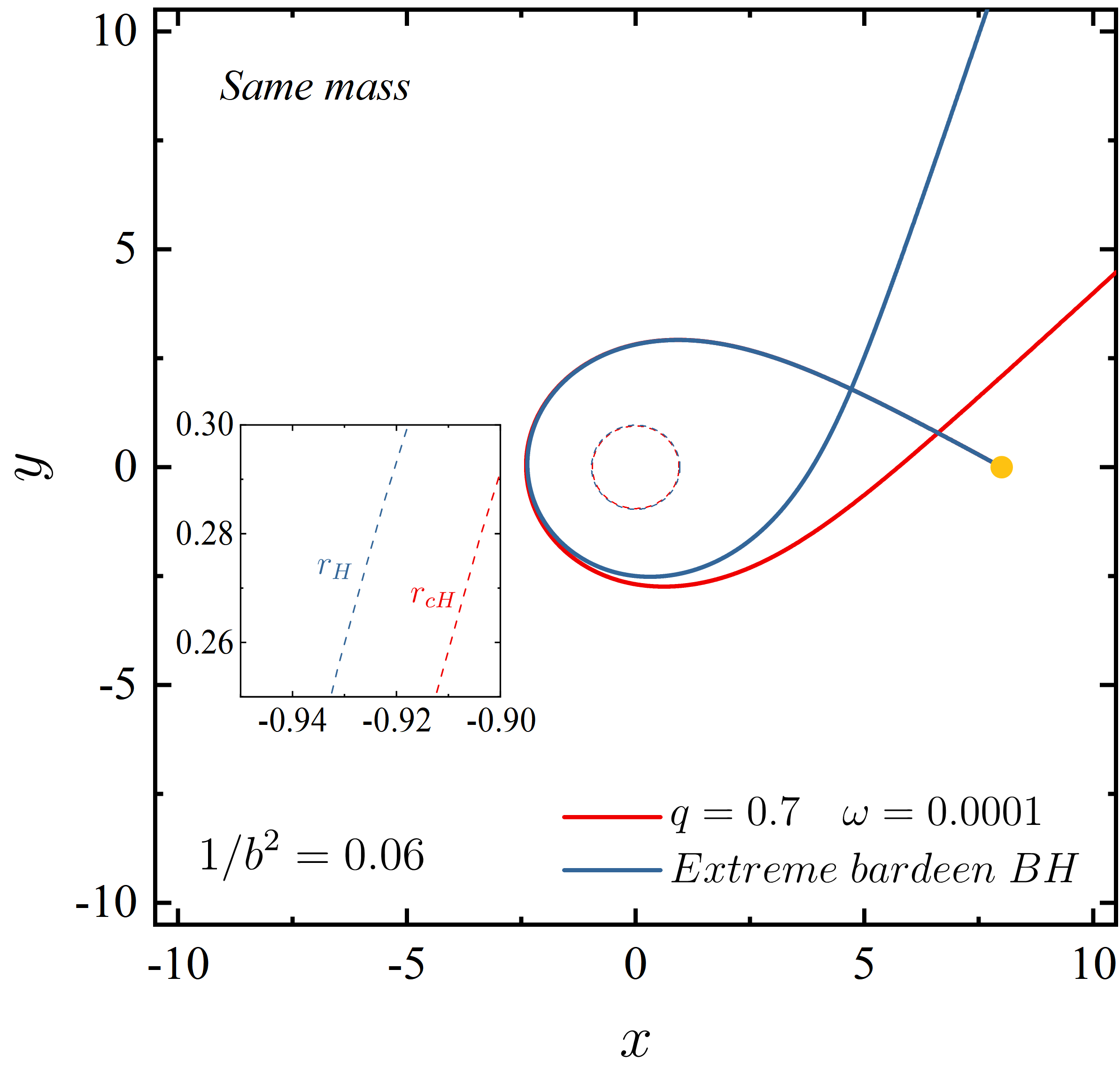}
   		} 	
            \subfigure{
			\includegraphics[height=.20\textheight,width=.20\textheight, angle =0]{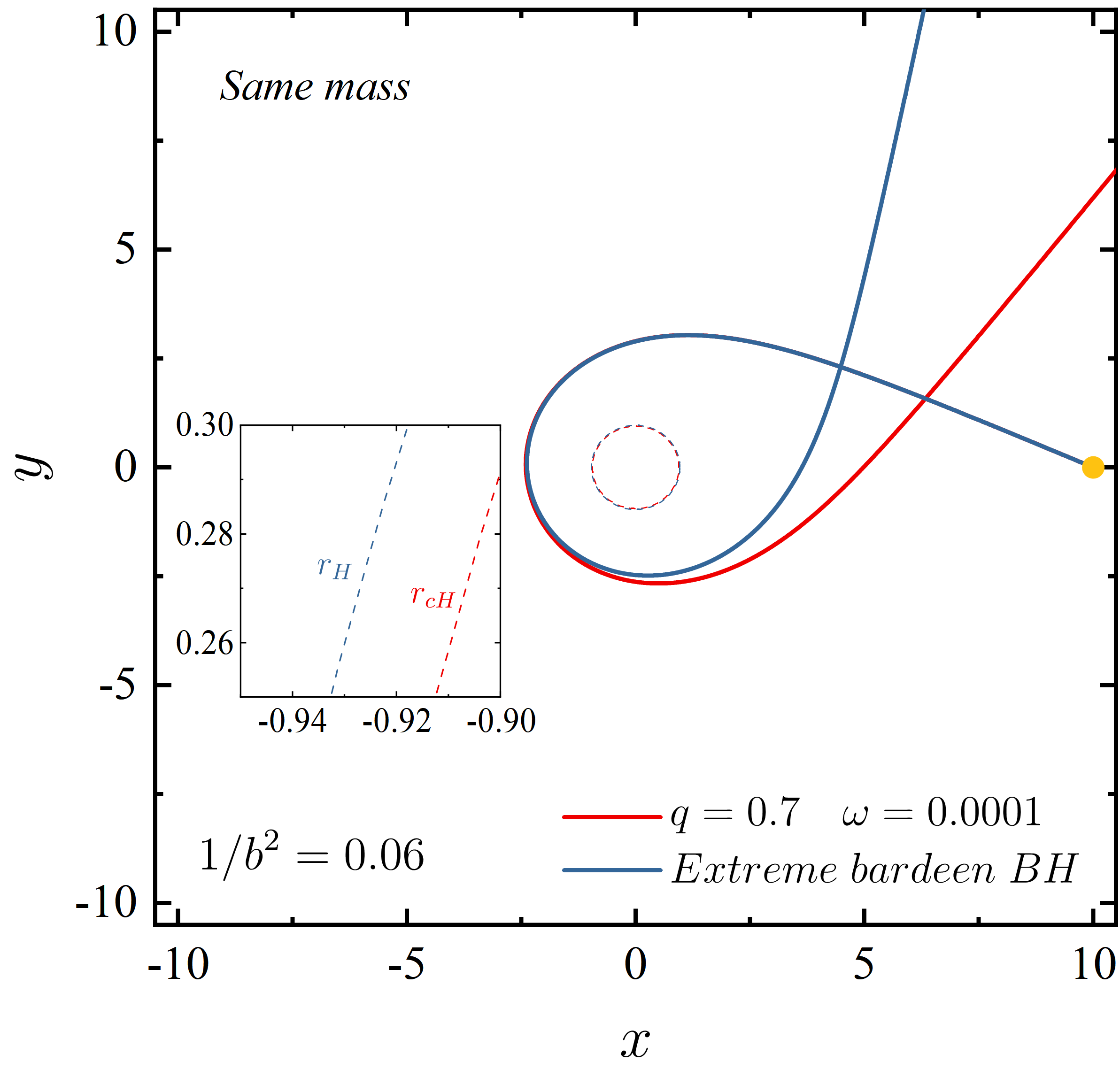}
   		} 
    \quad
        \subfigure{  
			\includegraphics[height=.20\textheight,width=.20\textheight, angle =0]{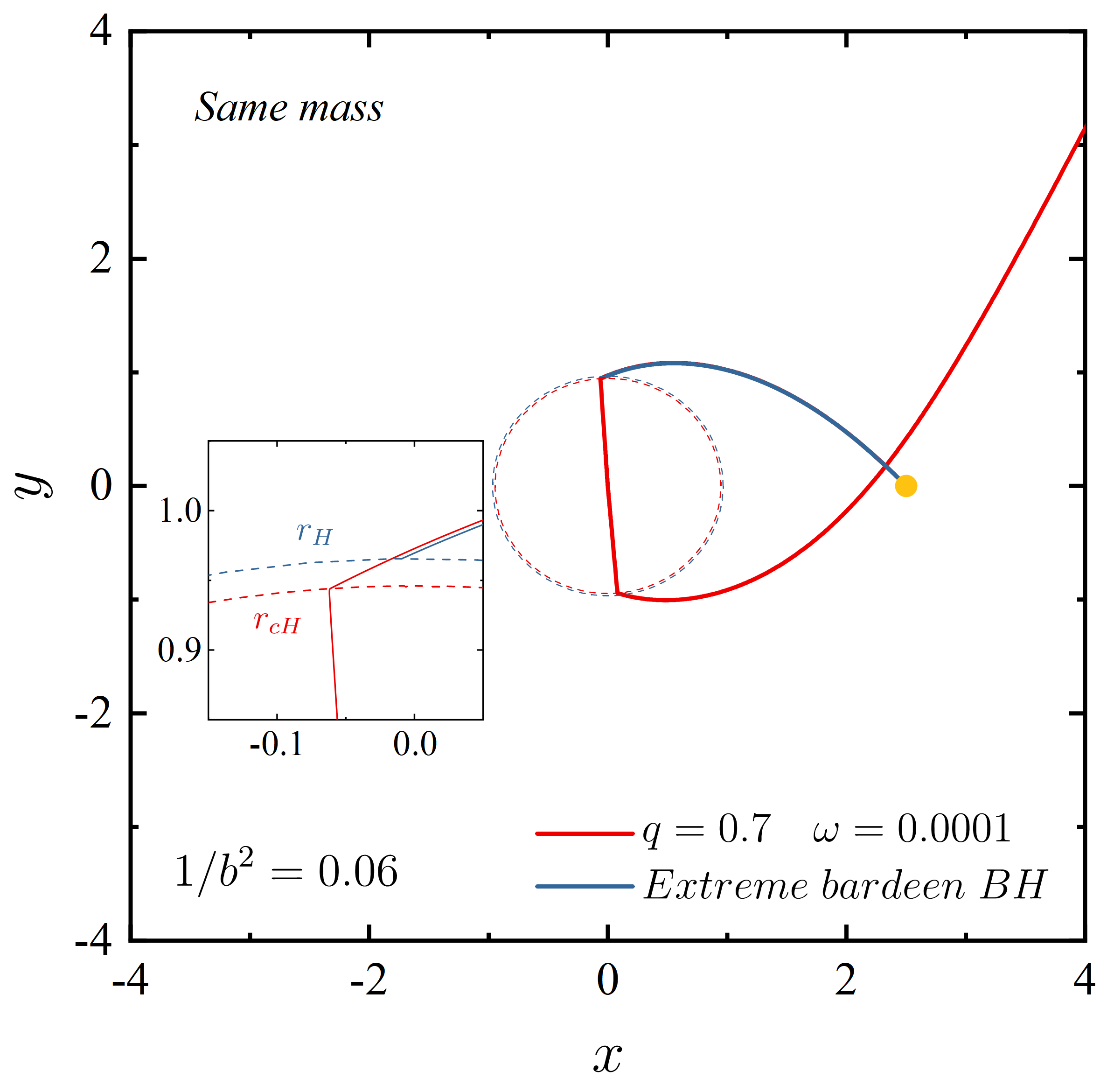}
		}
	\subfigure{
			\includegraphics[height=.20\textheight,width=.20\textheight, angle =0]{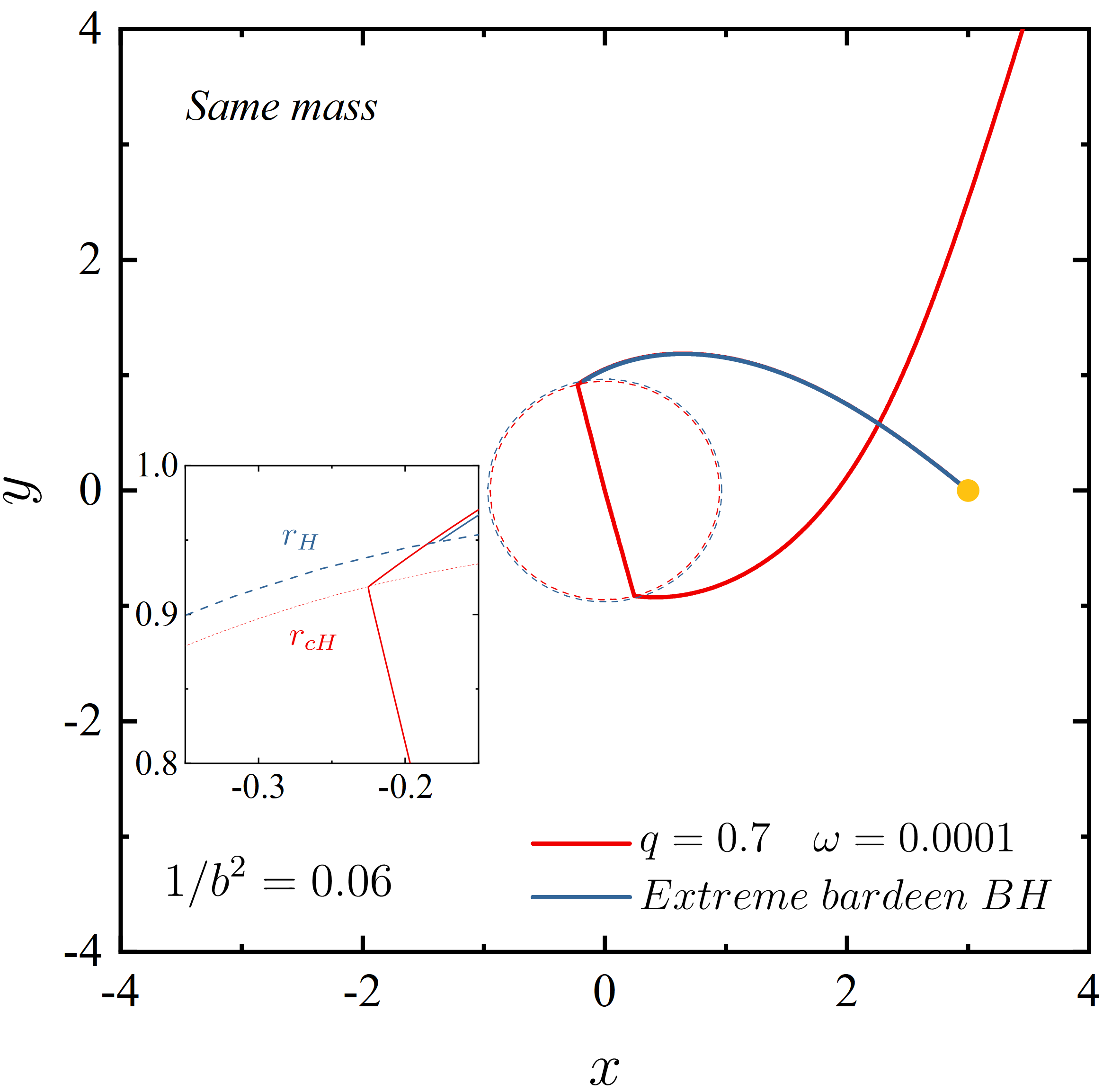}
   		} 	
        \subfigure{
			\includegraphics[height=.20\textheight,width=.20\textheight, angle =0]{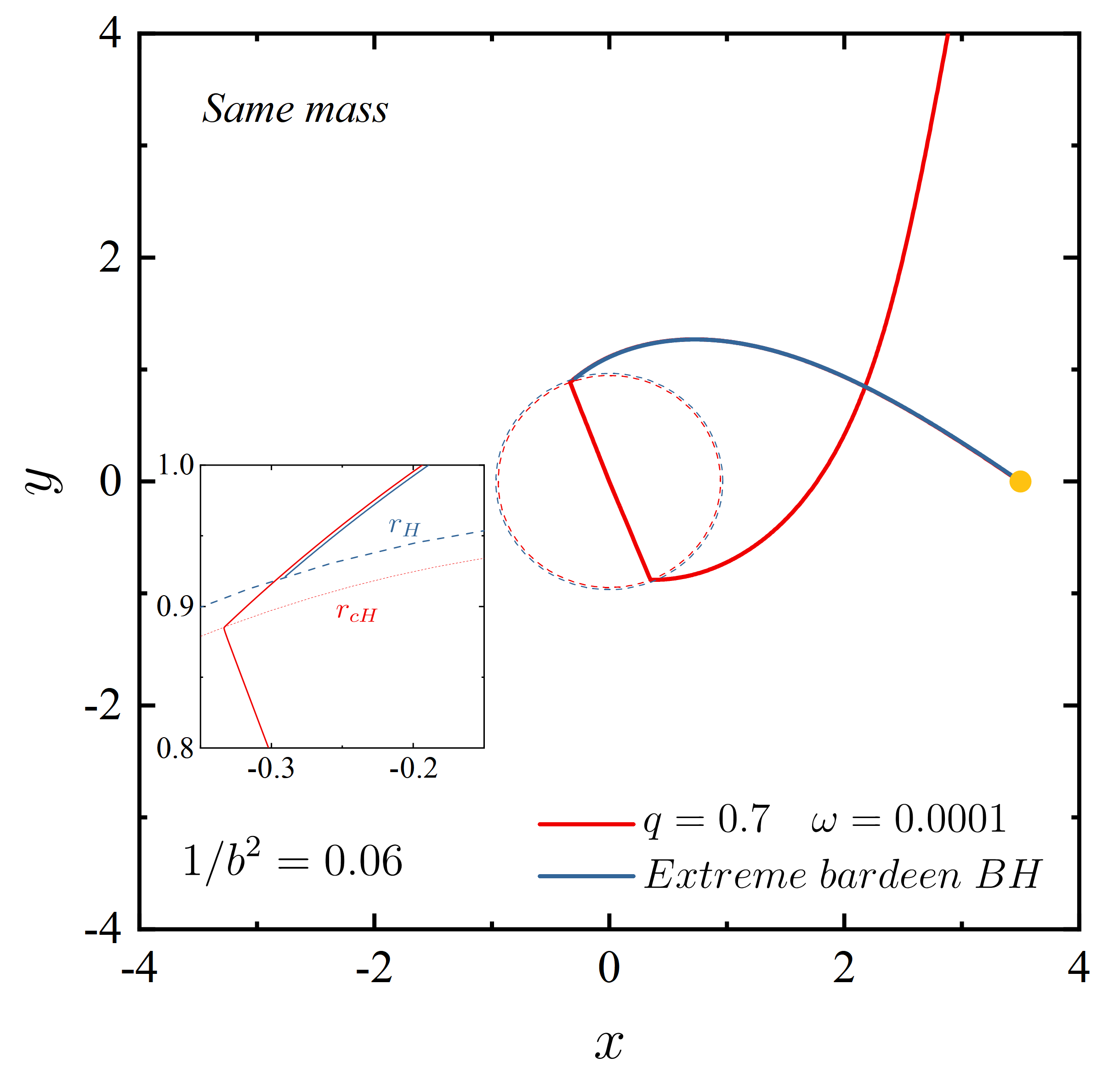}
   		}	
		\caption{The orbit of the photon in the background of the FBBS and the extreme Bardeen BH with same mass ($M=0.887018$). The parameters of the extreme Bardeen BH are $s=0.179461$ and $q=0.682827$. In the bottom panel, the dashed lines of different colors correspond to the event (or critical) horizons of different models.}
		\label{fig:samem}		
		\end{figure}
%
In the case of FBBSs, they all have two LRs, and the two LRs are located on either side of the critical horizon. We provide the radius of the LR and the critical horizon of FBBSs in Table~\ref{tab:rh}. It can be seen that these radii increase as the magnetic charge increases. To illustrate the properties of unbound photon orbits of these FBBSs, we present the unbound photon orbits of the BBSs with different frequencies in Fig.~\ref{fig:FNBHLR}, while Fig.~\ref{fig:EFFFNBHLR} shows the corresponding radial effective potentials. These BBSs have the same magnetic charge $q=0.7$ (illustrative) and relatively small frequency ($\omega=0.1$, $0.001$, $0.0001$). From Fig.~\ref{fig:FNBHLR}, it can first be seen that the orbit of the photon bends in two specific regions. As the frequency decreases, the two specific regions become smaller, and the orbit becomes sharper at the two regions. It is known from Fig.~\ref{fig:function} that the value of $(-g_{tt})_{min}$ and $g^{rr}_{min}$ decreases as the frequency decreases for the case of $q = 0.7$. Therefore, this, in fact, implies that the more ``frozen" the solution is, the sharper the deflections of the orbits formed in it will be. In particular, for the FBBSs with $\omega=0.0001$, as seen in the bottom panel, the photon orbit becomes a straight line after a sharp deflection in a tiny region near the critical horizon $r_{cH}$. It is not until the photon reaches $r=r_{cH}$ again that the orbit is drastically deflected again in a tiny region near $r_{cH}$ and then moves toward infinity.

	\begin{figure}[!htbp]
		\centering	
        \subfigure{  
			\includegraphics[height=.20\textheight,width=.20\textheight, angle =0]{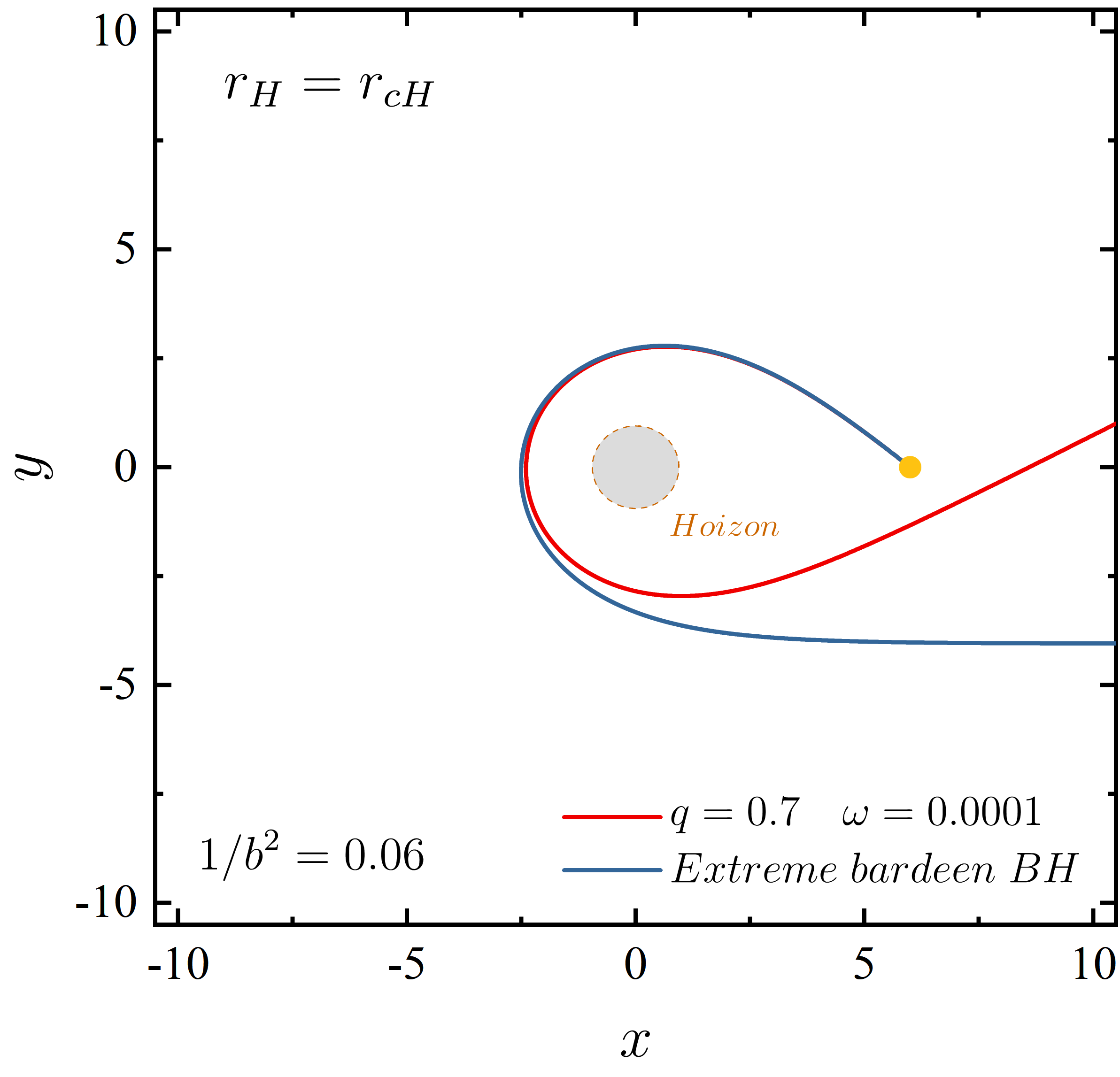}
		}
		\subfigure{
			\includegraphics[height=.20\textheight,width=.20\textheight, angle =0]{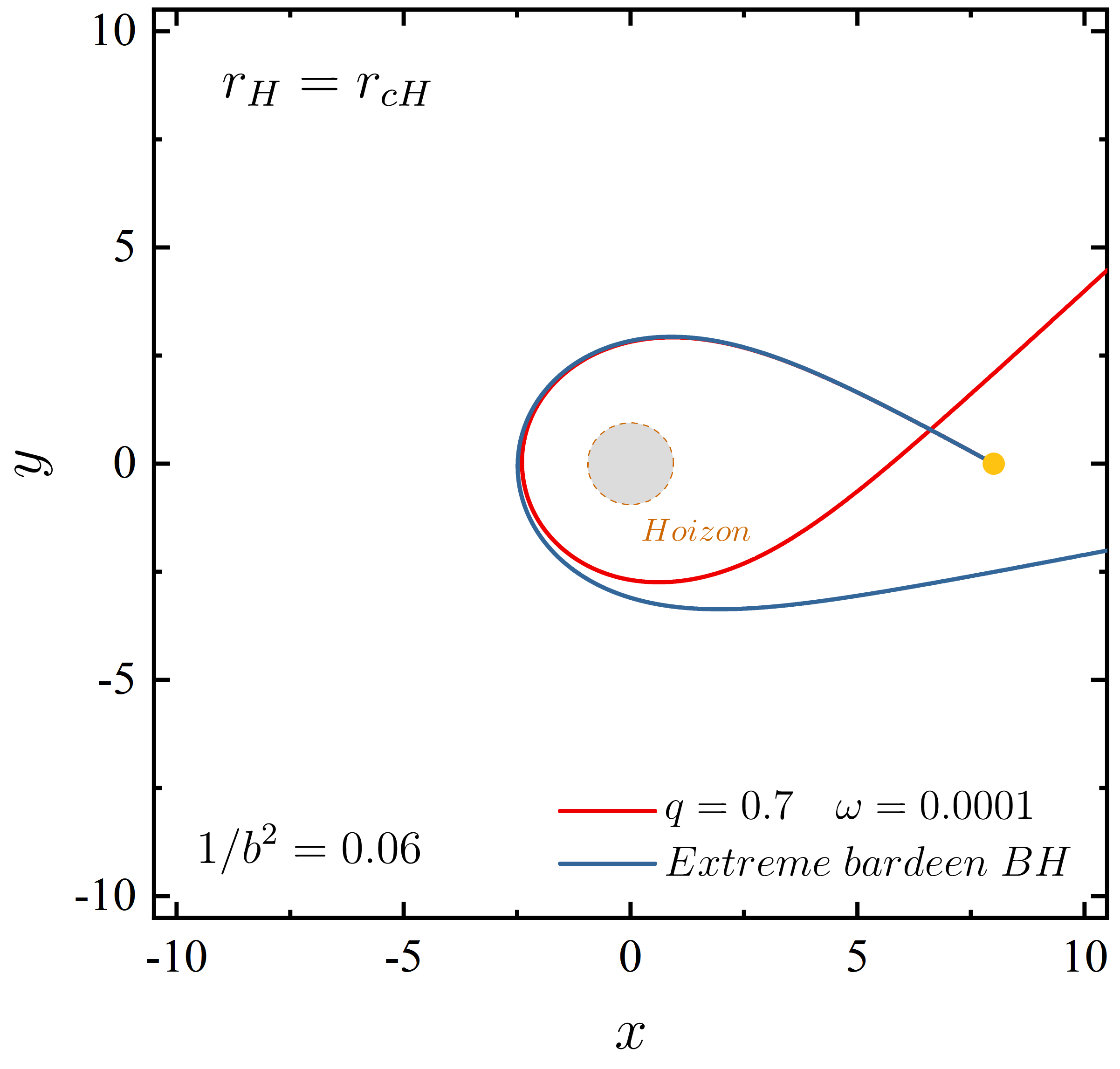}
   		} 	
            \subfigure{
			\includegraphics[height=.20\textheight,width=.20\textheight, angle =0]{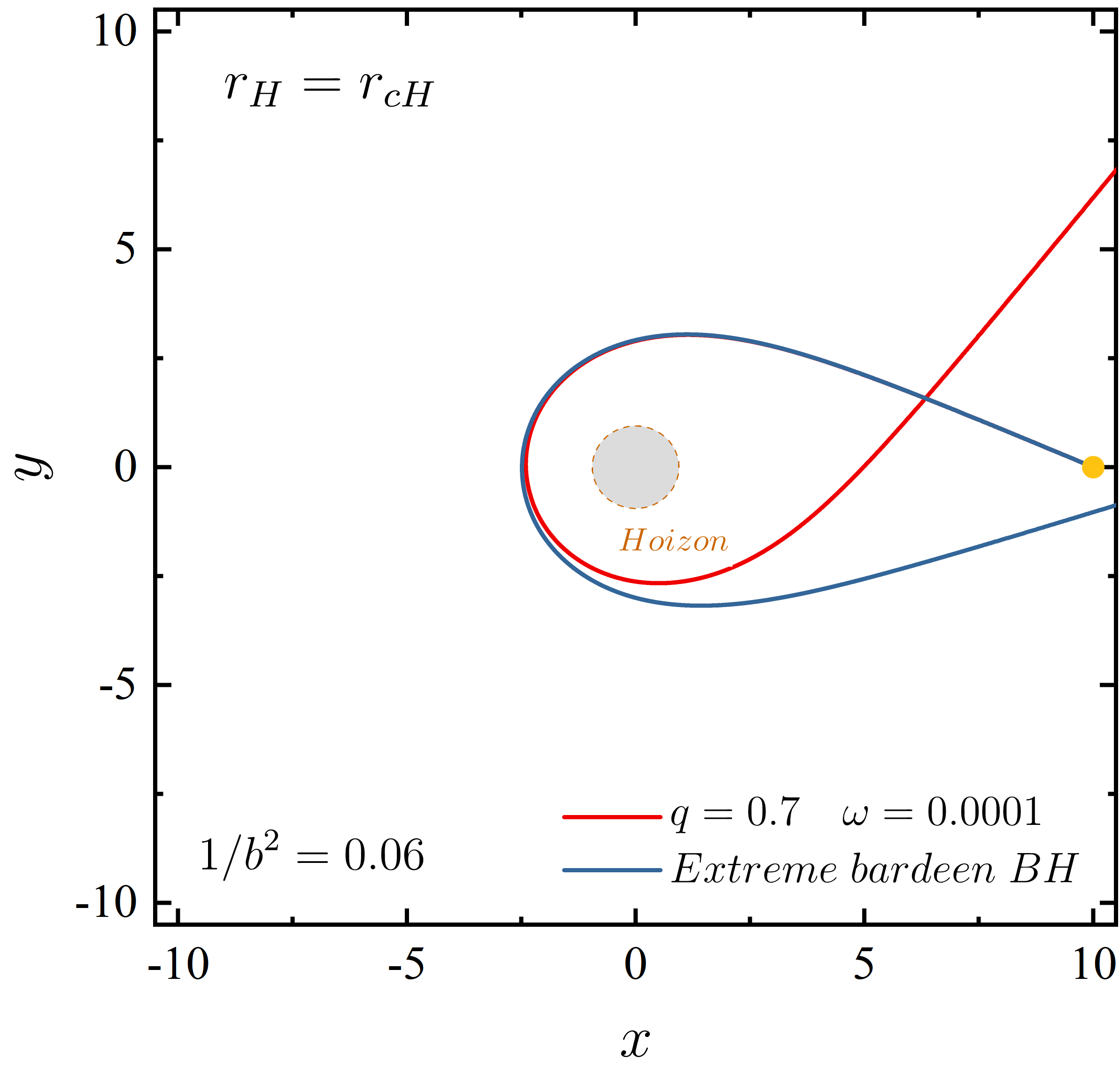}
   		} 
    \quad
        \subfigure{  
			\includegraphics[height=.20\textheight,width=.20\textheight, angle =0]{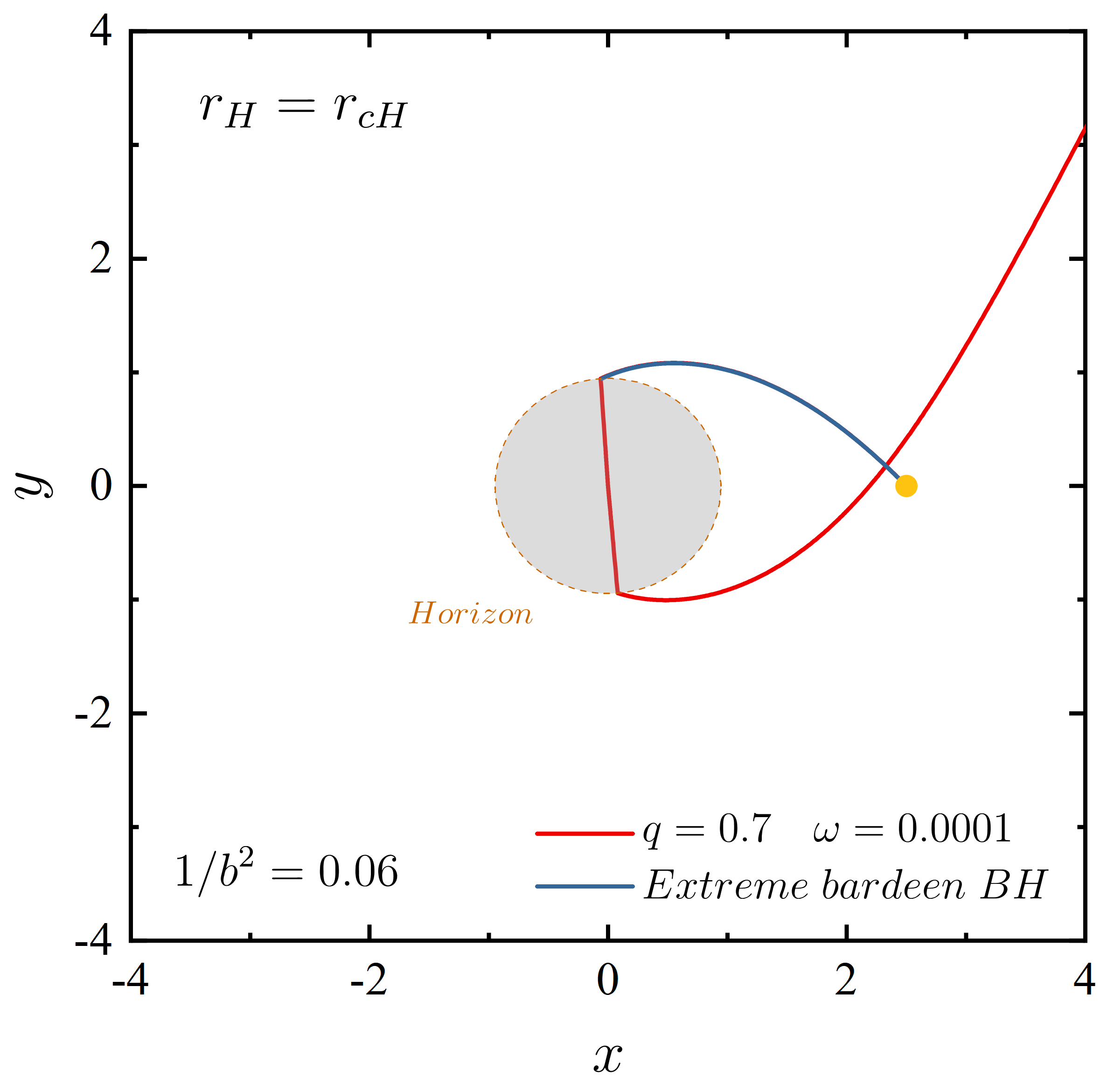}
		}
	\subfigure{
			\includegraphics[height=.20\textheight,width=.20\textheight, angle =0]{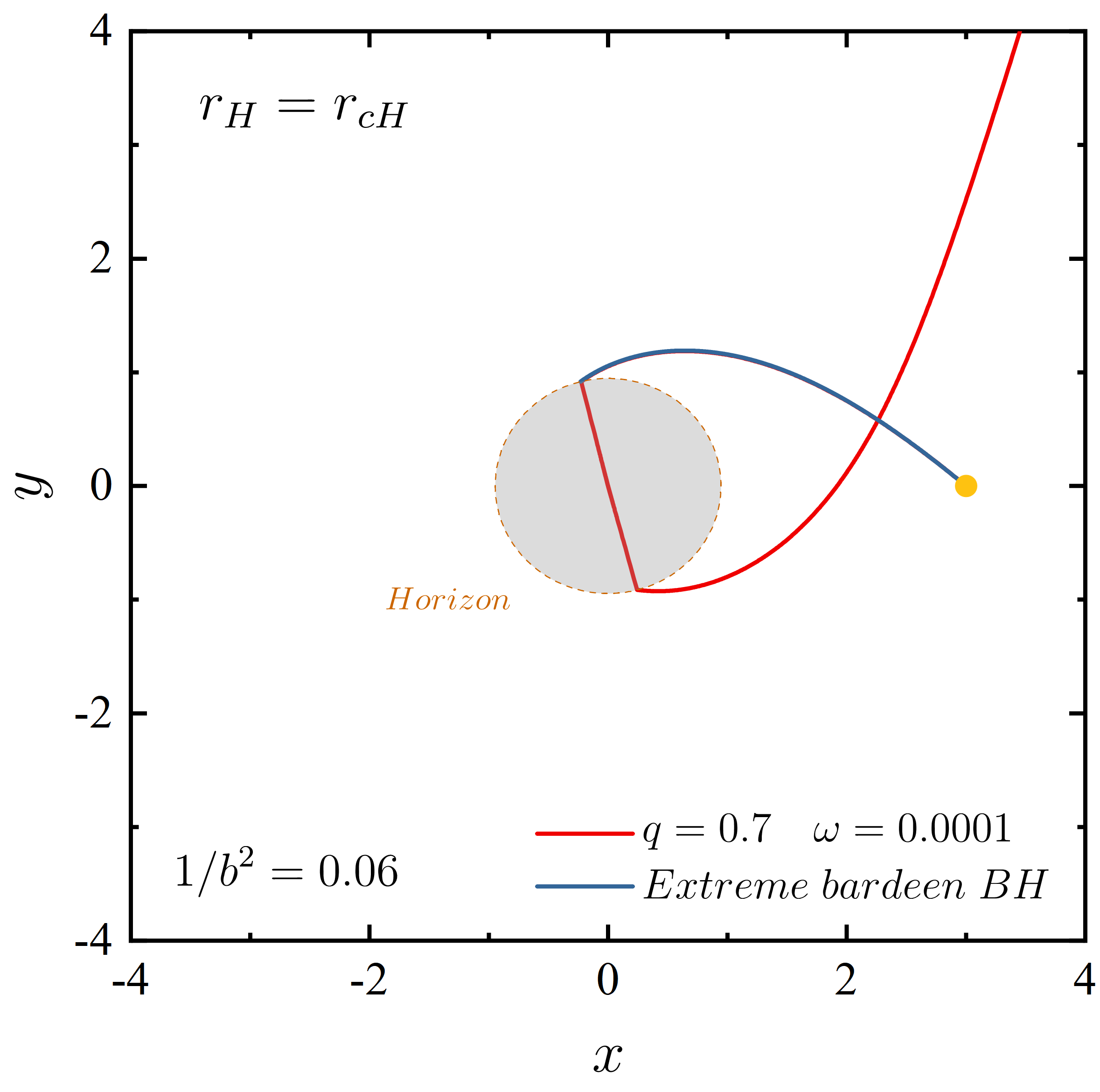}
   		} 	
        \subfigure{
			\includegraphics[height=.20\textheight,width=.20\textheight, angle =0]{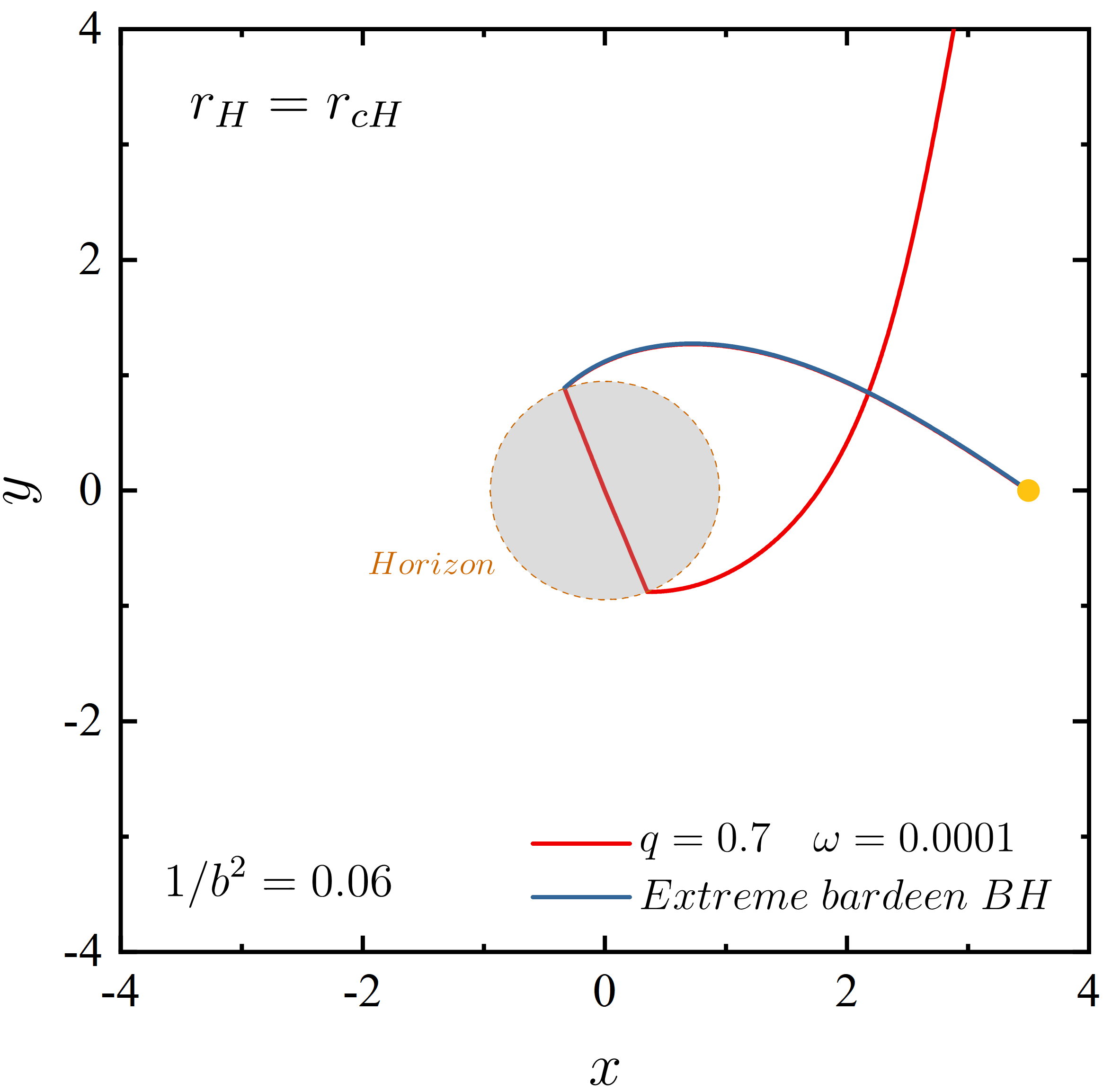}
   		}	
		\caption{The orbit of the photon in the background of the FBBS and the extreme Bardeen BH with the event horizon $r_{H}$ equal to the critical horizon $r_{cH}$. The parameters of the extreme Bardeen BH are $s=0.172261$ and $q=0.668989$.}
		\label{fig:sameh}		
		\end{figure}

As shown in Fig.~\ref{fig:samem} and Fig.~\ref{fig:sameh}, we show the orbits of photons in the background of the extreme Bardeen BHs and the FBBSs with $1/b^2=0.2$ or $1/b^2=0.06$, respectively. The yellow dots in these figures indicate the coordinate
 positions of the photons. In Fig.~\ref{fig:samem}, the mass of the extreme Bardeen BH is equal to that of the frozen star, and in Fig.~\ref{fig:sameh}, the radius of the event horizon $r_H$ of the extreme Bardeen BH is equal to the radius of the critical horizon $r_{cH}$ of the FBBS. It can be seen from the top panels of Fig.~\ref{fig:samem} and Fig.~\ref{fig:sameh} that the trajectories of photons are very close in the backgrounds of the FBBSs and Bardeen BHs at first, however, as the the photons spiral into the FBBS and Bardeen BH, the difference in the trajectories becomes more and more obvious. In addition, it can be seen from the bottom panels of Fig.~\ref{fig:samem} and Fig.~\ref{fig:sameh} that, unlike BHs, for FBBSs, photons can pass through the central region of FBBSs and then gradually move away from FBBSs. Therefore, although the spacetime geometry of the FBBS is very similar to that of the extreme BH, one can distinguish the FBBS and extreme BH with such kinds of orbits.

  	\begin{table}[!htbp]     
    \scriptsize 
	\centering 
	\begin{tabular}{c||c|c|c}
		\hline
		 $q$ & $r_{LR1}$ & $r_{LR2}$ &$r_{cH}$ \\  \hline  \hline
		$0.6$ & $0.4862$ & $1.3697$ & $0.5191$  \\  
		$0.62$ & $0.4588$ & $1.5073$ & $0.6427$  \\ 
		$0.64$ & $0.6295$ & $1.6375$ & $0.7262$  \\ 
  	$0.66$ & $0.7224$ & $1.7655$ & $0.7262$  \\ 
  	$0.68$ & $0.8109$ & $1.8289$ & $0.8021$  \\ 
		$0.7$ & $0.8988$ & $2.0204$ & $0.9461$  \\ 
		\hline
	\end{tabular}
 	\caption{$2^{nd}$ and $3^{rd}$ columns: the radius of inner LR ($r_{LR1}$) and outer LR ($r_{LR2}$) in the background of FBBSs; $4^{th}$ column: the critical horizon $r_{cH}$.}
	\label{tab:rh}
\end{table}

\begin{figure}[!htbp]
		\centering	
        \subfigure{  
			\includegraphics[height=.20\textheight,width=.20\textheight, angle =0]{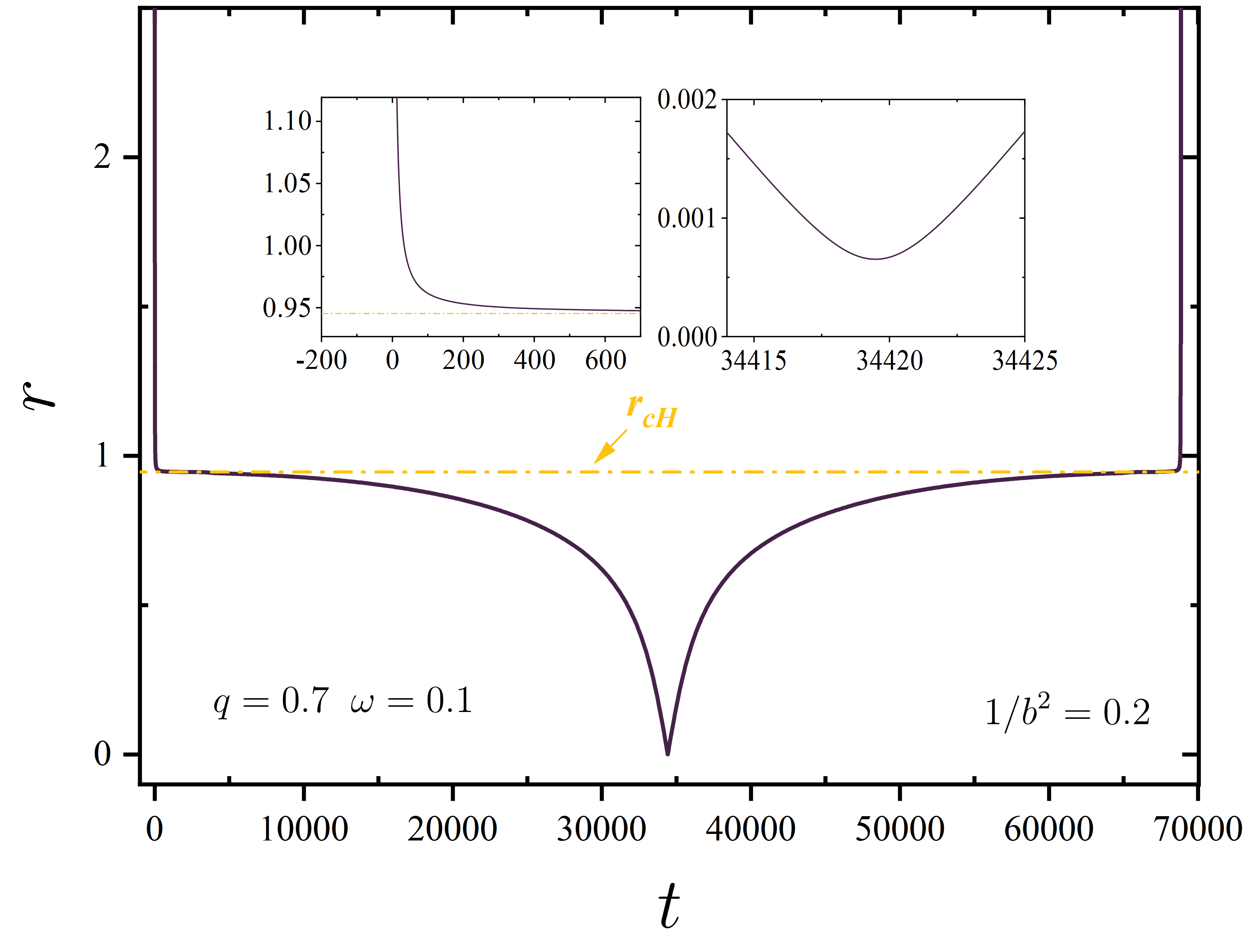}
		}
		\subfigure{
			\includegraphics[height=.20\textheight,width=.20\textheight, angle =0]{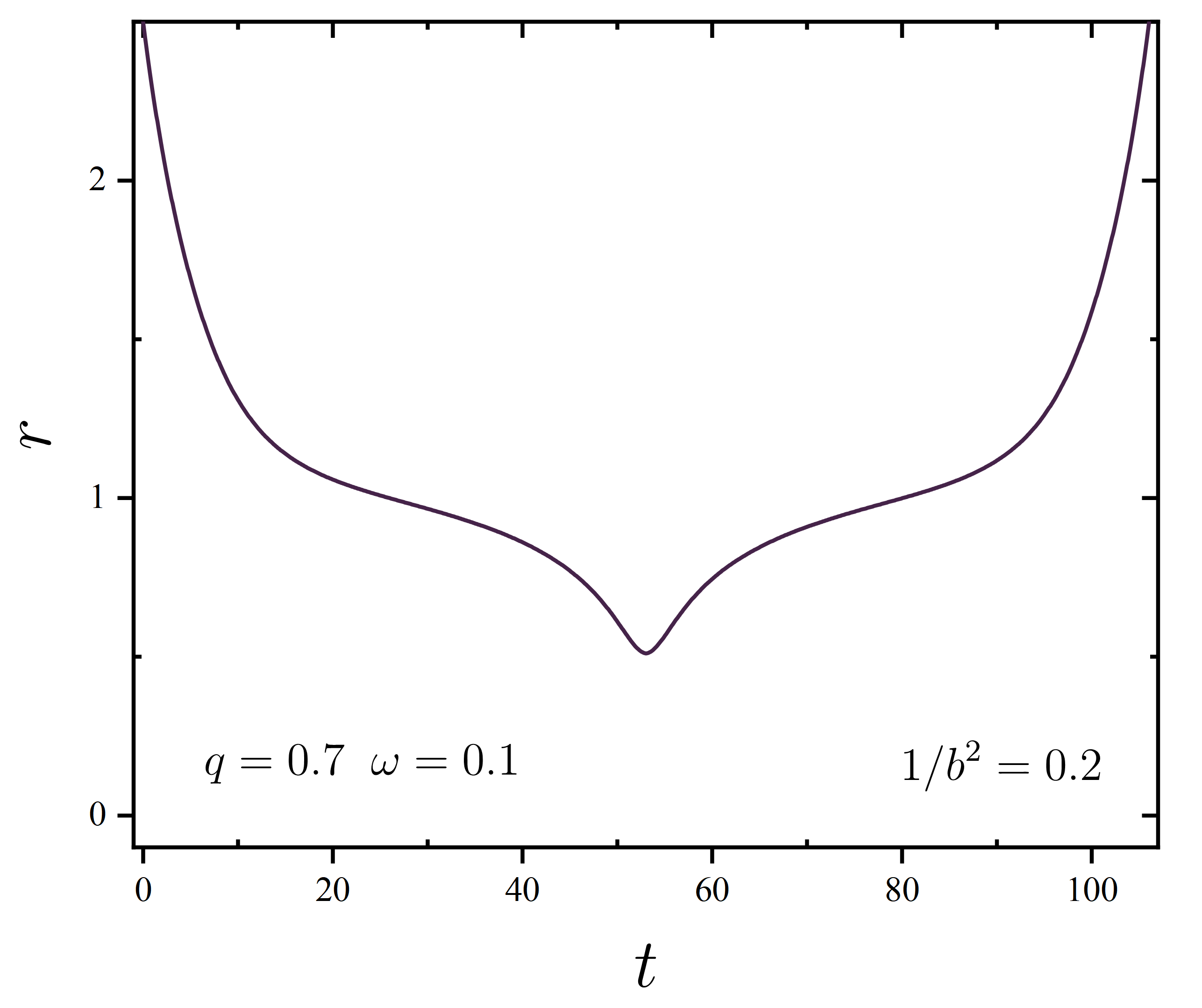}
   		} 	
            \subfigure{
			\includegraphics[height=.20\textheight,width=.20\textheight, angle =0]{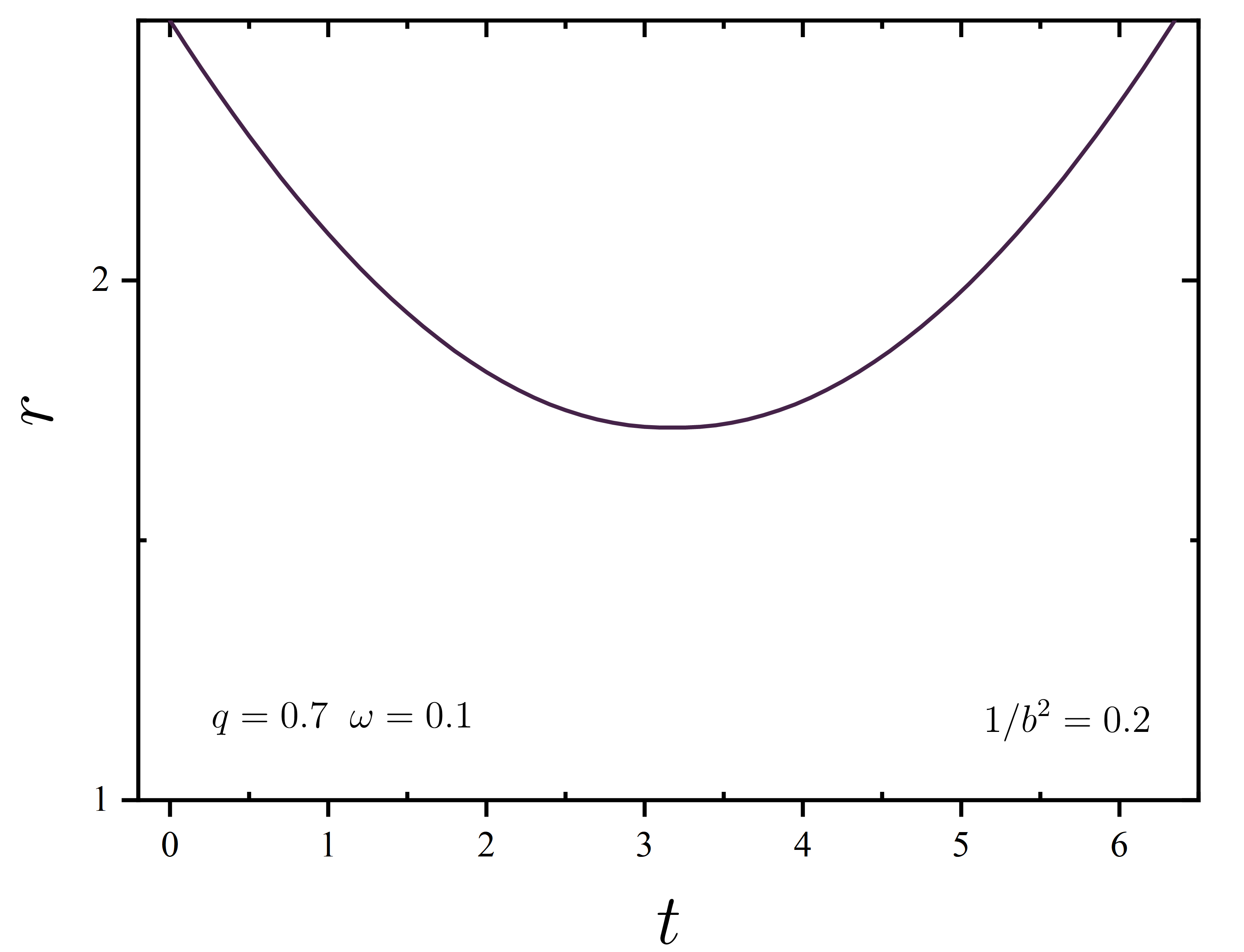}
   		} 
		\caption{Evolution of the radius of the null geodesic orbit in BBSs (left and middle) and BSs (right) with the LR.}
		\label{fig:RTUBHLR}		
		\end{figure}

Figure~\ref{fig:RTUBHLR} shows the evolution of the radial distance $r$ of the orbit as a function of time $t$ for the FBBS, general BBS, and the BS. From this figure, we can first see that for FBBSs (left panel), the radial distance inside $r_{cH}$ changes very slowly (especially near $r_{cH}$). In other words, the photon stays inside the critical Horizon of FBBs for a long time. In contrast, it can be seen that photons only take a short time to move the same distance in the background of general BBSs (middle panel) and BSs (right panel).

It is worth noting that for the frozen star solution, although the photon trajectories shown in Fig.~\ref{fig:RTUBHLR} can reach very close to $r = 0$ after entering the critical horizon, these photons never actually reach $r = 0$. This is because, according to equation (\ref{eq:latterenergy}), the radial effective potential of photons diverges to infinity at $r = 0$. Therefore, when $L\neq0$, a photon would require infinite energy and infinite velocity to reach $r = 0$. Only when $L = 0$ can a photon have a finite velocity at $r = 0$.

\begin{figure}[!htbp]
		\centering	
        \subfigure{  
			\includegraphics[height=.28\textheight,width=.30\textheight, angle =0]{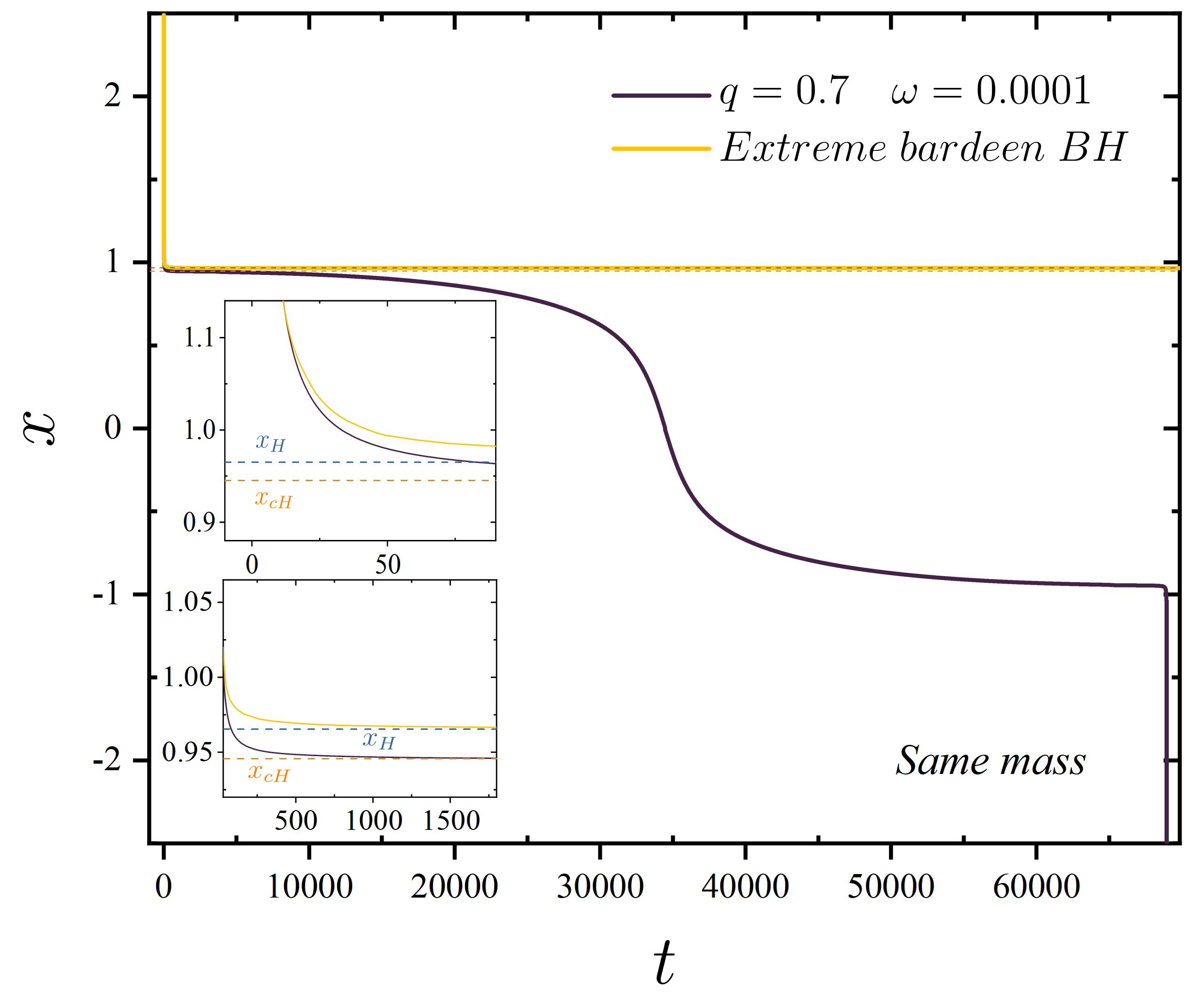}
		}
                \subfigure{  
			\includegraphics[height=.28\textheight,width=.30\textheight, angle =0]{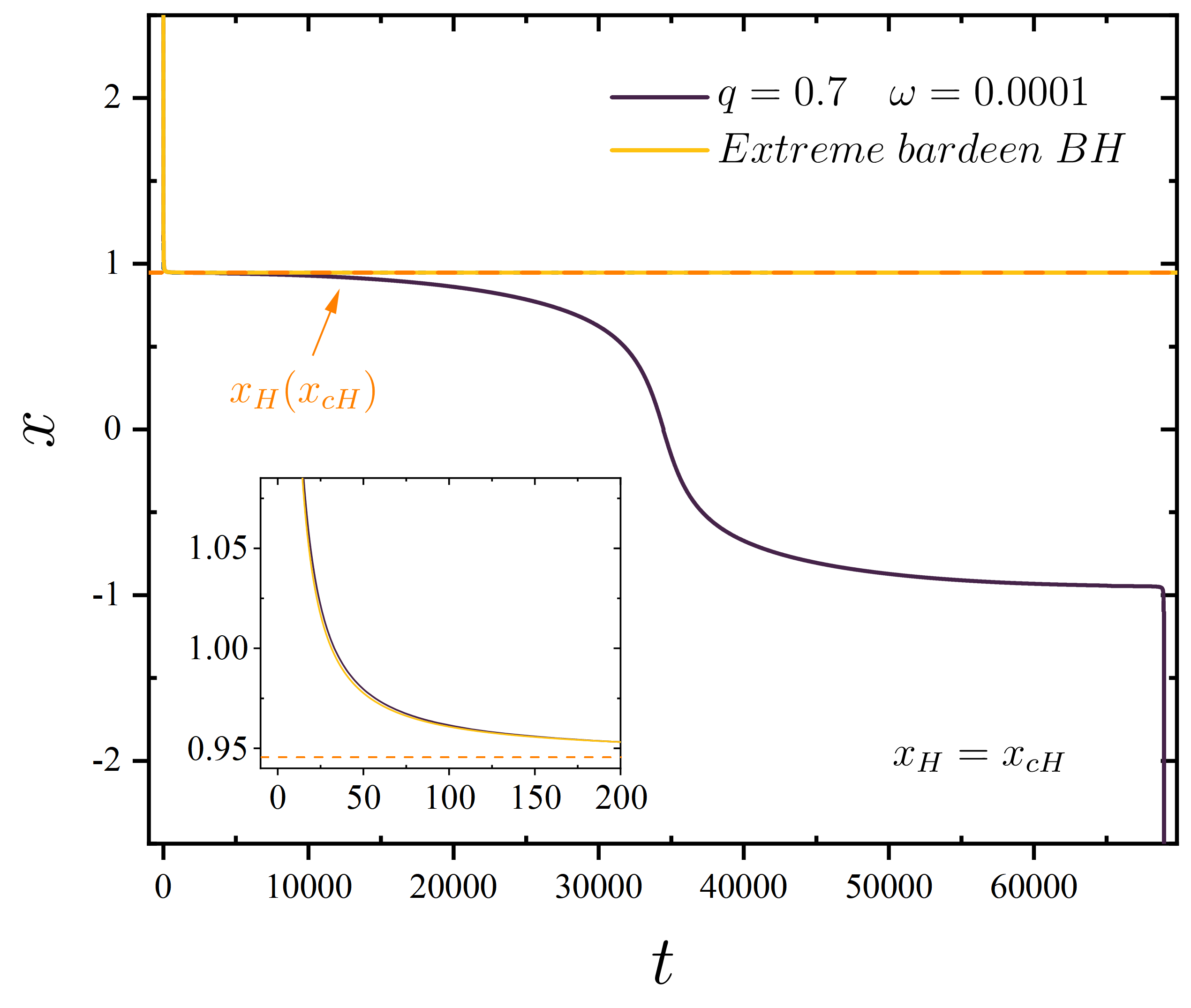}
		}

		\caption{The radius $r$ of the unbound orbit as a function of time $t$ around the FBBS and extreme Bardeen BH. For the left panel, the parameters of the Bardeen BH are $s = 0.179461$ and $q = 0.682827$, and for the right panel, the parameters are $s = 0.172261$ and $q = 0.668989$.}
		\label{fig:EFFFNBHLRL0}		
		\end{figure}

In the case of $L=0$ (or equivalently $b=0$), since the orbit can only be a straight line, without loss of generality, we can assume that the photon moves along the $x$-axis (therefore $|x|=r$). Figure~\ref{fig:EFFFNBHLRL0} exhibits the radial distance of the orbit of the FBBSs as a function of time in this case (the initial coordinate position of the photon is $(x=2.5$, $y=0)$). It can be seen that, although for an observer at infinity, the photon around FBBSs spend an extremely long time moving inside the critical horizon, the orbit of the photon in the background of the FBBS behaves differently from in the background of the Bardeen BH (yellow line) with the same mass (left panel) or the event horizon $x_{H}$ equal to the critical horizon $x_{cH}$ (right panel). Similarly to the case of $L\neq0$, the FBBS allows the photon to pass through the critical horizon $x=x_{cH}$ and the origin of coordinates, then exit through another critical horizon $x=-x_{cH}$, and finally move toward infinity.

\subsection{Bound orbit}

%
	\begin{figure}[!htbp]
		\centering	
        \subfigure{  
			\includegraphics[height=.20\textheight,width=.20\textheight, angle =0]{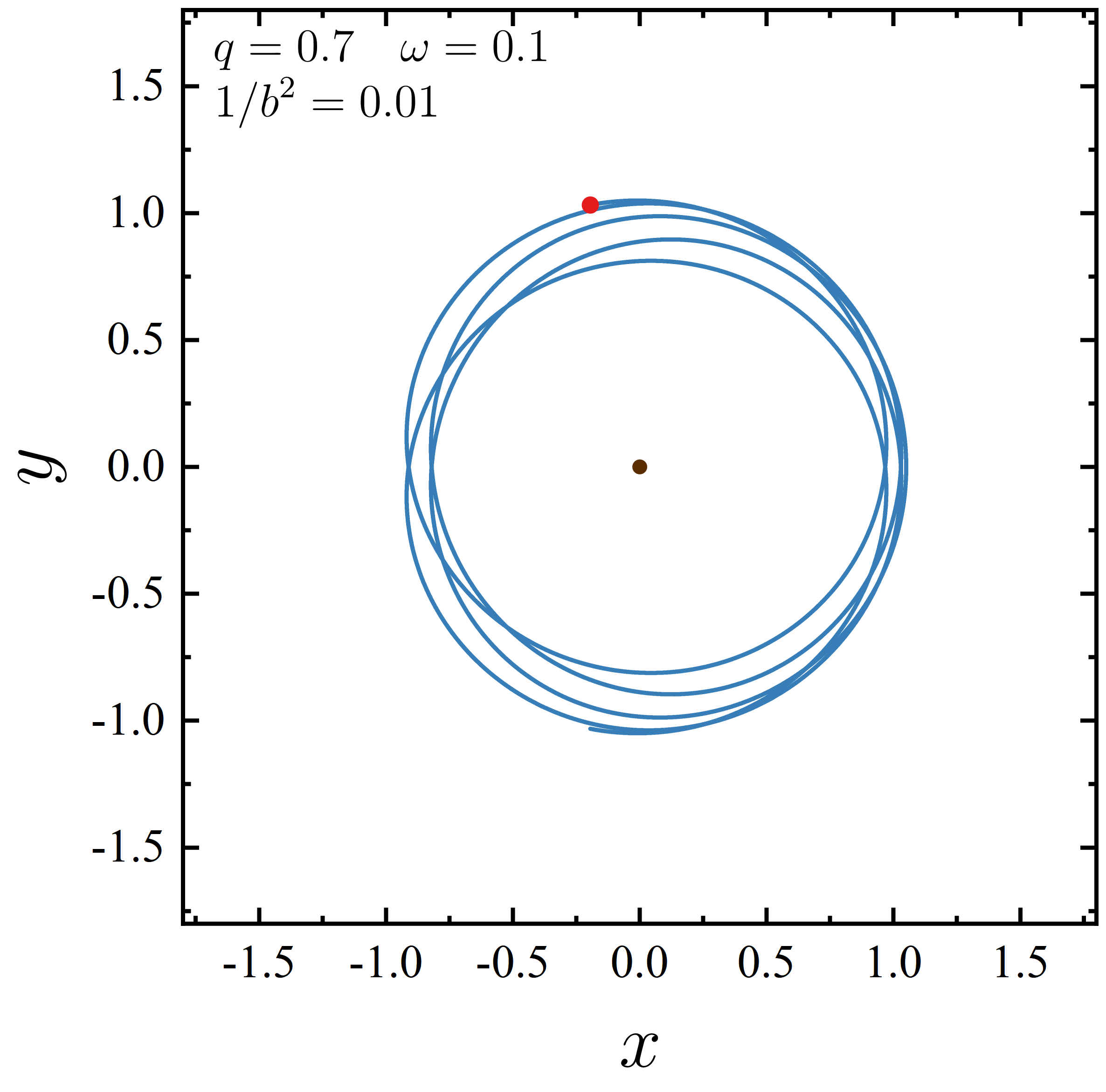}
		}
		\subfigure{
			\includegraphics[height=.20\textheight,width=.20\textheight, angle =0]{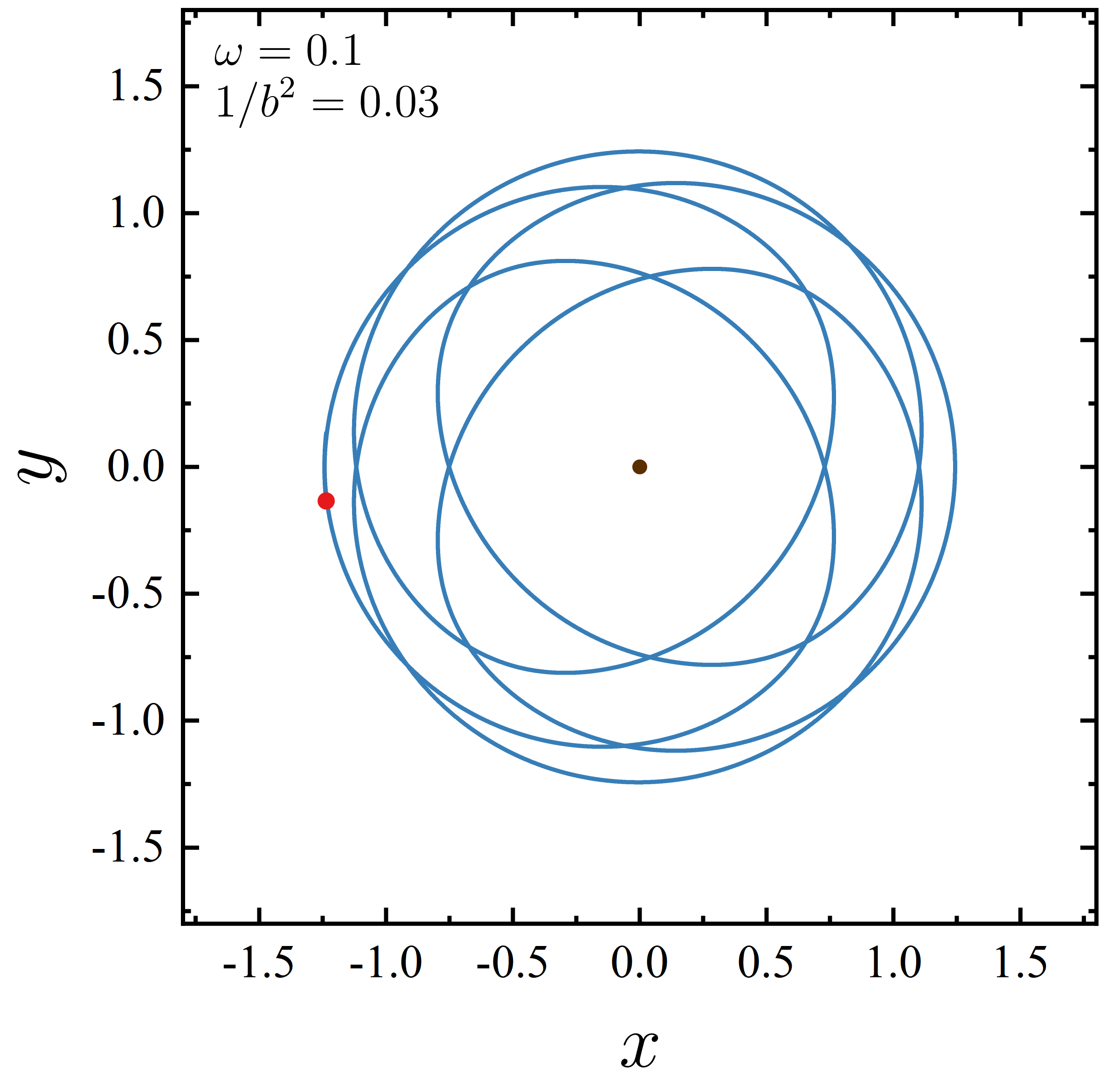}
   		} 	
            \subfigure{
			\includegraphics[height=.20\textheight,width=.20\textheight, angle =0]{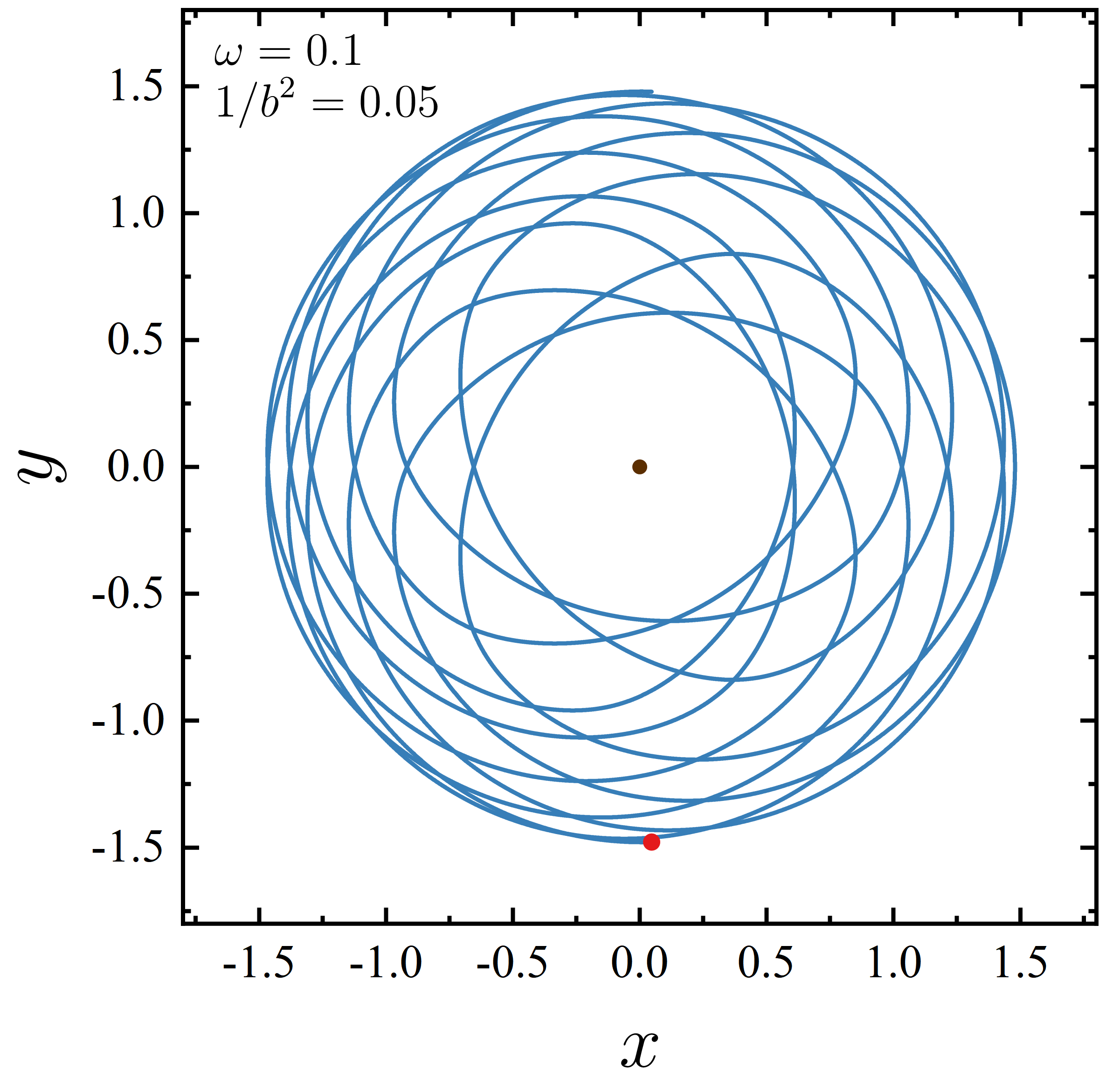}
   		} 
    \quad
        \subfigure{  
			\includegraphics[height=.20\textheight,width=.20\textheight, angle =0]{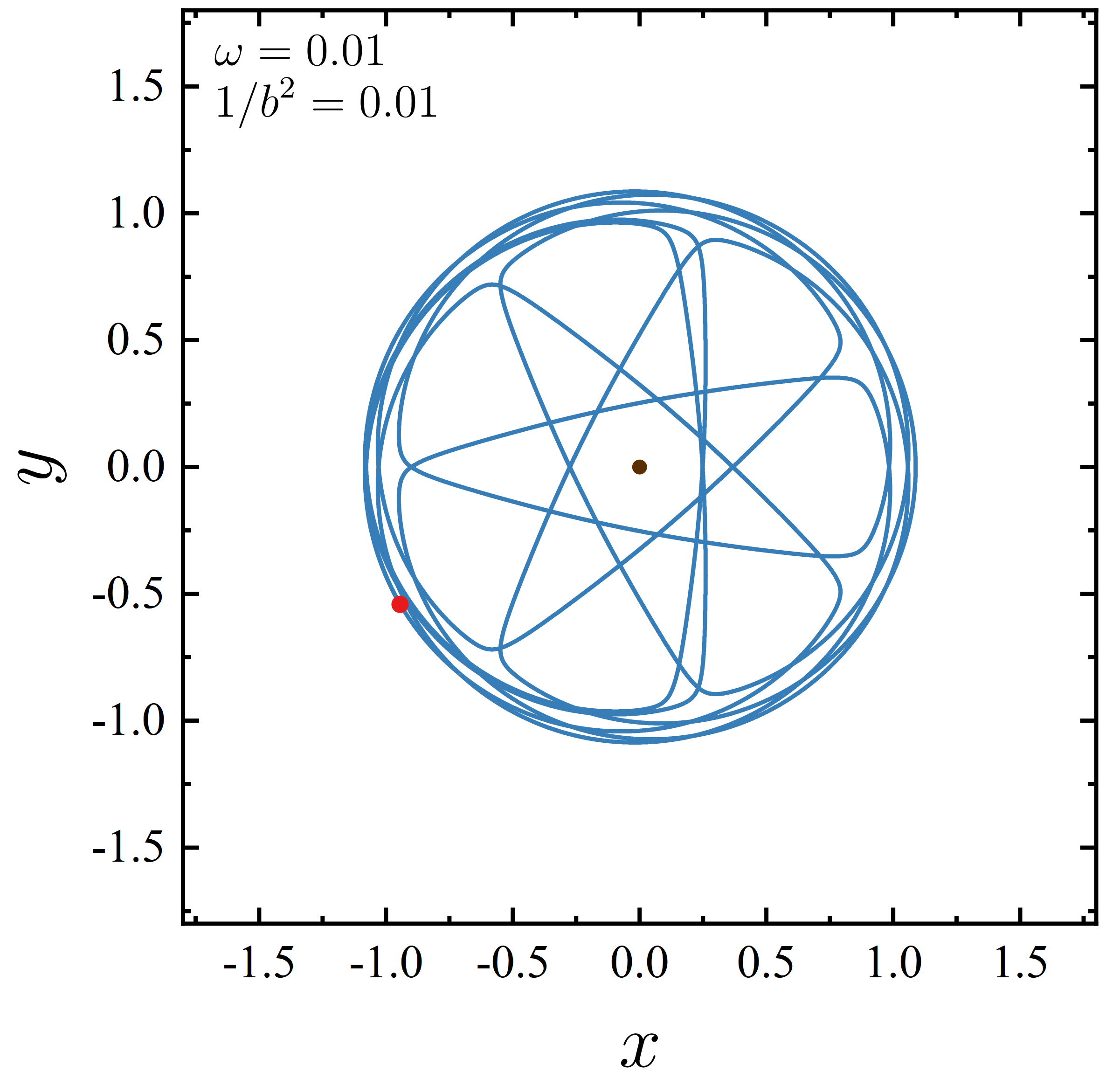}
		}
	\subfigure{
			\includegraphics[height=.20\textheight,width=.20\textheight, angle =0]{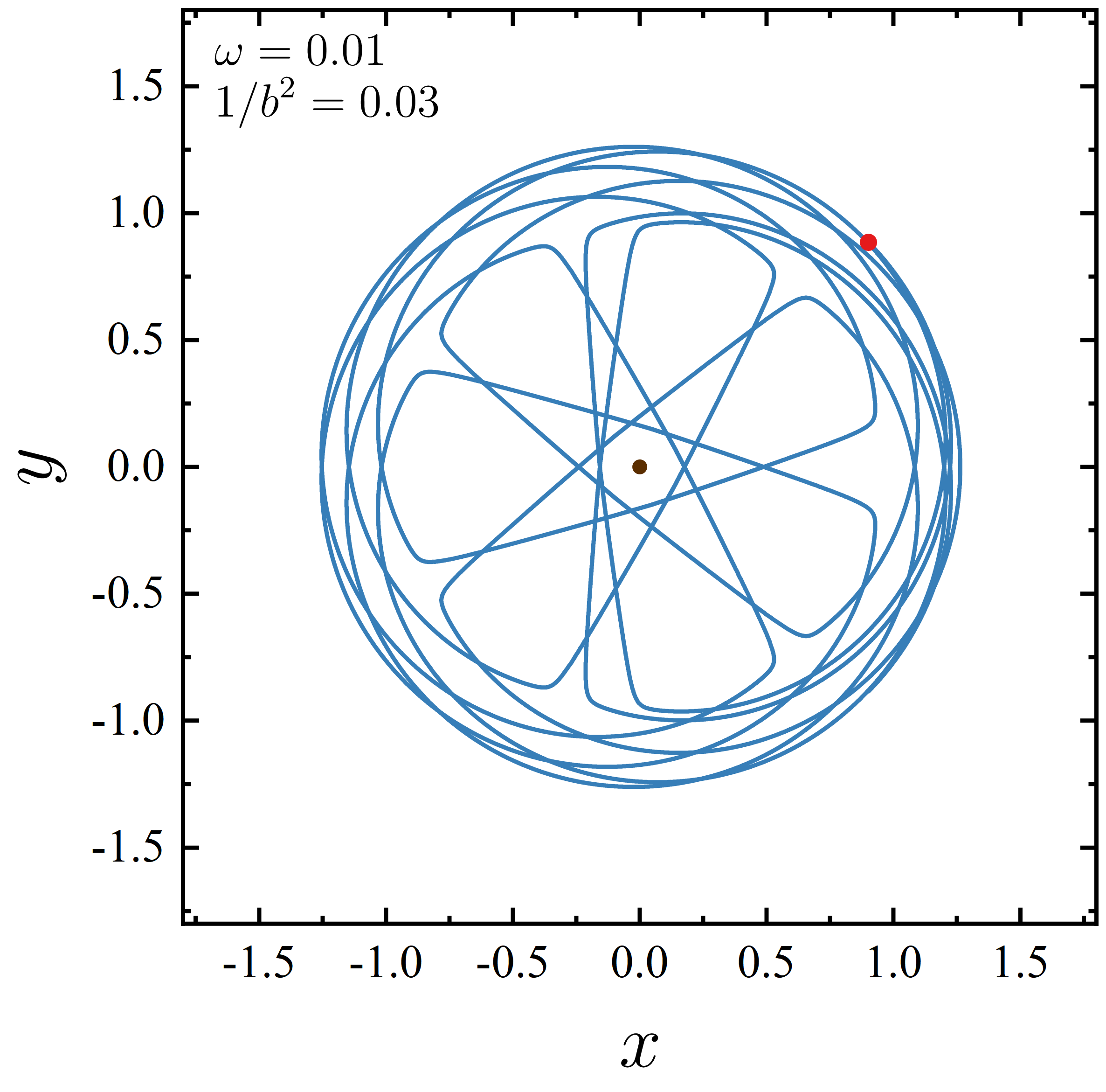}
   		} 	
        \subfigure{
			\includegraphics[height=.20\textheight,width=.20\textheight, angle =0]{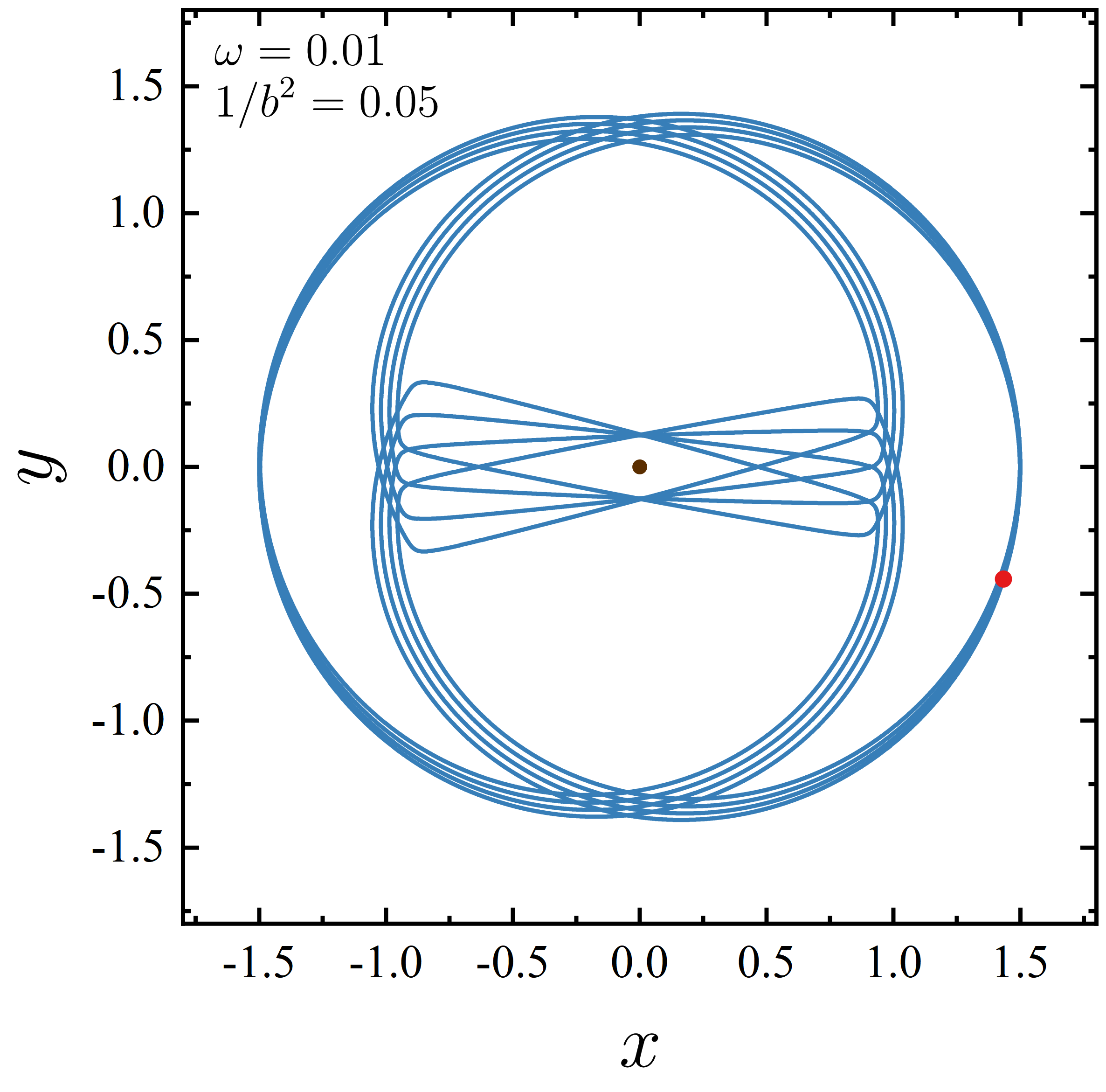}
   		}
            \quad
        \subfigure{  
			\includegraphics[height=.20\textheight,width=.20\textheight, angle =0]{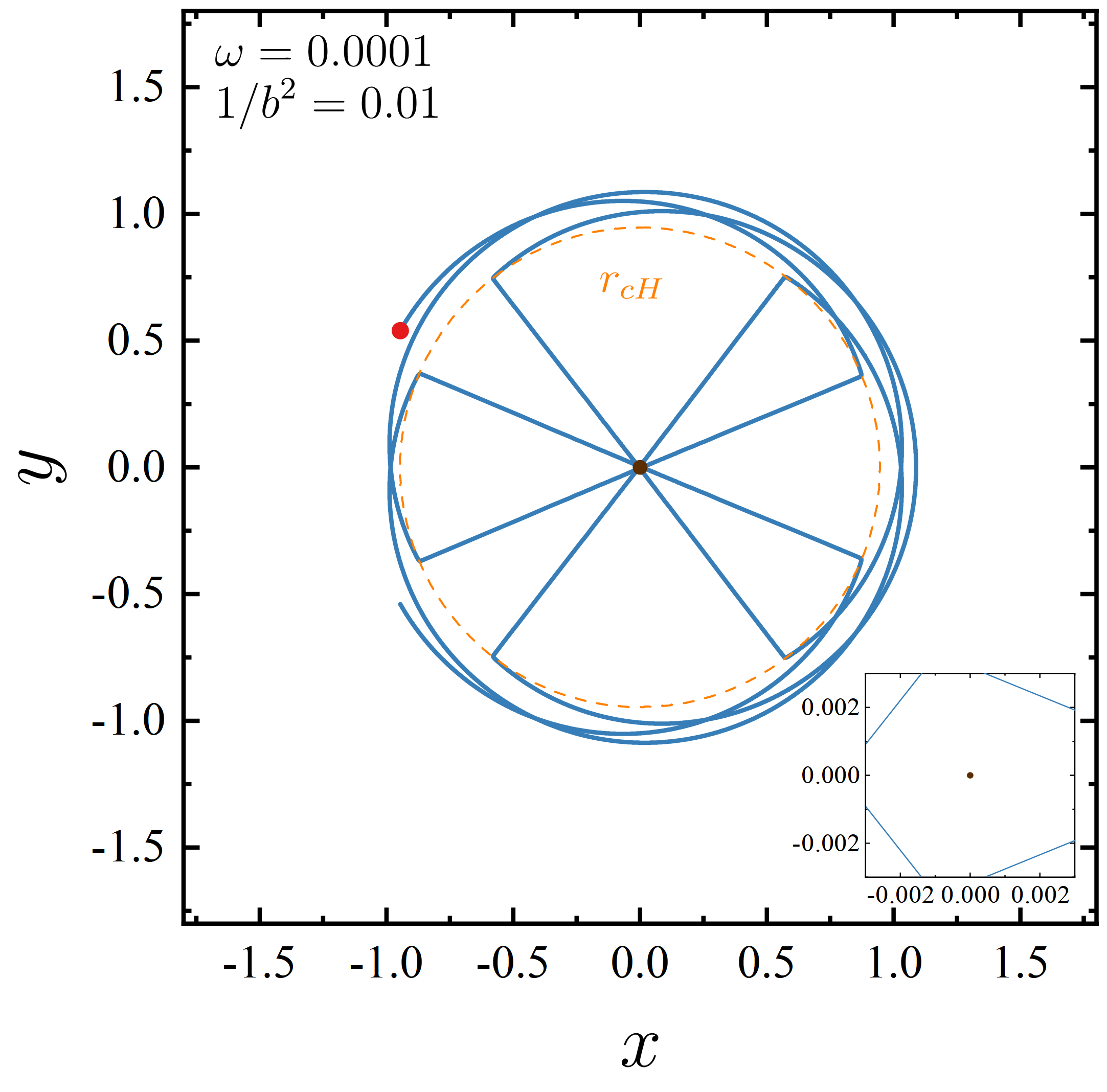}
		}
	\subfigure{
			\includegraphics[height=.20\textheight,width=.20\textheight, angle =0]{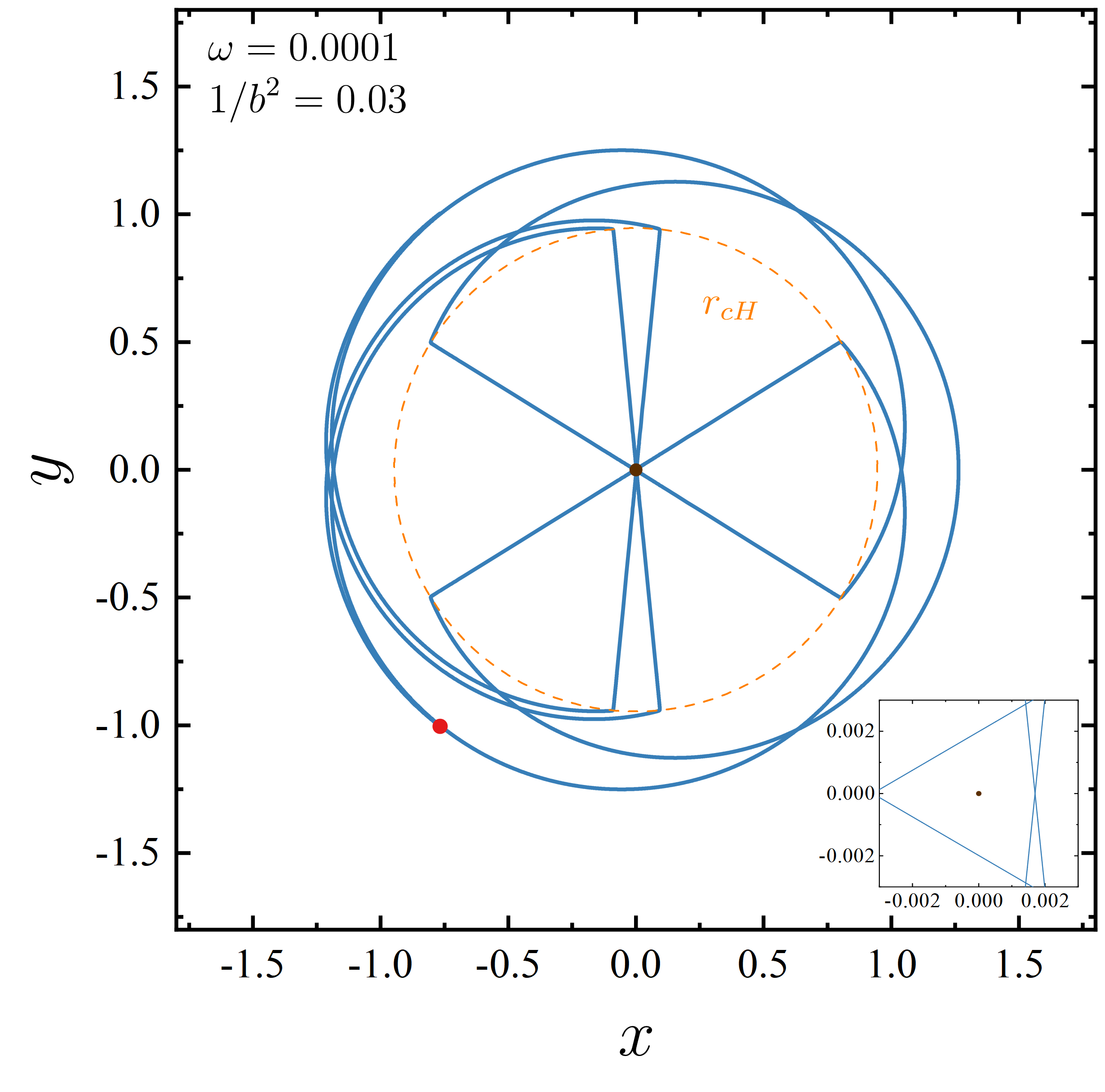}
   		} 	
        \subfigure{
			\includegraphics[height=.20\textheight,width=.20\textheight, angle =0]{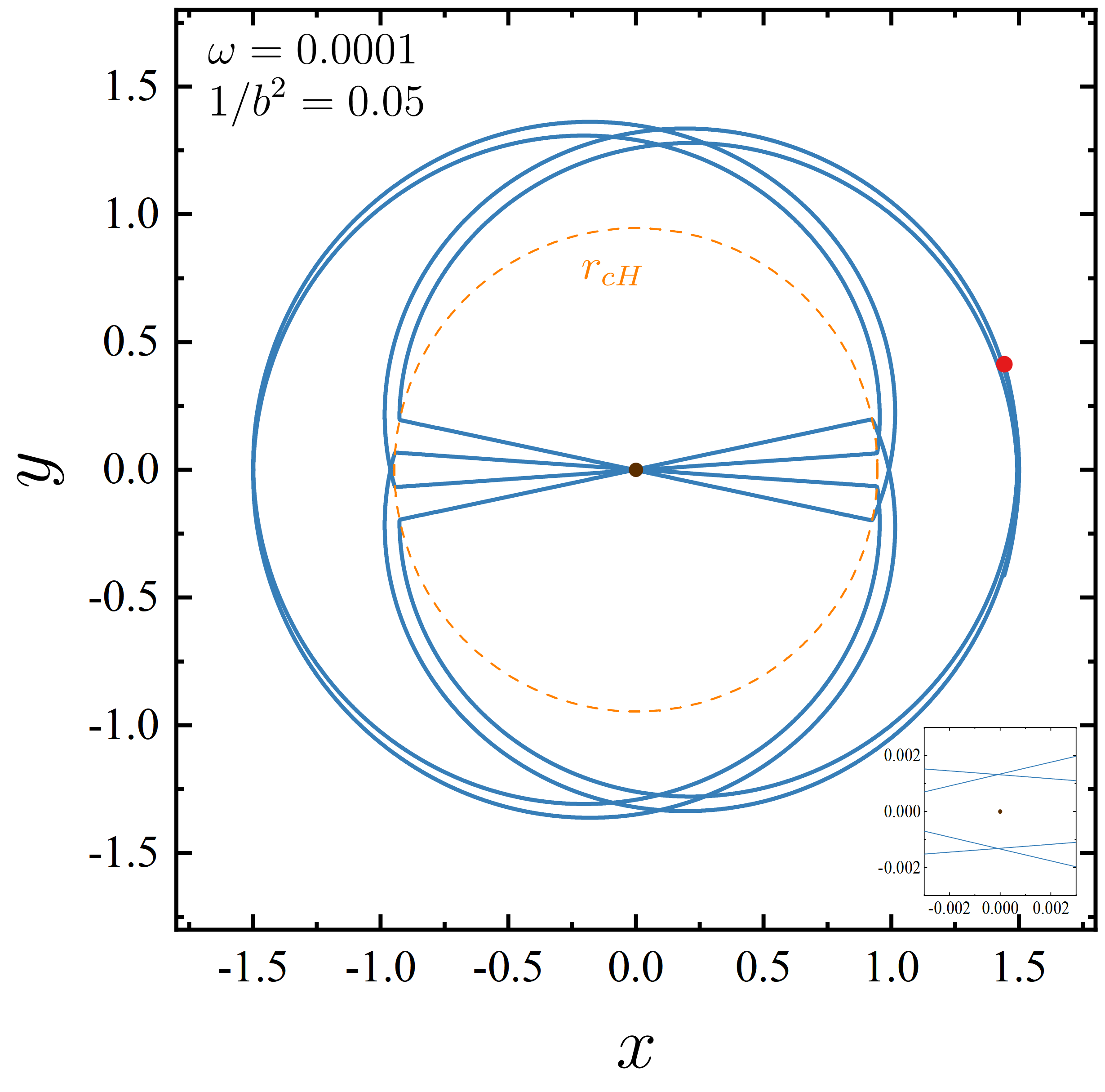}
   		}
		\caption{The bound orbits of the photon for different impact parameters of BBSs with $q=0.7$ and $\omega=0.1$ (top panel), $\omega=0.1$ (middle panel) and $\omega=0.0001$ (bottom panel)}
		\label{fig:FBNBHLR}		
		\end{figure}
%

 The photon oscillates between the periapsis and apoapsis, forming a bound orbit. As mentioned before, only the BBSs with the LR have bound orbits of photons. We take $q = 0.7$ as an example to show some bound orbits in Fig.~\ref{fig:FBNBHLR}. As seen in Fig.~\ref{fig:FBNBHLR}, the bound orbit is symmetric about the line connecting the origin of coordinates and either the periapsis or the apoapsis. Therefore, we have only shown a part of the trajectory, and further trajectories can be obtained based on symmetry. These orbits are influenced by the inverse of the impact parameter $1/b$. This can be seen more clearly from the diagram of the radial effective potential - see Fig.~\ref{fig:EFFFNBHLR} - as $1/b$ decreases, the distance between the two zeros (corresponding to periapsis and apoapsis) of the equation $V=1/b$ continuously decreases, and both move toward the origin of coordinates. That is, as the impact parameter increases, the region where the photon moves around the BBSs becomes smaller. In particular, as seen from the bottom panel of Fig.~\ref{fig:FBNBHLR}, for the FBBSs, its bound orbits are similar to the unbound orbits. The bound orbit will be sharply deflected near $r_{cH}$. Inside the critical horizon, the trajectories of photons form almost a straight line, and the photon can reach a position very close to $r=0$. 

	\begin{figure}[!htbp]
		\centering	
        \subfigure{  
			\includegraphics[height=.28\textheight,width=.30\textheight, angle =0]{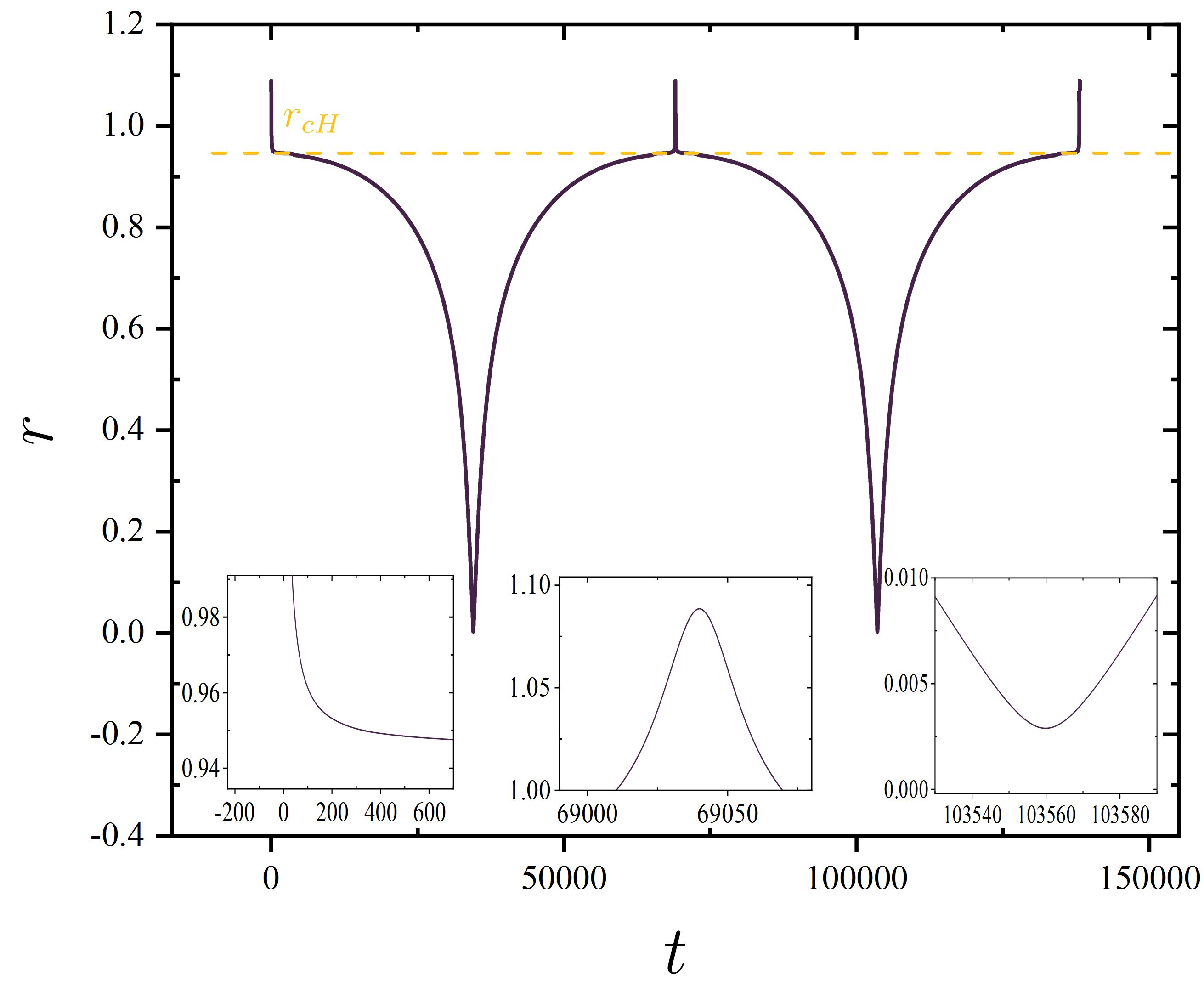}
		}

		\caption{Evolution of the radial distance of the bound orbit in FBBSs with $q=0.7$.}
		\label{fig:RTB070001001}		
		\end{figure}

Figure~\ref{fig:RTB070001001} shows the evolution of the radius of the bound orbit in the background of the FBBSs. We observe from this figure that the radial coordinate as a function of time is a periodic function. Similar to the case of unbound orbits, the photon spends an extremely long time moving inside the critical horizon. As a result, from the perspective of an observer at infinity, the photon's average radial velocity within the critical horizon is very small.

\section{CONCLUSION}\label{sec: conclusion}
In this paper, we studied the orbits of the photon (null orbit) around BBSs using the geodesic equation and the radial effective potential. We found that photons around BBSs can exhibit two types of orbits: bound orbits and unbound orbits. The impact parameter of the photon and the radial effective potential influence the type of orbit it forms around BBSs. Among the orbits formed by photons around BBSs, the circular orbits, i.e., the LR, are a particularly special type. Through the effective potential method, we discovered that BBSs, like some pure BSs, possess two LRs. When the magnetic charge is small, the first solution with the LR appears in the third branch of the $M(\omega)$ curve of the solution. As the magnetic charge increases, the solution can shift to the second branch or even to the first branch, and when the magnetic charge is sufficiently large, all the BBS solutions will possess the LRs.

It is worth noting that among these BBS solutions with the LR, the FBBSs are particularly special. From the perspective of an observer at infinity, the photon spends an extremely long time moving inside the horizon, allowing FBBSs to partially mimic BH models. However, comparing the orbit of photons in the background of the FBBS and Bardeen BH by letting the photons start from the same initial coordinate positions and the same four velocities, we found that the FBBSs are still different from BHs. When the photon does not cross the critical horizon, the difference between the trajectories of photons in the background of the FBBS and the Bardeen BH grows larger as the photons spiral into the FBBS and Bardeen BH. When the photon crosses the critical horizon, despite the sharp deflection of photon trajectories near the critical horizon, these FBBSs still remain a type of the BS model with the nonlinear electromagnetic field and do not possess the event horizon. Their null geodesics remain complete, allowing photons to travel throughout the entire spacetime.

Beyond the content in this work, it is worth mentioning that because the accretion disk composed of timelike particles has an influence on gravitational wave signals, the timelike orbit around the FBBSs will probably reveal more characteristics of FBBSs and have significant observational implications. Thus, we intend to investigate this aspect further in our future work.
\section*{ACKNOWLEDGEMENTS}
	This work is supported by the National Natural Science Foundation of China (Grants No.~12275110, No.~12105126 and  No.~12247101 ) and National Key Research and Development Program of China (Grant No. 2020YFC2201503).

\end{document}